\documentstyle[epsfig]{mn}

\newif\ifAMStwofonts

\ifoldfss
  \ifCUPmtlplainloaded \else
    \NewTextAlphabet{textbfit} {cmbxti10} {}
    \NewTextAlphabet{textbfss} {cmssbx10} {}
    \NewMathAlphabet{mathbfit} {cmbxti10} {} 
    \NewMathAlphabet{mathbfss} {cmssbx10} {} 
  \fi
  \ifAMStwofonts
    \ifCUPmtlplainloaded \else
      \NewSymbolFont{upmath} {eurm10}
      \NewSymbolFont{AMSa} {msam10}
      \NewMathSymbol{\upi}     {0}{upmath}{19}
      \NewMathSymbol{\umu}     {0}{upmath}{16}
      \NewMathSymbol{\upartial}{0}{upmath}{40}
      \NewMathSymbol{\leqslant}{3}{AMSa}{36}
      \NewMathSymbol{\geqslant}{3}{AMSa}{3E}

    \fi
  \fi
\fi 

\ifnfssone
  \newmathalphabet{\mathit}
  \addtoversion{normal}{\mathit}{cmr}{m}{it}
  \addtoversion{bold}{\mathit}{cmr}{bx}{it}
  \newmathalphabet{\mathbfit} 
  \addtoversion{normal}{\mathbfit}{cmr}{bx}{it}
  \addtoversion{bold}{\mathbfit}{cmr}{bx}{it}
  \newmathalphabet{\mathbfss} 
  \addtoversion{normal}{\mathbfss}{cmss}{bx}{n}
  \addtoversion{bold}{\mathbfss}{cmss}{bx}{n}
  \ifAMStwofonts
    \ifCUPmtlplainloaded \else
      \UseAMStwoboldmath
      \makeatletter
      \new@mathgroup\upmath@group
      \define@mathgroup\mv@normal\upmath@group{eur}{m}{n}
      \define@mathgroup\mv@bold\upmath@group{eur}{b}{n}
      \edef\UPM{\hexnumber\upmath@group}
      \new@mathgroup\amsa@group
      \define@mathgroup\mv@normal\amsa@group{msa}{m}{n}
      \define@mathgroup\mv@bold\amsa@group{msa}{m}{n}
      \edef\AMSa{\hexnumber\amsa@group}
      \makeatother
      \mathchardef\upi="0\UPM19
      \mathchardef\umu="0\UPM16
      \mathchardef\upartial="0\UPM40
      \mathchardef\leqslant="3\AMSa36
      \mathchardef\geqslant="3\AMSa3E
    \fi
  \fi
\fi 

\ifnfsstwo
  \DeclareMathAlphabet{\mathbfit}{OT1}{cmr}{bx}{it}
  \SetMathAlphabet\mathbfit{bold}{OT1}{cmr}{bx}{it}
  \DeclareMathAlphabet{\mathbfss}{OT1}{cmss}{bx}{n}
  \SetMathAlphabet\mathbfss{bold}{OT1}{cmss}{bx}{n}
  \ifAMStwofonts
    \ifCUPmtlplainloaded \else
      \DeclareSymbolFont{UPM}{U}{eur}{m}{n}
      \SetSymbolFont{UPM}{bold}{U}{eur}{b}{n}
      \DeclareSymbolFont{AMSa}{U}{msa}{m}{n}
      \DeclareMathSymbol{\upi}{0}{UPM}{"19}
      \DeclareMathSymbol{\umu}{0}{UPM}{"16}
      \DeclareMathSymbol{\upartial}{0}{UPM}{"40}
      \DeclareMathSymbol{\leqslant}{3}{AMSa}{"36}
      \DeclareMathSymbol{\geqslant}{3}{AMSa}{"3E}
    \fi
  \fi
\fi 

\ifCUPmtlplainloaded \else
  \ifAMStwofonts \else 
    \def\upi{\pi}
    \def\umu{\mu}
    \def\upartial{\partial}
  \fi
\fi

\title{Galaxy  number counts - V. Ultra-deep counts: The Herschel and Hubble 
Deep Fields}
\author[N. Metcalfe, et al.]
       {N. Metcalfe$^1$, T. Shanks$^1$, A. Campos$^2$, H.J. McCracken$^1$\thanks{Present address: LAS, Traverse du Siphon, Les Trois Lucs, F-13102 Marseille, France} and R. Fong$^1$ \\
        $^1$Physics Department, University of Durham, South Road, Durham DH1 3LE\\
        $^2$Instituto de Matematicas y Fisica Fundamental, CSIC, Spain}
\date{Accepted 2000.
      Received 2000;
      in original form 2000}

\pagerange{\pageref{firstpage}--\pageref{lastpage}}
\pubyear{2000}

\begin{document}

\maketitle

\label{firstpage}

\begin{abstract}

We present $u,b,r$ \& $i$ galaxy number counts and colours both from the
North and South Hubble Space Telescope Deep Fields and from the William
Herschel Deep Field. The latter comprises a $7'\times 7'$ area of sky
reaching $b\sim$28.5 at its deepest. Following Metcalfe et al. (1996) we show that simple Bruzual \& Charlot
evolutionary models which assume exponentially increasing star-formation
rates with look-back time and q$_0$=0.05 continue to give excellent fits to galaxy counts and colours in
the deep imaging data. With q$_0$=0.5, an extra population of `disappearing
dwarf' galaxies is required to fit the optical counts. 

We further find that the $(r-i):(b-r)$ colour-colour diagrams show
distinctive features corresponding to two populations of early- and
late-type galaxies which are well fitted by features in the Bruzual \&
Charlot models. The $(r-i):(b-r)$ data also suggest the existence of an
intrinsically faint population of early-types at $z\sim0.1$ with similar
properties to the `disappearing dwarf' population required if q$_0$=0.5.

The outstanding issue remaining for the early-type models is the
dwarf-dominated IMF which we invoke to reduce the numbers of $z>1$
galaxies predicted at $K<19$. For the spiral models, the main issue is
that even with the inclusion of internal dust absorption at the
A$_B=0.3$ mag level, the model predicts too blue $(u-b)$ colours for
late-type galaxies at $z\sim1$. Despite these possible problems, we
conclude that these simple models with monotonically increasing 
star-formation rates broadly fit the data to $z\sim3$.

We compare these results for the star formation rate history with those 
from the different
approach of Madau et al. (1996). We conclude that when the effects of
internal dust absorption in spirals are taken into account the results from
this latter approach are completely consistent with the $\tau$=9Gyr, exponentially
rising star formation rate density 
out to $z\approx3$ which fits the deepest, optical/IR galaxy
count and colour data.

When we compare the observed and predicted galaxy counts for UV dropouts
in the range $2\la z\la 3.5$ from the data of Steidel et al. (1999), Madau et
al. (1996) and new data from the Herschel and HDF-S fields, we find
excellent agreement, indicating that the space density of galaxies may not
have changed much between $z=0$ and $z=3$ and identifying the Lyman break 
galaxies
with the bright end of the evolved spiral luminosity function. Making the
same comparison for B dropout galaxies in the range $3.5\la z\la 4.5$ we find
that the space density of intrinsically bright galaxies remains the same
but the space density of faint galaxies drops by a factor of $\sim$5, 
consistent with the idea that L$^*$ galaxies were
already in place at $z\approx4$ but that dwarf galaxies may have formed
later at $3\la z\la 4$.

\end{abstract}

\begin{keywords}
galaxies: evolution - galaxies: photometry - cosmology: observations
\end{keywords}

\section{Introduction}

In four previous papers, Jones et al. (1991) (hereafter Paper I), Metcalfe et
al. (1991) (hereafter Paper II), Metcalfe et al. (1995a) (hereafter Paper III)
and McCracken et al. (2000) (hereafter Paper IV), we used photographic and CCD
data to study the form of the galaxy number-magnitude relation at both optical
($B\sim27.5$) and infra-red ($K\sim20$) wavelengths. In this paper we extend
our observations of the $B$-band field of paper III to even deeper magnitudes
($B\sim28$ mag), and over a much larger area of $7'\times 7'$, by exposing for a
total of 30 hours over five nights with the Tektronix CCD at the prime focus
of the 4.2 metre William Herschel telescope (WHT). We have also acquired
$\sim34$ hours of exposure in the $U$-band (to $U\sim27$ mag), 8 hours in the
$R$-band (to $R\sim26.5$ mag) and 5 hours in the $I$-band (to $I\sim25.5$ mag) on
the same field, again with the WHT. We refer to this field throughout as
the William Herschel Deep Field (WHDF).

Although not competing in terms of either collecting area or field of view
with these ground-based observations, the high resolution of the Hubble Space
Telescope offers the ability to image sub-arcsecond objects to much greater
depths than can be done from the ground. The Hubble Deep Field (HDF) project
(Williams et al. 1996) has provided such a dataset in the public domain, and
here we present our analysis of the $F300W$, $F450W$, $F606W$ and $F814W$ 
preliminary
release North (HDF-N) and South (HDF-S) fields, using similar data 
reduction techniques as for
the WHT data. Roughly speaking, these passbands correspond to $U$, $B$, $R$
\& $I$ (see section 4.2).
For unresolved objects the $3\sigma$ limit of $B\sim29.5$ mag on
the $F450W$ image is considerably deeper than our WHT frame. However, for
resolved galaxies this limit is much brighter. Indeed, for sources more
extended than the WHT `seeing' disk ($\sim 1.25''$ radius) the ground-based frame
is slightly deeper, making it an ideal complement to the HST images. A similar
situation exists between the WHT $u$ and the HDF $F300W$ frames, where, for a
source at the WHT seeing limit, the WHT is actually over $0.5$ mag deeper than
the HDF. The same is not true for the $F606W$ and $F814W$ images, which are
2-3 mag deeper (for typical galaxy sizes) than the deepest previous $R$ and
$I$-band galaxy counts (Smail et al. 1995, Hogg et al. 1997).

The modelling and interpretation of HDF-N data has been discussed briefly
elsewhere (Metcalfe et al. 1996). Here we discuss the data reduction
for both ground- and space-based data, and present the number counts, colour
distributions and models in greater detail. Finally we discuss the comparison 
between the models and  the data.

The optical - infra-red properties of the WHDF, based on infra-red imaging 
data to $K\sim20$ taken on the United Kingdom Infrared Telescope, and
data to $K\sim23$ on one smaller, $1.8'\times 1.8'$ area, are discussed
in Paper IV.

We now also have infra-red imaging data covering the whole of our WHDF to
$H\sim22.5$ mag  taken with the $\Omega$ Prime camera on the Calar Alto 3.5m
telescope. This data will be  discussed in a subsequent paper (McCracken et. 
al., in prep.).

\section{The Observations}

\subsection{New WHT observations}

Our data were taken during four observing runs at WHT prime focus; one of 5
nights in September 1994 during which all of the $b$-band data and about 3hrs
of $r$-band data were acquired, a further run of 4 nights in September 1995
during which the remaining $r$-band data and some of the $u$-band data were
collected, and 3 nights in both October 1996 and September 1997 when the rest
of the $u$-band exposures and all the $i$-band data were taken. Observing
conditions were much better in 1994, when 4 out of 5 nights were photometric,
than in 1995 or 1996, when the observations were affected by dust and poor seeing. The 1997
run, although not completely photometric, had good seeing of 
$\approx1''$ FWHM. For the 1994
and 1995 runs  we used the Tektronix CCD  camera at WHT prime focus, giving a
field size of $7.17'\times 7.17'$ at a scale of $0.42''$ per pixel. The gain
was set to $1.7e^-/ADU$ and the read-noise is about $7e^-$. However, in 1996
we switched to a Loral CCD, with $2048\times2048$ pixels at a scale of
$0.26''$ per pixel and a gain of $1.22e^-/ADU$. In 1997 we again used a Loral
CCD (although not the same chip as in 1996); this time the gain was set to
$1e^-/ADU$. These chips have a much higher (about a factor of three) short
wavelength response than the Tek CCD, as a result of which most of our
$u$-band signal comes from the 1996/7 data.

\begin{figure}
\begin{center}
\centerline{\epsfxsize = 3in
\epsfbox{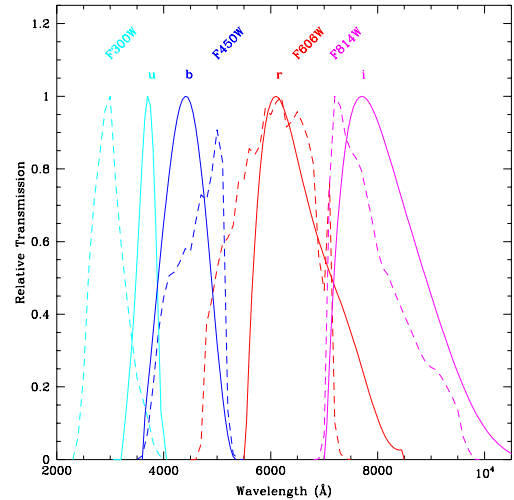}}
\caption{Filter/detector throughputs for the HDF (dashed lines) and
WHDF (solid lines) observations. The WHDF curves include the effects of
La Palma extinction.}
\label{fig:filters}
\end{center}
\end{figure}

The Harris $B$, $R$ and $I$ filters were used - these are much closer to the
standard photoelectric bands (throughout, any references to standard 
photoelectric $R$ and $I$ magnitudes imply those on the Kron-Cousins system)
than the KPNO filters used in papers II and III. The $U$ filter was a 50mm RGO
filter, which caused some vignetting at the edges of the CCD. 
The filter response curves can be seen in Fig. \ref{fig:filters}. To avoid
confusion we use lower-case $u$, $b$, $r$ \& $i$ to refer to the passbands
defined by the WHT filter/detector combination.

\setcounter{table}{0}

\begin{table*}
\begin{minipage}{140mm}
\caption{Details of the WHT and HDF images. The
WHT magnitudes are in $u_{ccd}$, $b_{ccd}$, $r_{ccd}$ and $i_{ccd}$ 
(section 3.1); the HDF-N \& S magnitudes are in the
$vega$ system (section 4.2); the STIS magnitudes are in the AB system. The
definition of effective exposure is given in section 2.1.}

\halign{\rm#\hfil&\hskip 10pt\hfil\rm#\hfil&\hskip 10pt\rm\hfil#\hfil&
\hskip 10pt\rm\hfil#\hfil&
\hskip 10pt\rm\hfil#\hfil&\hskip 10pt\rm\hfil#\hfil&
\hskip 10pt\rm\hfil#\hfil\cr

Frame&Area&Effective exposure&FWHM&$3\sigma$ measurement$^a$&$1\sigma$
isophote$^b$\cr
&deg$^2$&(sec)&($''$)&(mag)&(mag/arcsec$^2$)\cr
\cr
WHT $u$&$1.29\times10^{-2}$&121000&1.35&26.8&29.2\cr
WHT $b$&$1.35\times10^{-2}$&100000&1.25&27.9&30.3\cr
WHT $r$&$1.30\times10^{-2}$&27500&1.50&26.3&28.7\cr
WHT $i$&$1.45\times10^{-2}$&19000&1.20&25.6&26.5\cr
Co-added $b$&$7.9\times10^{-4}$&166000&1.35&28.2&30.6\cr
HDF-N $F300W$&$1.17\times10^{-3}$&170500&0.15&27.0&27.8\cr
HDF-N $F450W$&$1.17\times10^{-3}$&84600&0.15&29.1&29.8\cr
HDF-N $F606W$&$1.17\times10^{-3}$&94650&0.15&29.3&30.1\cr
HDF-N $F814W$&$1.17\times10^{-3}$&94000&0.15&28.4&29.2\cr
HDF-S $F300W$&$1.17\times10^{-3}$&140185&0.15&26.9&27.6\cr
HDF-S $F450W$&$1.25\times10^{-3}$&100950&0.15&29.1&29.8\cr
HDF-S $F606W$&$1.30\times10^{-3}$&81275&0.15&29.3&30.1\cr
HDF-S $F814W$&$1.29\times10^{-3}$&100300&0.15&28.4&29.2\cr
STIS Unfilt.&$1.4\times10^{-4}$&156000&0.15&30.3&31.1\cr}
\noindent$^a$ Inside an aperture with radius equal to the minimum radius given
in Table \ref{tab:parameters}
(magnitude is {\it total} magnitude of an unresolved object)

\noindent$^b$ Inside 1 arcsec$^2$ 

\label{tab:details}
\end{minipage}
\end{table*}

Exposures were centred on $00^h 19^m 59^s.6$ $+00^{\circ} 04' 18''$
(B1950.0). This field encompasses the 24 hour INT and 10 hour WHT
images from paper III, although the centre is about $1.5'$ north of
that of these previous exposures.  Individual $u$ and $b$-band
exposures were usually 2000s, with some reduced to 1000s on occasions
when the seeing was variable. For the $r$ band we used an exposure
time of 500s, and for the $i$ band between 300 and 500s. These short
exposures were necessary in order to avoid saturating the CCDs with
the background sky signal. To avoid excessive atmospheric extinction, 
no $u$-band
frames were taken at an airmass greater than $\sim1.5$, and no
$b$-band frames at greater than $\sim1.8$. The $r$ and $i$-band
frames, where extinction is less of a problem, were mostly taken at
higher airmass than this. The field was offset a few arcseconds in
different directions between nights in order to avoid dud pixels
always falling on the same place on the sky. The total exposure times
were 59.5ks (Tek) + 88ks (Loral) in $u$, 112ks in $b$, 27.5ks in
$r$ and 21.5ks in $i$. The effective exposure times, defined as the
equivalent number of seconds at the zenith in photometric conditions
(with the Loral CCD for the $u$), are  given in Table
\ref{tab:parameters}. The seeing was good in 1994 and 1997, always
less than $2''$ FWHM, and around $0.9''$ on some nights, but ranged
from $1''-5''$ in 1995 and $1.3''-2.0''$ in 1996. The images with the
worst seeing were not used in the data reduction (see section
3.1). Standard stars were taken at the beginning and end of each
night, as were twilight flat-fields. Standards were also occasionally
taken during the night, although we were also able to use our image
frames to monitor atmospheric conditions.

\subsection{The Hubble Deep Fields}

For the HDF-N we use the $F300W$, $F450W$, $F606W$ 
and $F814W$
preliminary `drizzled' WFPC2 images released into the public domain on 1996
January 15 (Williams et al. 1996). These are flat-fielded, cosmic-ray
rejected, sky-subtracted stacked images, resampled to a scale of
$0.04''$/pixel. Each of the three wide-field WFPC2 chips were analysed 
separately. We made no attempt to use the data from the PC chip. 
Note that some 25\% of the observations
are not included in the $F450W$ images and an optimal weighting scheme was not
used for stacking these data, nor for the $F300W$ data. 
For the HDF-S we use
the combined WFPC2 images released in November 1998. We have also analysed
the unfiltered STIS QSO field image.  For unresolved objects the effective 
FWHM is $\sim0.15''$ (see Williams et al. 1996).

\section{William Herschel Deep Field}
\subsection{Data reduction}

A detailed description of our data reduction procedures can be found in Paper
III. Briefly, a master bias frame is formed from a median of many individual
bias exposures and normalised to the bias level in the overscan region of each
exposure. This bias frame is then subtracted. The resulting frames are then 
trimmed to remove the
overscan and divided by a master flat-field, formed from the median of all the
flat fields in the relevant filter throughout the run. Cosmic rays are removed
by comparing each pixel in each frame with the mean and standard deviation in
that pixel over all other exposures (where necessary, normalised and
re-scaled) taken that night. Any pixels more than $4.5\sigma$ away from that
mean are replaced by the mean of the surrounding $5\times5$ pixels in the
image being cleaned. Finally, each image is inspected visually on the computer
 and any remaining blemishes (e.g. satellite trails) removed interactively.
All the individual images are then registered with one another to the nearest
integer pixel shift by doing a pixel-pixel cross-correlation between images.
They are then added together. Frames on which the seeing was equal to or 
greater than $2.0''$ FWHM were discarded.

The final $b$-band image is $998\times991$ pixels ($7.0'\times6.9'$) and has a
sky surface brightness of 21.9 mag/arcsec$^2$. The seeing is $\sim1.25''$ FWHM 
(as in papers II and III this is judged by comparison with a simulated 
Moffat profile (Moffat 1969)). 

The final $r$-band image has $986\times966$ pixels
($6.9'\times6.8'$) and has a sky brightness of 20.0 mag/arcsec$^2$. There was 
a slight rotational shift between the 1994 and 1995
data which was corrected for using the shear technique described in paper III.
The seeing on this frame is $\sim1.5''$ FWHM.

The 1995, 1996 and 1997 $u$-band data were reduced separately. The 1995 data
were then resampled to $0.26''$/pixel and, along with the 1996 data,
shear-rotated slightly in order to match the 1997 data. There was a problem
with the Loral CCD used in 1997 in that it had a $\sim5\%$ non-linear response
at low signal levels. This posed a particular problem for $u$-band deep field
data frames, where even in 2000 sec exposures the sky levels were low.
Extensive exposure tests were carried out with a dome lamp to enable a
correction factor as a function of count rate to be determined. This was then
applied to the data. As a check on this procedure one standard star field was
repeated with a series of exposure times, and detailed comparisons made
between stars covering a range of magnitudes on our field in 1997 and 1995/6.
In fact, the uncorrected and corrected standards gave the same zero-point to
within $0.01$ mag (as expected, as most of the light from these stars comes at
high count rates where the non-linear correction is negligible). However, the
correction did improve the relation between the 1995/6 and 1997 data frames,
removing a $\sim0.05$ mag non-linearity between bright and faint stars. The
three datasets were then stacked together (with appropriate weighting to
allow for the different gain factors, signal strength and sky levels) and all
the image analysis was done on this final combined frame, which encompasses
$1593\times1582$ pixels ($6.9'\times6.9'$) with a `seeing' of
$\approx1.35''$ FWHM. The sky brightness was 21.5 mag/arcsec$^2$.

The 1996 and 1997 $i$-band data were also reduced separately. The
non-linearity in the 1997 Loral CCD was not a problem for the $i$ data, as the
sky levels were well into the linear regime (and as we found for $u$ data the
zero-point from the standards was not affected). As a result no correction was
applied. However, an extra problem with all the $i$-band data was the presence
of fringing. To remove this it was necessary to create a master fringe frame by
taking the median of all the data-frames. This was then scaled to suit each
individual frame (by trial and error) and subtracted.  A small amount of 
rotation and pixel rescaling ($\sim$1\%) were required before 
the two runs could be stacked together. The final $i$-band image 
has $1622\times1747$ pixels
($7.0'\times7.6'$) and a sky brightness of 18.9 mag/arcsec$^2$. It has the best
seeing of all our data, with a FWHM of only $1.2''$.

\begin{figure}
\begin{center}
\centerline{\epsfxsize = 3.5in
\epsfbox{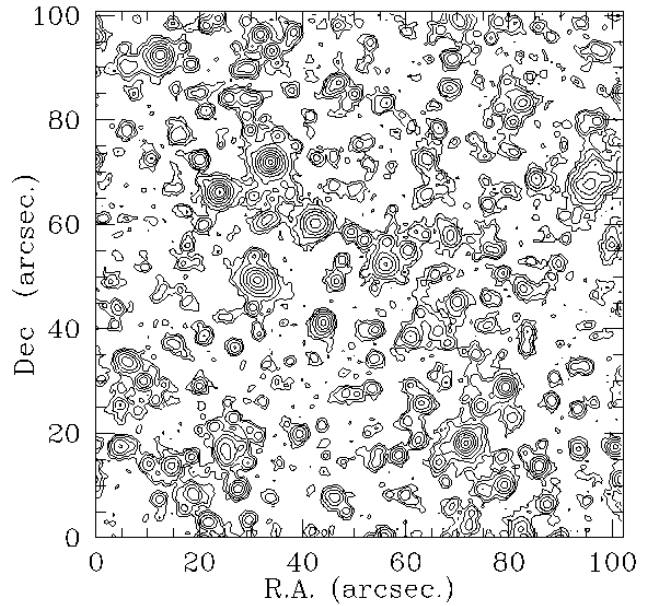}}
\caption{Isophotal contour plot for the 50 hour co-added $b$-band data. 
Contours start at 29.4 mag/arcsec$^2$ and rise in 0.75 mag steps.}
\label{fig:contour}
\end{center}
\end{figure}

Both the present $b$-band data and our 26hr INT prime focus CCD data from
paper III cover the $\sim2'$ diameter area observed for 10hrs with the WHT
auxiliary focus CCD camera (paper III). By stacking all three together in the
common area we have been able to create an ultra-deep image. To achieve this
the INT and WHT auxiliary images were rotated, again using the shear technique
described in paper III, and resampled to the same pixel scale as the WHT prime
image. The three were then registered to one pixel accuracy and added. The
resulting image is $243\times240$ pixels ($1.7'\times1.7'$) (as the WHT
auxiliary image was circular, the effective exposure time is lower in the
corners). The ratios of number of electrons detected in each separate image
from an object are; 3.06 : 1.03 : 1.0 (WHT prime : WHT aux. : INT prime). The
stacked image is therefore equivalent to $1.7\times$ the WHT prime exposure,
i.e. $\sim46$hrs of 4-m time, and is $\sim0.3$ mag deeper. Fig.
\ref{fig:contour} shows a sky-subtracted surface brightness contour plot of 
this co-added image, with a
geometric progression of levels starting at 1000ADU ($29.4$ mag/arcsec$^2$)
and rising by factors of two.

Although not adding to the discussion of counts and colours in this
paper, this small field has the advantage of having been the subject 
of a 30 hour $K$-band exposure taken with IRCAM3 camera (which has an almost
identical field of view) at UKIRT. 
Thses data, and a detailed analysis of the $(b-K)$ colour
distribution, has been presented elsewhere (Paper IV).

Table \ref{tab:details} lists the parameters of all our final 
images.

\subsection{Calibration}

Zero-points for our data were provided by observations of equatorial
standard star fields from Landolt (1992). Mostly they were observed close 
to the meridian. The CCD field of view is large enough to see $5\sim10$
standards on a Landolt field in a single exposure.

As all our WHDF frames are on the same field we adopt the procedure of
calibrating all the frames (in each run) in one band relative to ones taken
on the meridian on photometric nights. This is done by comparing
magnitudes for a selection of bright (but not saturated) objects. 
The standard stars observations then provide an absolute calibration. As the
airmass of the standards and of the WHDF on the meridian are very similar, 
this way we are not sensitive to airmass extinction coefficients. In
fact, by fitting a $\sec$ $z$ law to the relative magnitudes of objects 
on our WHDF field throughout a night's observations, we were able to derive 
atmospheric extinction coefficients appropriate to each run and hence
apply relative airmass corrections to those standards not observed at the
meridian.

Where possible, colour equations measured during the run are used to convert 
the standards to the
`natural' CCD bands (which agree with standard $UBRI$ for objects with zero
colour) and the magnitudes and colours we measure for galaxies are in these
`natural' systems, which we designate $u_{ccd}$, $b_{ccd}$,
$r_{ccd}$ and $i_{ccd}$ (although, as we shall see below, these 
are very close to the standard photoelectric bands).

The $b_{ccd}$ and $r_{ccd}$ magnitudes from the Tek CCD used in 1994
suffered a minor problem 
in that our observed standard star magnitudes appeared to be correlated
with exposure time (at the level of a few hundredths of a magnitude). After
consultation with the RGO (T. Bridges, priv. comm.) it became apparent
that a small additive shutter timing correction was necessary for the WHT 
prime focus camera. The magnitude of this correction was dependent on 
zenith angle, being $\sim +0.05$s for most of our standards.  
Applying this significantly reduced the scatter between the observed and
catalogued magnitudes, and the following 
colour equations were determined from $n$ observations of 28 stars:

$$b_{ccd}=B-0.012(B-R)\hskip0.25in(n=73)$$
$$r_{ccd}=R\hskip0.25in(n=65)$$

\begin{figure}
\begin{center}
\centerline{\epsfxsize = 3.5in
\epsfbox{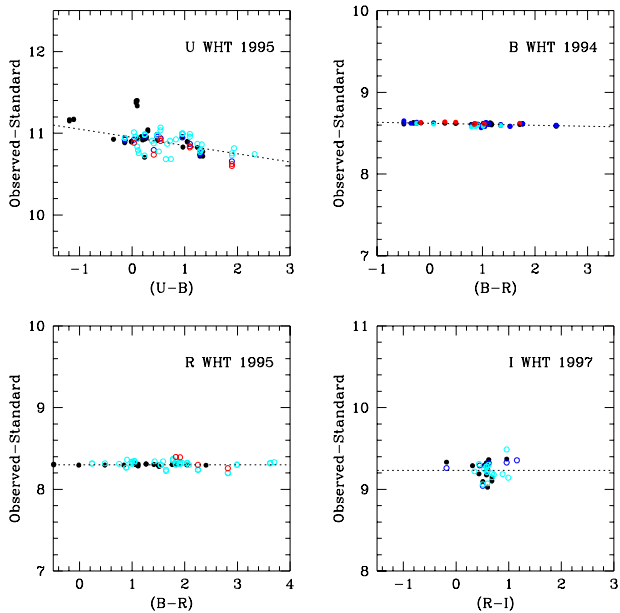}}
\caption{Example standard star calibration plots (catalogue colour $versus$
observed - catalogue magnitude) for the WHT 
$u$, $b$, $r$ and $i$ bands. The dotted lines indicate the adopted colour
equations.
}
\label{fig:calibration}
\end{center}
\end{figure}

The $rms$ scatter about the $B$ and $R$ relations was $\pm0.014$ mag. 
The range in colour covered was $-0.5<(B-R)<2.5$. Fig \ref{fig:calibration}
shows examples of the data on which these relations were based.

In 1995 only one night was demonstrably photometric and this was used to 
calibrate our data independently from the 1994 observations. Again we
found $r_{ccd}=R$, with a scatter of $\pm0.009$ mag from measurements of 23 
standards.The $r_{ccd}$ 
zero-point deduced from this gave magnitudes on the WHDF which
agreed to within $0.02$ mag in the mean from those found in 1994. Note that
no shutter correction was applied to this data, as a series of test exposures 
ranging from 1s to 20s were taken sequentially on the same field during 
photometric conditions and no dependence of the zero-point on exposure time 
was present down to the $0.01$ mag level. 

The data from the $u_{ccd}$ and $i_{ccd}$ standards (whether taken with Tek or
Loral CCD between 1995
and 1997) show a significantly higher scatter than the $b_{ccd}$ or 
$r_{ccd}$ data. The reasons for this are not clear; airmass corrections are 
not the cause because, as already noted, all the standards were taken at 
very similar airmass - also the effect exists between standards on the same 
CCD image and is not due to 
variations from night to night (or within a single night). 

All the 1995 $u$-band data were tied to the one
photometric night, although frequent standards were taken every night (the
maximum variation in zero-point from night-to-night was $0.2$ mag), and we were
able to utilise most of these to derive an approximate colour equation of

$$u_{Tek}\approx U-0.1(U-B)\hskip0.25in(n=96)$$

\noindent valid for $-1.5<(U-B)<2.5$, with a scatter of $\pm0.08$ mag, excluding
two standards whose magnitudes were confirmed by repeat measurements 
to be $\sim0.4$ mag off the mean line. Fig \ref{fig:calibration} shows these
data.

The 1996 and 1997 $u$-band standards taken with the Loral CCD showed a
similar scatter with respect to the Landolt magnitudes of $\sim\pm0.10$ mag. 
This excludes the two standards which 
were discrepant in the 1995 Tek data, which were again found to be off by 
$\sim0.4$ mag. Significantly, if we compare frames with several standards on 
taken both in 1996 and 1997, the scatter between the 1996 and 1997 data was 
only $\pm0.03$ mag. A similar exercise between fields in common to
1995 and 1996/7 datasets gave $\pm0.06$ mag. 
The Loral data do not show a colour term as large as the $0.1(U-B)$ found for
the Tek data, and the scatter prevents any reliable estimate of a shallower
relation. However, the standards on fields in common between the Tek and Loral
observations indicate that there is a significant colour term between  
these two CCDs. We estimate this as 

$$u_{Tek}-u_{Loral}\approx-0.07(U-B)\hskip0.25in(n=25)$$

\noindent which reduces the scatter to $\pm0.04$ mag.
As the 1996/7 data dominate the stacked image in terms of signal-to-noise we
adopt a colour equation of

$$u_{ccd}\approx U-0.03(U-B)$$

\noindent as appropriate for our final dataset (for $-1.5>(U-B)>2.5$).

Although there is clearly some uncertainty in the $u_{ccd}$ zero-point, when
the 1995, 1996 and 1997 WHDF data were calibrated and reduced independently, 
the magnitude scales agreed to within $0.03$ mag. 

The scatter in the $i_{ccd}$-band standards ($\pm0.10$ mag) appears partly due 
to non-photometric variations on some nights (especially in 1997) at the 
$0.05$ mag level. There is also some evidence of inaccuracies in some of the 
fainter Landolt standards on the fields we used (and many of the brighter 
standards were unusable due to saturation), as
standards on one field taken in both 1996 and 1997 show
the same offsets with respect to the Landolt magnitudes (up to $0.2$ mag, with
no correlation with colour), and a scatter of $<0.01$ mag between the
two years. Fig. \ref{fig:calibration} shows the data
from 1997.
As a result of these uncertainties, and the short colour range in $(V-I)$,
we make no attempt to fit a colour equation but just assume $i_{ccd}=I$.

Fortunately, in 1995 we observed an area encompassing the WHDF in the 
$I$-band with a Tek CCD on the INT 2.5m telescope. These independent data show
an offset of only $\sim0.04$ mag with the WHT data, giving us confidence in
our zero-point (although  any possible problems with the standard
magnitudes would affect these data as well). 

\begin{table*}
\begin{minipage}{140mm}
\caption{Parameters used in the WHT and HDF image analysis. As
in Table \ref{tab:details}, WHT magnitudes are in the natural ccd system,
whilst the HDF magnitudes are on the $vega$ system.}
\halign{\rm#\hfil&\hskip 10pt\hfil\rm#\hfil&\hskip 10pt\rm\hfil#\hfil&
\hskip 10pt\rm\hfil#\hfil&
\hskip 10pt\rm\hfil#\hfil&\hskip 10pt\rm\hfil#\hfil\cr
Frame&Limiting isophote&Limiting isophotal&Minimum radius&Kron multiplying&
Correction to total\cr
&(mag/arcsec$^2$)&magnitude&($''$)&factor&(mag)\cr
\cr
WHT $u$&29.75&28.0&1.30&1.40&0.32\cr
WHT $b$&31.0&29.0&1.25&1.40&0.34\cr
WHT $r$&29.0&27.3&1.4&1.50&0.30\cr
WHT $i$&28.0&26.8&1.25&1.44&0.28\cr
Co-added $b$&32.0&30.0&1.375&1.50&0.33\cr
HDF-N $F300W^a$&&&&&\cr
HDF-N $F450W$&28.5&29.5&0.35&2.0&0.11\cr
HDF-N $F606W$&28.7&29.7&0.35&2.0&0.11\cr
HDF-N $F814W$&27.7&28.9&0.35&2.0&0.11\cr
HDF-S $F300W^a$&&&&&\cr
HDF-S $F450W$&28.5&29.5&0.35&2.0&0.11\cr
HDF-S $F606W$&28.7&29.7&0.35&2.0&0.11\cr
HDF-S $F814W$&27.7&28.9&0.35&2.0&0.11\cr
STIS Unfilt.&28.50&31.0&0.35&2.0&0.11\cr
}
\noindent$^a$ $F450W$-band detections used for the $F300W$-band - see text.
\label{tab:parameters}
\end{minipage}
\end{table*}

\subsection{Image analysis}

Our procedure is very similar to that used in Papers I-IV (note that the data
in each passband are reduced independently); the sky background
is removed and images are detected isophotally using the limits given in Table
\ref{tab:parameters}. These images are then removed from the frame, replaced
by a local sky value, and the resulting frame smoothed heavily before being
subtracted from the original. This produces a very flat background. The
isophotal detection is then repeated. In order to reduce false detections,
images whose centres lie only two pixels apart are recombined into one single
image. A Kron-type pseudo-total magnitude is then calculated for each image,
using a local value of sky.

One complication with both our $u$-band and $b$-band exposures was the
presence of scattered light (at the 1\% level in $b$ and 10\% in $u$) on
one side of the field. This had a sharp edge which proved difficult to
remove by the above background flattening technique. It therefore
proved necessary to remove this interactively from the frame.

Table \ref{tab:parameters} shows the limiting isophotes 
and magnitudes for our isophotal detection routine and the minimum radius,
radius multiplying factor (see paper III) and correction to total 
magnitude for our 
Kron magnitudes. As in our previous papers the minimum radius is set to be 
that for an unresolved image of high signal-to-noise, and the correction to 
total is the light outside this minimum radius for such an image. 
Our measurement limits (table \ref{tab:details})
give the total magnitudes of unresolved objects which are a
$3\sigma$ detection inside the minimum radius.

As in paper III we measure fixed aperture colours. For all the WHT data we use
an aperture of $1.5''$ radius, and correct the measured colours for the
difference in `seeing' between the different bands (this is a small effect,
and is estimated from inspection of stellar profiles). Images detected in one
band are matched with those detected in the other within $\sim1''$ of the
position in the first band. Any multiple matches which occur are decided by
visual inspection. If either image is below the $3\sigma$ magnitude 
measurement limit for that band then no colour is measured.

We can compare $(B-R)$ colours between our new WHT data and
our INT results in Paper III on an object by object basis. We find a scatter
in ($B-R$) rising from $\sim\pm0.1$ mag for $22<B<24$ mag to $\sim\pm0.5$ mag for
$25<B<26$ mag. The much noisier INT data will, of course, dominate the
scatter.

\subsection{Star/galaxy separation}

Star-galaxy separation was done on the $b$-band frame using the 
difference between
the total magnitude and that inside a $1''$ aperture, as described in paper
II. This enabled us to separate to $b\sim24$ mag. Some additional very red
stars were identified from the $r$ and  $i$ frames in similar fashion.

\begin{figure}
\begin{center}
\centerline{\epsfxsize = 3.5in
\epsfbox{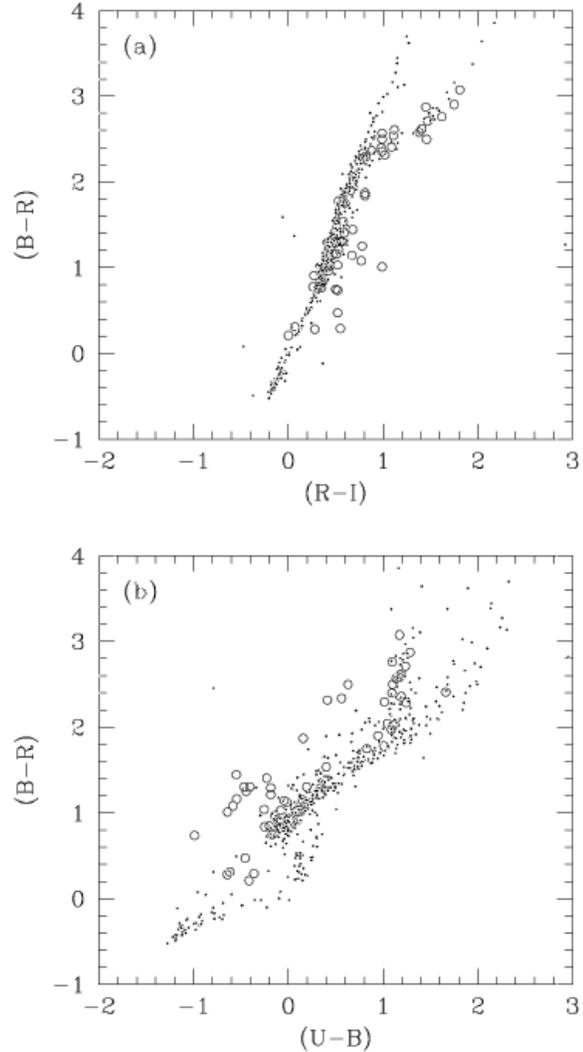}}
\caption{(a) $(R-I):(B-R)$ colour-colour 
diagrams for objects identified as stars on our data (open circles) compared
with the Landolt (1992) observations of standard stars (dots). Our 
magnitudes have been corrected to the standard passbands using the colour
equations in section 3.2; (b) as (a) but now for $(U-B):(B-R)$}
\label{fig:standard stars}
\end{center}
\end{figure}

The colours of the stars can be used as an external, additional check on the
accuracy of our calibration. Fig. \ref{fig:standard stars} shows the 
$(U-B):(B-R)$ and $(R-I):(B-R)$ colour-colour diagrams for the 
unsaturated stars on our frames. Also shown are the colours of the 
Landolt (1992) standard stars.
The agreement between the stellar loci is at the $0.1$ mag level or better for
all the colours. We note that we have no stars on the giant branch and that
nearly all our stars are type G or later. There are, however, quite a sizeable
number of stars which lie well away from the expected locus, in the sense that
for their $(B-R)$ colour they are too blue in $(U-B)$ and, sometimes, too 
red in
$(R-I)$. Visual inspection reveals that most of these are isolated objects (so
not errors due to confusion), and their appearance concurs with their
automated identification as stellar. They tend to be amongst the fainter
stars identified. These may be very compact galaxies,
subdwarfs or QSO's.

\subsection{Comparison with our previous published data}

\begin{figure}
\begin{center}
\centerline{\epsfxsize = 3.5in
\epsfbox{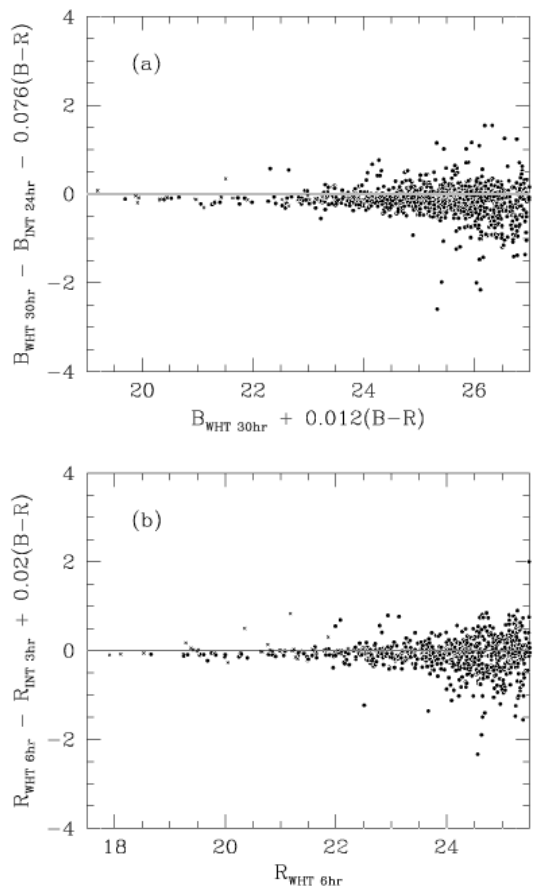}}
\caption{A comparison between magnitudes from this work and 
those from our previous observations on part of this field: (a) WHT $B$-band 
against
INT 24hr $B$-band, (b) WHT $R$-band against INT 3hr $R$-band. Colour equations
have been used as indicated on the axes.
}
\label{fig:mag comparison}
\end{center}
\end{figure}

In Fig. \ref{fig:mag comparison} we show the comparisons between our $B$
and $R$ INT data from paper III and our new data on this field. Colour
corrections have been included to put all the data onto the standard
photoelectric system. We find for the mean offset and {\it rms} scatter for
over 500 objects

$$B(wht) - B(int) = -0.10\pm0.20\hskip0.25in B<25$$
$$R(wht) - R(int) = -0.06\pm0.19\hskip0.25in R<23.5$$

It is clear that there is significant offset between the new and old data, in
the sense that the new magnitudes are brighter.

We have made a detailed comparison with our original short exposures on this
field (paper II), both for aperture magnitudes and for Kron total magnitudes,
in order to determine the source of this discrepancy. It would appear that
the major problem comes from an error in the exposure time recorded in 
the header of one of the original $B$ CCD frames on
this field described in Paper II. 
This accounts for $0.06$ mag, and propagates to the
Paper III data, as this was calibrated using the results from Paper II. The
remaining offset in $B$, and that in $R$, appears to be a combination of 
uncertainties in the zero-pointing the Paper III data using the Paper II
results, slight
errors in the correction to total magnitudes in paper III and genuine 
calibration disagreements. All these effects are individually only a few
hundredths of a magnitude, and within their 
respective error limits, but unfortunately appear to have all summed 
in the same direction.

From the identified sources of error, we believe that the
magnitude scales for both the WHT and INT data of paper III should be
brightened by $0.08$ mag in $B$ and $0.04$ mag in $R$. 
Whenever subsequently referred
to, these data have been corrected by this amount.
As the counts in Paper II were an average of 12 fields, and the timing 
error was only present on the one frame, these
results are unaffected at the $0.01$ mag level.

\begin{table}
\caption{Results of adding artificial stars to the real WHT
data frames; (a) for the 30hr $b$-band frame; (b) for the 
co-added $b$-band frame;
(c) for the 6hr $r$-band frame; (d) for the stacked $u$-band data; (e) for the
stacked $i$-band data.}
\halign to\hsize{%
\hfil\rm#\hfil&\hskip 10pt\hfil\rm#\hfil&\hskip 10pt\rm\hfil#\hfil\cr
True Magnitude&Measured Magnitude&Detection rate(\%)\cr
\noalign{\hbox{(a)}}
25.75&$25.60\pm0.18$&85\cr
26.25&$26.11\pm0.20$&79\cr
26.75&$26.64\pm0.25$&70\cr
27.25&$27.23\pm0.31$&59\cr
27.75&$27.83\pm0.46$&43\cr
\noalign{\hbox{(b)}}
25.75&$25.59\pm0.21$&90\cr
26.25&$26.11\pm0.21$&78\cr
26.75&$26.62\pm0.24$&71\cr
27.25&$27.16\pm0.35$&63\cr
27.75&$27.77\pm0.39$&50\cr
28.25&$28.16\pm0.48$&46\cr
\noalign{\hbox{(c)}}
23.75&$23.67\pm0.13$&88\cr
24.25&$24.15\pm0.20$&86\cr
24.75&$24.74\pm0.27$&78\cr
25.25&$25.35\pm0.41$&70\cr
25.75&$25.78\pm0.39$&60\cr
\noalign{\hbox{(d)}}
23.50&$23.42\pm0.11$&99\cr
24.00&$23.94\pm0.10$&96\cr
24.50&$24.46\pm0.17$&94\cr
25.00&$24.96\pm0.16$&94\cr
25.50&$25.50\pm0.26$&93\cr
26.00&$26.05\pm0.34$&86\cr
26.50&$26.53\pm0.41$&63\cr
\noalign{\hbox{(e)}}
22.75&$22.69\pm0.13$&94\cr
23.25&$23.14\pm0.18$&93\cr
23.75&$23.68\pm0.19$&86\cr
24.25&$24.19\pm0.29$&85\cr
24.75&$24.79\pm0.41$&75\cr
25.25&$25.37\pm0.51$&48\cr
}
\label{tab:artificial}
\end{table}

\subsection{Completeness corrections and simulations}

\begin{figure*}
\begin{center}
\centerline{\epsfxsize = 5in
\epsfbox{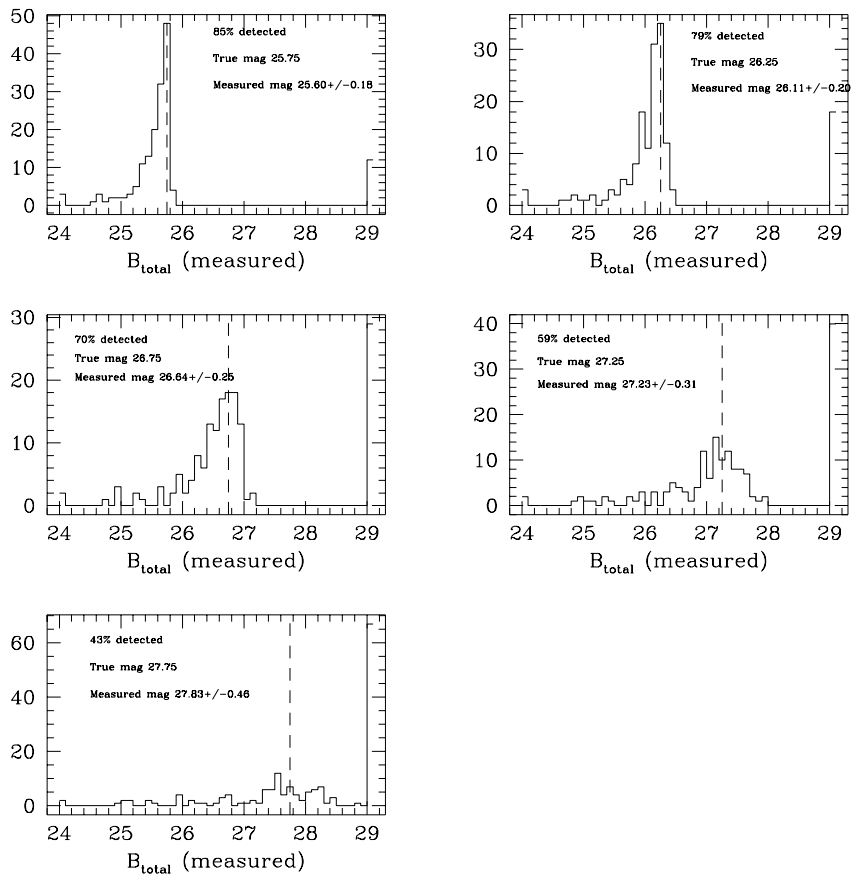}}
\caption{The results of adding simulated stars of known 
magnitude to the real WHT 30 hour $b$-band frame. Histograms of measured 
magnitude are shown for 162 simulations of stars of each of five known 
magnitudes (indicated by the dashed vertical lines). Non-detections are 
placed in the faintest bin.
}
\label{fig:simulations}
\end{center}
\end{figure*}

The deeper the ground-based number-counts are pushed the more important the
corrections for confusion and incompleteness become. The magnitude limits
in Table \ref{tab:details} are just theoretical calculated values for
isolated images and are based on measured sky noise, and take no account of the
efficiency of the detection procedure or the effects of confusion. To make a
better estimate of the completeness of our counts we have added numerous
artificial stars
of various known magnitudes to our real data frames and subjected them to
the normal
data reduction procedure. Fig. \ref{fig:simulations} shows the distribution 
of measured magnitudes for these stars for the 30hr $b$-band data. 
Table \ref{tab:artificial} gives the
mean magnitudes, scatter, and detection rate for this data and for the stacked
$b$-band frame and the $r$-, $u$- and  $i$-band frames. 
Note that an image is considered undetected if
it is merged with another image and the combined brightness is a factor
two or more greater than its true magnitude, or if it is not found within
$\pm2$ (Tek) or $\pm3$ (Loral) pixels of its true position.
As expected the detection rates in the real data drop as the magnitude becomes
fainter.
This is almost entirely due to images being merged with other, brighter images.
Hence the higher detection rates in both the $r$-band and the
$i$-band data, where the density of
objects is lower, despite the similar signal-to-noise ratios.

\begin{figure}
\begin{center}
\centerline{\epsfxsize = 3.5in
\epsfbox{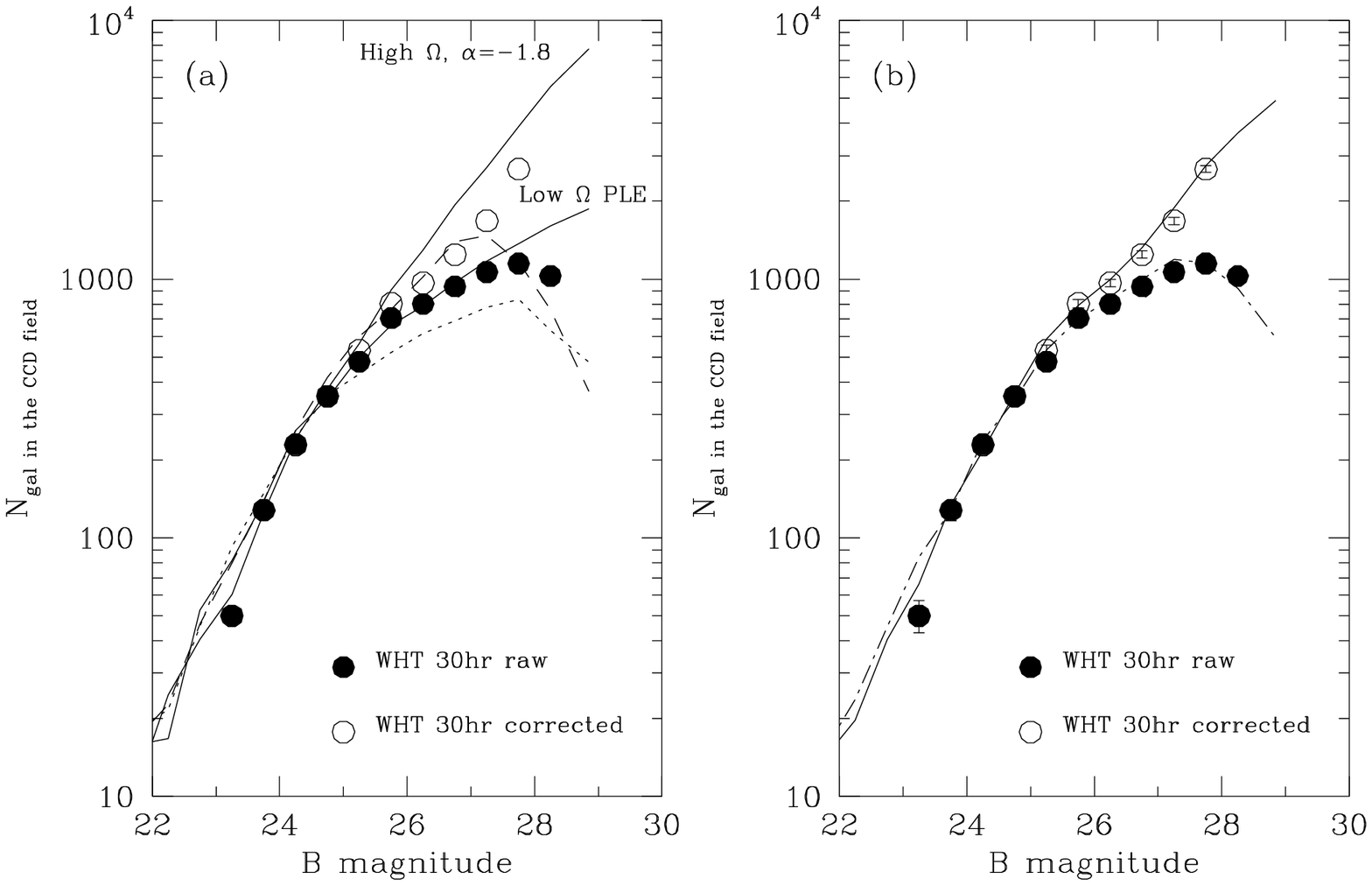}}
\caption{(a) Closed and open circles represent raw and corrected (on the basis
of the simulated star results) galaxy counts from the WHT $b$-band data 
compared
with true (solid line) and measured (dot-dashed lines) counts from the two
simulations discussed in Paper III. (b) As for (a) but now showing our best
fit simulation, where the true galaxy count has been adjusted to give a
measured count in agreement with the data.
}
\label{fig:simulated counts}
\end{center}
\end{figure}

Although the above procedure gives a reasonable estimate of the measurement of
real images it tells us little about whether we are losing low surface
brightness, extended galaxies, and nothing about the number of false
detections we may pick up. Some of these may be genuine noise spikes, but most
are actually caused by the effect of noise causing the software to deblend
single images into multiple components. The best way to account for all these
effects is to create full simulated CCD images, as described in paper III, for
each evolutionary cosmological model of interest and run the image detection
and analysis software on these. The results can then be compared with the real
data. Ideally this would require a knowledge not only of how galaxies evolve
in luminosity and number, but also in morphology. In practice, such detailed
information is not available; however, much can be inferred by treating
galaxies as ideal bulges and disks and we have therefore re-created the two
evolving $B$-band simulations described in paper III, with the appropriate
parameters for our 30hr WHT data (note that in order to simulate the
noise correctly the simulations include a
contribution from sources fainter than the measurement threshold). 
Fig. \ref{fig:simulated counts} displays the
true and measured counts from the two models, together with that from our real
data, both raw and corrected for the detection rates in Table
\ref{tab:artificial}. Two points emerge from this. First, the corrections
based simply on the detection rates appear very close to those inferred from
the full simulations. To demonstrate this further we show in Fig.
\ref{fig:simulated counts} the results of a simulation in which we have
adjusted the true count to be very close to the real count as implied by the
detection rates. The measured count from this simulation is in reasonably good
agreement with the raw data, apart from the faintest points where the model
falls off faster than the data. We suspect this is a limitation of the models
rather than an indication that the count slope suddenly rises again. Second,
the true counts appear to lie somewhere between our two models. In particular,
that model favoured in Paper III, with the steep luminosity function slope at
high redshift, is now seen to overpredict the number of galaxies.

The corrections applied to the counts for the 30hr $b$-band frame are 
based on this
simulation. For the other bands, i.e the co-added $b$-band frame, where with 
such a small area the corrections are particularly susceptible to the actual
arrangement of images, and the $u$-, $r$- and $i$-band frames which are not 
quite as deep,
we appeal to the agreement between the $b$-band detection rate corrections and
those from the full simulations and use the detection rate corrections
determined for each band (Table \ref{tab:artificial}) to correct the counts.

None of the above tests account for the problem of the halos of bright images,
which tend to be broken into many faint images by isophotal detection
algorithms. To some extent the measurement of a local sky for the Kron
magnitudes reduces this effect, but it still proved necessary to inspect all
images around the bright stars and galaxies in our frames by eye and remove
detections judged to be false. The effect this has on the counts is small
(always $<10\%$), but it is much more important when attempting to measure the
clustering of galaxies.

\section{The Hubble Deep Fields}

\subsection{WFPC2 Data Reduction and Image Analysis}

The HDF-N and HDF-S images were reduced in similar fashion, except where 
noted.
 
In order to facilitate image detection, we rebinned the images into $0.08''$
pixels and trimmed off (HDF-N) or blanked-out (HDF-S) the regions of lower
signal-to-noise around the edge of the frames caused by the dithering
technique used in the observations. We then followed a similar procedure to
our WHT reductions. A 3-D polynomial of up to $3^{rd}$ order was fitted to the
sky background and subtracted from the data. An isophotal image detection
algorithm was then run over the data and a smoothed version of the data with
the detected images removed was subtracted from the original. The detection
algorithm was then re-run on this flat-background frame. We used the
zero-points issued with the data-release as the initial basis for our
photometry.

Normally we would then use our isophotal detections as a basis for measuring
total magnitude using our Kron-style algorithm. However, the HDF data,
especially at the shorter wavelengths where for most galaxies we are looking
far into the rest-UV, suffers from the `problem' that most of the bright
spiral and irregular galaxies are broken into numerous bright knots, each
detected as separate images by the software. It was therefore necessary to
re-assemble these galaxies, and we chose to do this by visual inspection on
the frames of all pairs of objects whose centres were closer than $0.35''$. On
average, this resulted in $\sim100$ re-assembled objects per frame, composed
of an average of 4 sub-images, out of about 2500 detected objects. Inevitably
it is ambiguous as to whether some images should be re-assembled or not - many
of them do not resemble `normal' galaxies. In making our judgements we have
borne in mind that an angular separation of $0.1''$ never exceeds a true
separation of $\sim400$pc for $q_0=0.5$ and is unlikely to reach even twice
this for low $q_0$. It is difficult to conceive of images this close as separate
entities.

Once the `broken' images have been reassembled we run our Kron magnitude
software to determine total magnitudes. Tables \ref{tab:details} and
\ref{tab:parameters} give details of the magnitude limits and isophotal and
Kron parameters adopted for the HDF fields.With our choice of minimum radius
($0.35''$) and multiplying parameter (2.0) we should detect about 90\% of the
light from resolved or unresolved objects. However, this choice of minimum
radius is not optimal for a star (for which a $0.15''$ radius would result in
a $0.5$ magnitude improvement in detection limit), but a compromise between
signal-to-noise and the desire not to underestimate the magnitudes of faint,
resolved galaxies. Many of these are likely to have their radii set to the
minimum value just due to the noise in calculating the Kron radius.

As the $F300W$ frames are much less deep than the others (due to the poor UV
response of the WFPC2 CCDs) it was decided to use the $F450W$ image detections
as input to the Kron magnitude routine for these frames, rather than those
from the $F300W$ image.

For an unresolved image, the total magnitudes which give a $3\sigma$ 
detection inside our minimum radius of $0.35''$ are given in Table
\ref{tab:details}. We emphasise the point that these limits will be brighter
for resolved galaxies.

Although the PC data is included in the HDF-S frames (and went through the
reduction procedure), we exclude this area from our final sample due to the
much lower signal-noise on this chip. The final areas for the HDF-S and
combined HDF-N data are listed in Table \ref{tab:details}. These include a
small loss caused by images being too close to the edge of the fields to be
measured.

Due to the vastly improved resolution and pixel scale the HDF images do not
suffer from confusion losses due to crowding in the way the ground-based data
does, and no correction has been applied. Incompleteness does occur at the
faintest magnitudes due to the offset and scatter between our isophotal and
total magnitudes, and due to isophotal effects. These are discussed in
Section 4.3.

As with the ground-based data we use fixed apertures to measure colours 
(section 3.3), but with a much smaller radius of $0.35''$. No relative `seeing' 
correction is required.
To match images we adopt a slightly different strategy to that for the WHT
data, in that we use the positions of the images detected in one band as input
to the Kron magnitude measuring routine on the frames in the other bands (with
the one exception of the $F300W$ band, where, as noted above, we already 
use the
$F450W$ detections). The reason for this is that, due to the high resolution of
the HDF and the irregular nature of many of the images, the positions of
images detected in one band may not coincide precisely with those detected
independently in another. One consequence of this is that we attempt a
measurement of the colour for all the images detected in a particular band,
irrespective of whether the corresponding object is below the detection limit
in the other bands.

\subsection{WFPC2 Magnitude Systems}

To compare the HST results with ground-based data it is necessary to make 
some form of conversion from WFPC2 magnitudes (throughout this paper we
zeropoint the HDF data onto the $vega$ system, defined so that an A0 star
has zero colour) into the standard
$U, B, R \& I$ bands. This can only be done approximately, as none 
of the HDF filters are particularly close to their more standard counterparts,
and in general exact colour equations have not been measured. Note that
for our cosmological models we always use the correct filters and do not 
rely on these transforms.

As a starting point we adopt the synthetic colour transforms of 
Holtzman et al. (1995), except for $F814$ where we use their observed values. 
For $F450W$, $F606W$ and $F814W$ (we
defer discussion of the $F300W$ band until the end of this section) the
three Holtzman et al. (1995) equations we use are
\begin{eqnarray*}
B&=&F450_{vega}+0.23(B-V)-.003(B-V)^2\\
V&=&F606_{vega}+.254(V-I)+.012(V-I)^2\\
I&=&F814_{vega}-.062(V-I)+.025(V-I)^2
\end{eqnarray*}

\noindent
However, we really need these relations in terms of HDF colour, not ($B-V$)
and ($V-I$), and to introduce a conversion for the $R$ band. The 
$I:F814_{vega}$  relation can be deduced in terms of ($F606-F814$)$_{vega}$ 
from the above equations, but to proceed 
further we use the following two 
{\it approximate}
relations between the standard photoelectric bands which we
have derived from the Landolt (1992) list of standard stars
\begin{eqnarray*}
(V-R) &\approx& 0.58(B-V) \\
(V-I) &\approx& 1.95(V-R)
\end{eqnarray*}
\noindent These relations are accurate to no better than 0.05 mag, but this 
is adequate for our purposes as this error gets multiplied by the colour 
coefficients, most of which are small. All five 
equations apply only for stars with $(B-V)\la1.4$ and $(V-I)\la2.0$. 

Combining all the above (approximating second order terms),
we can deduce
$$(B-V)\approx0.94(F450-F606)_{vega}$$
\noindent and
$$(V-I)\approx1.44(F606-F814)_{vega}$$
\noindent and hence,
\begin{eqnarray*}
&B \approx F450_{vega}+0.22(F450-F606)_{vega}\\
&R \approx F606_{vega}-0.37(F606-F814)_{vega}\\
&I \approx F814_{vega}-0.07(F606-F814)_{vega}
\end{eqnarray*}
\noindent For the median colour of the HDF galaxies (with $F814_{vega}<28$) 
these approximate to
\begin{eqnarray*}
&B \approx F450_{vega}+0.1\\
&R \approx F606_{vega}-0.1\\
&I \approx F814_{vega}
\end{eqnarray*}

Note that the $F450W$ colour equation is almost identical to that used 
in Papers
I, II and III for the photographic $b_j$ band. Hence 
$$b_j\approx F450_{vega}.$$ 
\noindent The F606W band is midway between the $V$ and
$R$ passbands, and so has a large colour transform to either. Here we
have chosen to convert to $R$, as there are very few published $V$-band
counts.

The $F300W$ observations are more of a problem. This band is significantly
shorter in wavelength than photoelectric $U$ and the Holtzman et al. synthetic
colour transformation is large (varying by $\ga1$ mag for $-1 < (U-B) < 1$)
and may be multi-valued at $(U-B)\sim0.0$. According to Holtzman et al. there
is also some question of the reliability of the WFPC synthetic transforms for
such short wavelength filters. The best we can do in these circumstances is to
take an approximate linear fit of the form 
$$U\sim F300_{vega}-0.75(U-B)$$ 
\noindent which reproduces the Holtzman et al. results to
within $\sim0.3$ mag for $-1.5 < (U-B) < 1$. Combined with our $B$-band
transformation this implies \begin{eqnarray*}
U\sim F300_{vega}-0.43(F300-F450)_{vega} +\\
0.09(F450-F606)_{vega} 
\end{eqnarray*}
\noindent For galaxies with colours near the median of our distributions 
this roughly translates to
$$U\sim F300_{vega}+0.4$$
We would hope that our $B, R$ and $I$ transforms are accurate to $\sim0.1$ mag,
which is sufficient for the number-counts and colour distributions. Any
comparison between HDF and ground-based $U$ can only be described as 
approximate.

\subsection{WFPC2 Completeness}

\begin{figure}
\begin{center}
\centerline{\epsfxsize = 3.5in
\epsfbox{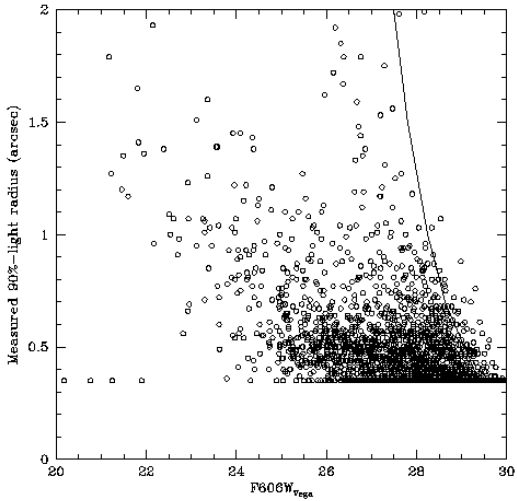}}
\caption{Measured radius encompassing 90\% of the light against
magnitude for the HDF-N $F606W$ data. Object to the left of the solid line 
would give a $>3\sigma$ measurement inside this radius.}
\label{fig:kron radii}
\end{center}
\end{figure}

The combination of ultra-high resolution but fairly unspectacular surface
brightness limits can potentially lead to problems in measuring resolved
images. To counter this we emphasise that our adopted magnitude limits are
quite conservative. For unresolved images the highest signal-to-noise would be
achieved inside a radius of only $0.15''$. By choosing a minimum radius of
$0.35''$ we are sacrificing $\sim0.5$ mag in depth for such objects, in the
hope of better measuring extended galaxies. Even so, the $3\sigma$ limit for a
galaxy with a Kron radius of $1''$ will be $\sim 1$ mag brighter than that
listed in Table \ref{tab:details}. As an example, Fig. \ref{fig:kron radii}
shows the distribution of our measured Kron radii (containing $\sim90$ per
cent of the light) for the HDF-N $F606W$ data. 
The solid line shows the $3\sigma$
magnitude limit as a function of Kron radius. Although at the faintest
magnitudes the majority of
measured galaxies have Kron radii less than $0.6''$, there is
a significant tail stretching out to $\sim 1''$.

We also have to consider the scatter between the isophotal magnitudes used in
the detection routine (which were limited as in Table \ref{tab:parameters})
and the Kron magnitudes finally adopted. As a result of this the Kron
magnitudes always have a brighter completeness limit than the isophotal 
detections.

Taking these effects into account we consider the following are reasonable
completeness limits for our counts; 
$F300_{vega}\sim27$; $F450_{vega}\sim28.5$; $F606_{vega}\sim28.5$; 
$F814_{vega}\sim27.5$ (for HDF-N) and
$F300_{vega}\sim26.5$; $F450_{vega}\sim28.0$; $F606_{vega}\sim28.0$; 
$F814_{vega}\sim27.0$ (for HDF-S).
We analysed the HDF-S data to a slightly brighter magnitude limit than the 
HDF-N due to the presence of areas of lower signal to noise at the joins
between the three three individual WFPC chips.

As a check, we have run the simulation used to correct the WHT $b$-band data
(section 3.6) with parameters appropriate to the HDF-N $F450W$ data. 
This suggests
that for $F450W_{vega}\ga29$ we start to lose significant numbers of
disk dominated (i.e. low surface brightness) objects, but that at brighter
magnitudes the measured count is a good representation of the true count.
However, the size of images (which is dependent on cosmology) is a much more 
important
factor in the HDF simulations than in those for the WHT, and it is not
clear how accurate our simulations are in this respect; for now we are 
confident that corrections to our magnitude scale or counts will not 
exceed $20\%$.

We have also run the detection/analysis procedure on a negative image of
the HDF-S $F450W$ data. This enables us to look for spurious noise detections,
which may add to the count. We find less than a 5\% contribution from
such images to the differential count at $B\sim28$, and nearly all those
which are found lie in the areas of reduced signal-to-noise around the 
edge of the field. We therefore conclude that such detections are not a
significant problem.

\subsection{Comparison with other reductions}

There are several other versions of galaxy counts based on the HDF frames.
Ferguson (1998) has presented a comparison of our HDF-N data with that of
Williams et al. (1996) and Lanzetta et al. (1996). Clear differences exist
between the datasets, all three of which used different photometry packages to
analyse the HDF images.

\begin{figure}
\begin{center}
\centerline{\epsfxsize = 4.5in
\epsfbox{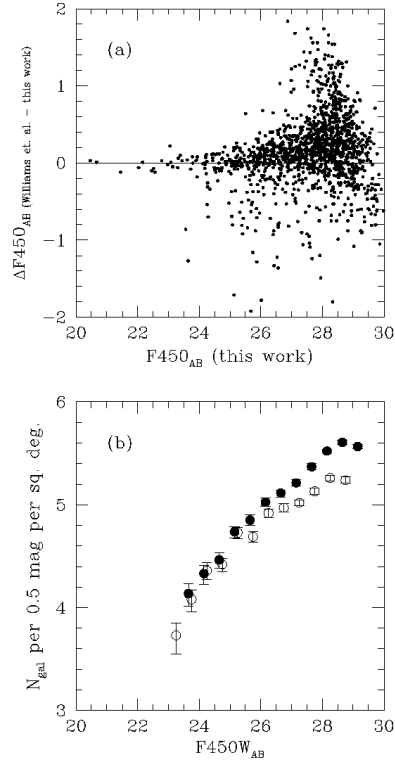}}
\caption{(a) The difference between our magnitudes and Williams et al. (1996) 
(on their AB system) for all objects detected by us on the HDF-N $F450W$ 
frames. 
(b) A comparison of the galaxy counts from the above frames - solid dots, this
work, open circles, Williams et al..}
\label{fig:stsci comparison}
\end{center}
\end{figure}

Here we compare in detail our HDF-N data with those of Williams et al. (1996),
as they show the largest discrepancy with ours. Fig. \ref{fig:stsci
comparison} shows an object by object magnitude comparison between the two
datasets for the $F450W$ band, and a comparison of the resulting counts. It
is clear that there is a systematic magnitude offset and that our count slope
is steeper than theirs, resulting in a  difference of almost a factor two in
the surface density of objects at the faintest magnitudes. Similar differences
exist in the other 3 bands. As pointed out by Ferguson (1998) there are two
reasons for this discrepancy. First, as noted above our magnitudes become
systematically brighter as they become fainter. The exact offset is difficult
to gauge as the magnitude difference distribution has broad wings and becomes
skewed, but, as an example, the peak in 
$(F606_{this\ paper} - F606_{Williams\ et\ al})
\sim 0.05$ at $F606_{AB}\sim25$, $\sim0.10$ at $F606_{AB}\sim27$ and 
$\sim0.20$ at $F606_{AB}\sim28$.
The affect of this on the counts is actually quite small, generally $<10\%$.
Second, we find objects which Williams et al. apparently do not detect at all.
This appears to account for the majority of the difference between the
datasets. A visual inspection of these images leads to the conclusion that
virtually all are merged into adjacent images in the Williams et al. data. We
suspect Williams et al.'s claim that merging is not a significant problem is
optimistic, particularly as we find that even if we include all their split
objects we still find a significant number of galaxies merged in their
reductions but not in ours. As an example, in frame 2 in the range
$F606_{AB}=27-28$ Williams et al. find 142 images with their fully-merged
criteria (parent objects only) and 172 with no merging at all (only daughter
objects). In the same range we find 197 images. Of these 49 are missing from
their fully-merged sample (at any magnitude), but 31 are still missing from
their unmerged dataset. Of these, visual inspection reveals 25 are still
merged with close neighbours.

As at some level the definition of what is one galaxy split into two and what
is two close galaxies is subjective, it is not possible to say who is `right'
and who is `wrong'. However, this ambiguity in the HDF counts should be
borne in mind when comparing with theoretical models.

\subsection{HDF-S STIS Imaging Data}
The Hubble
Deep Field South STIS, unfiltered, 43hr image is to date the deepest image ever
taken, being about $\sim1$ mag deeper that the $F606W$ images (although only
covering $\sim50''\times50''$). The passband is effectively defined by the CCD
and is $\sim4000$\AA~ wide (FWHM) centred on $\sim6000$\AA~. 
We have analysed the
data in similar fashion to the WFPC2 HDF images. Details of the parameters used
are give in Table \ref{tab:parameters}. Areas around the central quasar and the
bright star on one edge of the frame were removed (and replaced by zero signal)
before analysis, and the area of the frame was adjusted to take this into
account. We adopted the AB zero point of 26.39 given on the HDF-S WWW pages.
As the passband is very different from conventional filters we make no
attempt to convert these magnitudes into any other system.

\section{Galaxy Counts and Colours}

\subsection{U-band counts}

\begin{table}
\caption{$u$-band differential galaxy counts from the WHT
data. Corrected counts are based on measurement of artificial stars.}
\halign to\hsize{%
\hfil\rm#\hfil&\hskip 10pt\hfil\rm#&\hskip 10pt\rm\hfil#&
\hfil\rm#\hfil\cr
Magnitude&\multispan2\hskip 10pt\hss Raw N$_{gal}\hss$&Corrected
N$_{gal}\hss$\cr
($u_{ccd}$)&(per frame)&($\deg^{-2}$)\hss&($\deg^{-2}$)\hss\cr
\noalign{\vskip10pt}
20.75-21.25&15\hskip 10pt&1160\hskip 10pt&1160\cr
21.25-21.75&13\hskip 10pt&1010\hskip 10pt&101\cr
21.75-22.25&17\hskip 10pt&1320\hskip 10pt&1320\cr
22.25-22.75&41\hskip 10pt&3180\hskip 10pt&3180\cr
22.75-23.25&69\hskip 10pt&5350\hskip 10pt&5350\cr
23.25-23.75&158\hskip 10pt&12250\hskip 10pt&12250\cr
23.75-24.25&260\hskip 10pt&20150\hskip 10pt&20900\cr
24.25-24.75&338\hskip 10pt&26200\hskip 10pt&27500\cr
24.75-25.25&504\hskip 10pt&39100\hskip 10pt&41700\cr
25.25-25.75&597\hskip 10pt&46300\hskip 10pt&50100\cr
25.75-26.25&682\hskip 10pt&52900\hskip 10pt&67600\cr
26.25-26.75&815\hskip 10pt&63200\hskip 10pt&100000\cr
}
\label{tab:u counts}
\end{table}

\begin{table}
\caption{Differential galaxy counts from the HDF-N and HDF-S $F300W$ 
fields.}
\halign to\hsize{%

\hfil\rm#\hfil&\hskip10pt\hfil\rm#\hfil&\hskip10pt\hfil\rm#\hfil&
\hskip10pt\hfil\rm#\hfil&\hskip10pt\hfil\rm#\hfil&\hskip10pt\hfil\rm#\hfil\cr

Magnitude&\multispan2 \hskip15pt\hfil N$_{gal}\hfil$&
Magnitude&\multispan2 \hskip10pt\hfil N$_{gal}\hfil$\cr

($F300_{vega}$)&(total)&($\deg^{-2}$)&
($F300_{vega}$)&(total)&($\deg^{-2}$)\cr
HDF-S&&&HDF-N\cr
23.5-24.0&19&15000 & 23.5-24.0&31&26500\cr
24.0-24.5&31&24400 & 24.0-24.5&47&40200\cr
24.5-25.0&47&37000 & 24.5-25.0&55&47000\cr
25.0-25.5&65&51200 & 25.0-25.5&91&77800\cr
25.5-26.0&84&66100 & 25.5-26.0&108&92300\cr
26.0-26.5&114&89700 & 26.0-26.5&137&117100\cr
26.5-27.0&169&133000 & 26.5-27.0&164&140200\cr
27.0-27.5&183&144100 & 27.0-27.5&231&197000\cr
}
\label{tab:uhst counts}
\end{table}

\begin{figure*}
\begin{center}
\centerline{\epsfxsize = 4.75in
\epsfbox{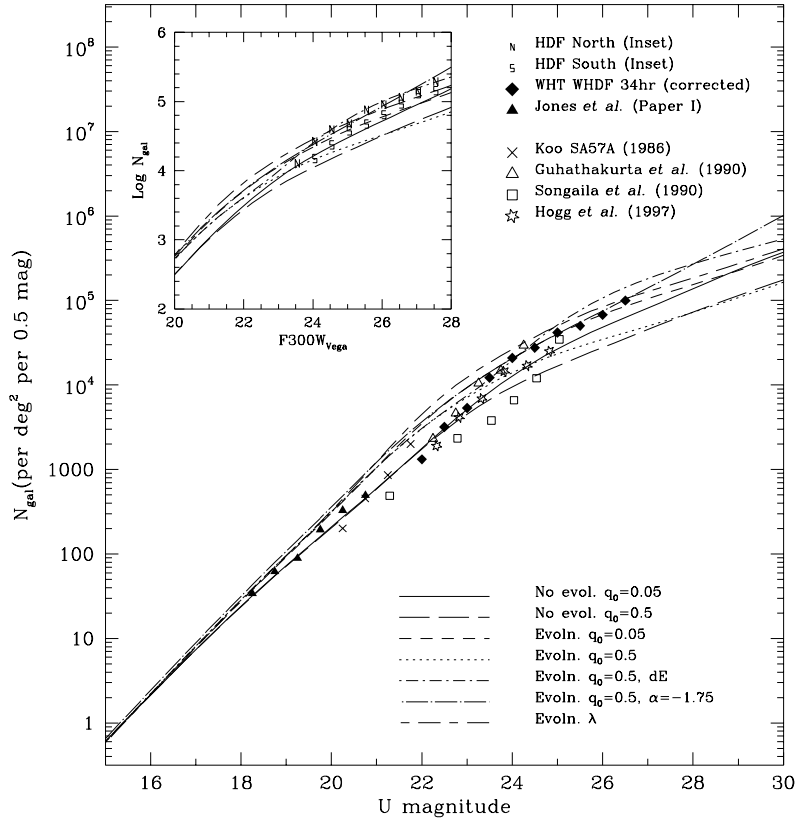}}
\caption{$U$-band differential ground-based count compilation, together with
the HDF-N and HDF-S $F300W$ counts (inset). The models 
discussed in the text 
are shown for comparison. Note that the models shown in the inset are 
calculated for the $F300W$ filter, whilst those on the main figure are 
appropriate to $u_{ccd}$.}
\label{fig:u counts}
\end{center}
\end{figure*}

There are very few $U$-band counts in the literature. Fig. \ref{fig:u counts}
shows the differential $U$-band galaxy count for those we have been able to
find, together with our WHT data from this paper (throughout, we assume
galactic extinction of $E_{B-V}=0.02$ for the WHDF, as in Paper
III). All the ground-based magnitudes have been converted to standard 
photoelectric $U$.
The agreement between the ground-based counts is good, apart from those
of Songaila et al. (1990) which are somewhat low. These data were taken 
through a non-standard $U$-filter, and the offset to standard $U$
we have used (from Hogg et al. 1997) must be rather uncertain.
Overall
there seems little change in the slope of $d\/log(N)/dm\sim0.4$ from $U\sim18$
to $U\sim25$. Faintward of this there is some indication that the counts 
are becoming shallower. 
We also show as an inset our HDF $F300_{vega}$ counts. 
The approximate equivalent $U$-limit for the HDF
data is $27.5$ mag. These appear to have quite a similar slope to the 
ground-based data, although the $F300W$ band is much further into the
ultra-violet than the ground-based $U$ and the K-corrections for galaxies will
be very different. In particular, the Lyman-$\alpha$ forest and the
Lyman-limit start to dim the flux reaching the F300W band at lower redshifts
than for the $U$ band.

Table \ref{tab:u counts} gives our WHT counts, whilst 
Table \ref{tab:uhst counts} lists our HDF counts as a function of 
$F300_{vega}$.

\subsection{B-band counts}

\begin{table}
\caption{$b$-band differential galaxy counts from the 30hr WHT
frame. Corrections to the counts are based on the simulated CCD frame
discussed in the text.}
\halign to\hsize{%
\hfil\rm#\hfil&\hskip 10pt\hfil\rm#&\hskip 10pt\rm\hfil#&
\hfil\rm#\hfil\cr
Magnitude&\multispan2\hskip 10pt\hss Raw N$_{gal}\hss$&Corrected
N$_{gal}\hss$\cr
($b_{ccd}$)&(per frame)&($\deg^{-2}$)\hss&($\deg^{-2}$)\hss\cr
\cr
22.5-23.0&44\hskip 10pt&3375\hskip 10pt&3375\cr
23.0-23.5&50\hskip 10pt&3825\hskip 10pt&3825\cr
23.5-24.0&128\hskip 10pt&9825\hskip 10pt&9825\cr
24.0-24.5&229\hskip 10pt&17575\hskip 10pt&17575\cr
24.5-25.0&352\hskip 10pt&27000\hskip 10pt&27000\cr
25.0-25.5&480\hskip 10pt&36850\hskip 10pt&40400\cr
25.5-26.0&706\hskip 10pt&54175\hskip 10pt&61000\cr
26.0-26.5&801\hskip 10pt&61475\hskip 10pt&73500\cr
26.5-27.0&939\hskip 10pt&72075\hskip 10pt&95500\cr
27.0-27.5&1064\hskip 10pt&81650\hskip 10pt&129500\cr
27.5-28.0&1149\hskip 10pt&88175\hskip 10pt&205000\cr}
\label{tab:b counts}
\end{table}

\begin{table}
\caption{$b$-band differential galaxy counts from the co-added
INT and WHT frames (equivalent to 46hrs WHT exposure). Corrections
are based on the artificial star results.}
\halign to\hsize{%
\hfil\rm#\hfil&\hskip 10pt\hfil\rm#&\hskip 10pt\rm\hfil#&
\hfil\rm#\hfil\cr
Magnitude&\multispan2\hskip 10pt\hss Raw N$_{gal}\hss$&Corrected
N$_{gal}\hss$\cr
($b_{ccd}$)&(per frame)&($\deg^{-2}$)\hss&($\deg^{-2}$)\hss\cr
\cr
25.0-25.5&38\hskip 10pt&55100\hskip 10pt&55100\cr
25.5-26.0&46\hskip 10pt&66700\hskip 10pt&74100\cr
26.0-26.5&33\hskip 10pt&47800\hskip 10pt&61700\cr
26.5-27.0&47\hskip 10pt&68100\hskip 10pt&95500\cr
27.0-27.5&69\hskip 10pt&100000\hskip 10pt&158500\cr
27.5-28.0&63\hskip 10pt&91300\hskip 10pt&182000\cr
28.0-28.5&79\hskip 10pt&114500\hskip 10pt&251200\cr}
\label{tab:bdeep counts}
\end{table}

\begin{table}
\caption{Differential galaxy counts from the HDF-N and HDF-S $F450W$ fields.}
\halign to\hsize{%
\hfil\rm#\hfil&\hskip10pt\hfil\rm#\hfil&\hskip10pt\hfil\rm#\hfil&
\hskip10pt\hfil\rm#\hfil&\hskip10pt\hfil\rm#\hfil&\hskip10pt\hfil\rm#\hfil\cr

Magnitude&\multispan2 \hskip15pt\hfil N$_{gal}\hfil$&
Magnitude&\multispan2 \hskip10pt\hfil N$_{gal}\hfil$\cr

($F450_{vega}$)&(total)&($\deg^{-2}$)&
($F450_{vega}$)&(total)&($\deg^{-2}$)\cr
HDF-S&&&HDF-N\cr
23.5-24.0&14&11200&23.5-24.0&18&15400\cr
24.0-24.5&22&17600&24.0-24.5&24&20500\cr
24.5-25.0&38&30400&24.5-25.0&34&29100\cr
25.0-25.5&52&41600&25.0-25.5&64&54700\cr
25.5-26.0&79&63200&25.5-26.0&87&74400\cr
26.0-26.5&90&72000&26.0-26.5&129&110300\cr
26.5-27.0&122&97600&26.5-27.0&149&127400\cr
27.0-27.5&158&126400&27.0-27.5&191&163300\cr
27.5-28.0&235&188000&27.5-28.0&279&238500\cr
28.0-28.5&299&232000&28.0-28.5&389&332500\cr
28.5-29.0&232&185600&28.0-28.5&485&414600\cr
}
\label{tab:bhst counts}
\end{table}

\begin{figure*}
\begin{center}
\centerline{\epsfxsize = 4.75in
\epsfbox{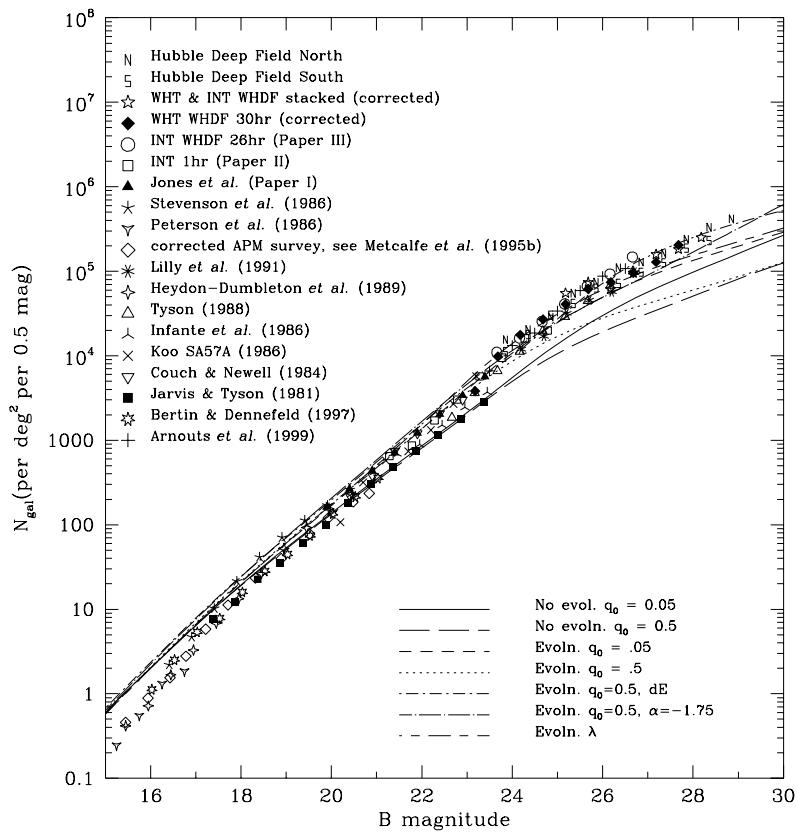}}
\caption{$B$-band differential count compilation. The models discussed in
the text are shown for comparison, calculated for our ground-based $b_{ccd}$ 
passband.}
\label{fig:b counts}
\end{center}
\end{figure*}

Fig. \ref{fig:b counts} shows a compilation of published $B$-band galaxy
counts over the range $15\la B\la 29$, including the corrected counts 
from our new WHT data (for the whole WHDF and for the co-added field) 
and the HDF $F450W$ counts. 
All the counts have been adjusted to be on 
the standard photoelectric $B$-band system. For our HDF counts the difference 
between $F450_{vega}$ and $B$ is $0.1$ mag - section 4.2. 
As in Paper III we have
endeavoured to plot all the datasets only to their respective $3\sigma$
limits. All the datasets have been corrected for
galactic extinction. It can be seen immediately that our WHT and HDF counts
are in good agreement, suggesting that the WHT confusion corrections are quite
accurate.

With the improved statistics and signal-to-noise over Paper III it is clear
that the conclusion we drew previously is still 
supported and that the count slope
has flattened at the faintest magnitudes. However, the slope is somewhat
shallower that before, $d\log N/dm_b\sim0.25$, which, if this reflects the
slope of the faint end of the luminosity function at high redshift as we
suggested in Paper III, corresponds to $\alpha\sim-1.6$. There is still no
sign that the counts have stopped rising, even at $B\sim29$ mag, and the total
number of galaxies exceeds $1.5\times10^6$/sq. deg.

Our results for the WHDF, the WHT+INT co-added frame and for the 
HDF $F450_{vega}$ are given in
Tables \ref{tab:b counts}, \ref{tab:bdeep counts} and \ref{tab:bhst counts}.

\subsection{R-band counts}

\begin{table}
\caption{$r$-band differential galaxy counts from the 6hr WHT
frame. Corrections are based on artificial stars.}
\halign to\hsize{%
\hfil\rm#\hfil&\hskip 10pt\hfil\rm#&\hskip 10pt\rm\hfil#&
\hfil\rm#\hfil\cr
Magnitude&\multispan2\hskip 10pt\hss Raw N$_{gal}\hss$&Corrected
N$_{gal}\hss$\cr
($r_{ccd}$)&(per frame)&($\deg^{-2}$)\hss&($\deg^{-2}$)\hss\cr
\noalign{\vskip10pt}
21.0-21.5&23\hskip 10pt&1825\hskip 10pt&1825\cr
21.5-22.0&41\hskip 10pt&3275\hskip 10pt&3275\cr
22.0-22.5&59\hskip 10pt&4700\hskip 10pt&4700\cr
22.5-23.0&101\hskip 10pt&8050\hskip 10pt&8050\cr
23.0-23.5&154\hskip 10pt&12275\hskip 10pt&12275\cr
23.5-24.0&224\hskip 10pt&17875\hskip 10pt&20400\cr
24.0-25.5&373\hskip 10pt&29750\hskip 10pt&34700\cr
24.5-25.0&493\hskip 10pt&39300\hskip 10pt&50100\cr
25.0-25.5&696\hskip 10pt&55500\hskip 10pt&79400\cr
25.5-26.0&833\hskip 10pt&66425\hskip 10pt&109600\cr}
\label{tab:r counts}
\end{table}

\begin{table}
\caption{Differential galaxy counts from the HDF-N and HDF-S $F606W$ 
fields.}
\halign to\hsize{%

\hfil\rm#\hfil&\hskip10pt\hfil\rm#\hfil&\hskip10pt\hfil\rm#\hfil&
\hskip10pt\hfil\rm#\hfil&\hskip10pt\hfil\rm#\hfil&\hskip10pt\hfil\rm#\hfil\cr

Magnitude&\multispan2 \hskip15pt\hfil N$_{gal}\hfil$&
Magnitude&\multispan2 \hskip10pt\hfil N$_{gal}\hfil$\cr

($F606_{vega}$)&(total)&($\deg^{-2}$)&
($F606_{vega}$)&(total)&($\deg^{-2}$)\cr

HDF-S&&&HDF-N\cr
23.5-24.0&28&21500&24.0-24.5&35&29900\cr
24.0-24.5&35&26900&24.0-24.5&35&29900\cr
24.5-25.0&50&38400&24.5-25.0&51&43600\cr
25.0-25.5&80&61400&25.0-25.5&89&76100\cr
25.5-26.0&112&86000&25.5-26.0&125&106800\cr
26.0-26.5&145&111400&26.0-26.5&153&130800\cr
26.5-27.0&194&149000&26.5-27.0&198&169200\cr
27.0-27.5&258&198200&27.0-27.5&257&219600\cr
27.5-28.0&338&259600&27.5-28.0&388&331600\cr
28.0-28.5&447&361000&28.0-28.5&497&424700\cr
28.5-29.0&361&277300&28.5-29.0&509&435000\cr
}
\label{tab:rhst counts}
\end{table}

\begin{figure*}
\begin{center}
\centerline{\epsfxsize = 4.75in
\epsfbox{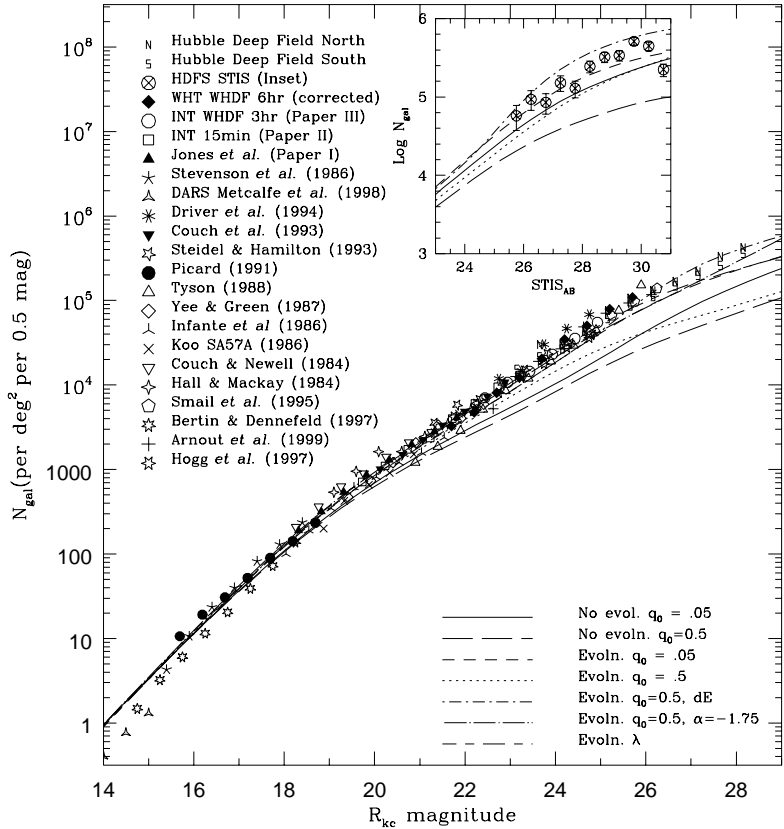}}
\caption{$R$-band differential count compilation, together with the HDF-S
STIS count in the unfiltered passband (inset). 
The models discussed in the text are 
shown for comparison - for ground-based $r_{ccd}$ in the main figure and 
specifically  for the STIS passband in the inset.}
\label{fig:r counts}
\end{center}
\end{figure*}

The $R$-band count compilation, shown in Fig. \ref{fig:r counts}, has changed
significantly since Paper III. Apart from our new WHT data, the deepest
ground-based $R$-band counts published are those of Smail et al. (1995) and 
Hogg et al. (1997), based on observations in good seeing on the Keck telescope.
However, HDF data provide the major change, extending about two magnitudes
fainter than the ground-based data. All data are plotted on the photoelectric 
$R$ system. To get to $R$ the HDF $F606_{vega}$ measurements have been 
shifted by $-0.1$ mag as discussed in section 4.2. 

As with the $B$ band, the counts continue to increase, with a slope
$d\/log(N)/dm\sim0.37$ for $20\la R\la 26$, but with evidence of a change to a
shallower slope faintward of this. This occurs where the HDF data take over
from the ground-based data, and could be an artifact of the fact that $F606W$
is
actually midway between $V$ and $R$, and so may not have a slope appropriate
to $R$. However, the mean $(F606-F814)$ colour for the HDF data is fairly
constant faintward of $R\sim24$, suggesting that any change in count slope
between $F606W$ and $R$ is going to be minimal. The integral $R$ counts
reach $\sim2\times10^6$deg$^{-2}$. We note that the faint end slope is very
similar to the $B$ data, with $d\log N/dm_R \sim 0.25$.
The count data are presented in Tables \ref{tab:r counts} 
and \ref{tab:rhst counts}.

Also shown inset in Fig. \ref{fig:r counts} are the HDF-S STIS counts. 
These probe even deeper than the HDF $F606$ counts and appear to continue
with a similar slope of $\sim0.25$. Note that the apparent turn-over in the faintest bin
is due to incompleteness.

\subsection{I-band counts}

This is the first time we have included an $I$-band count in this series of
papers. The WHDF counts are detailed in Table \ref{tab:i counts}, whilst the
HDF counts in $F814_{vega}$ are listed in Table \ref{tab:ihst counts}.
From section 4.2 we see that $F814_{vega}\sim I$.

\begin{figure*}
\begin{center}
\centerline{\epsfxsize = 4.75in
\epsfbox{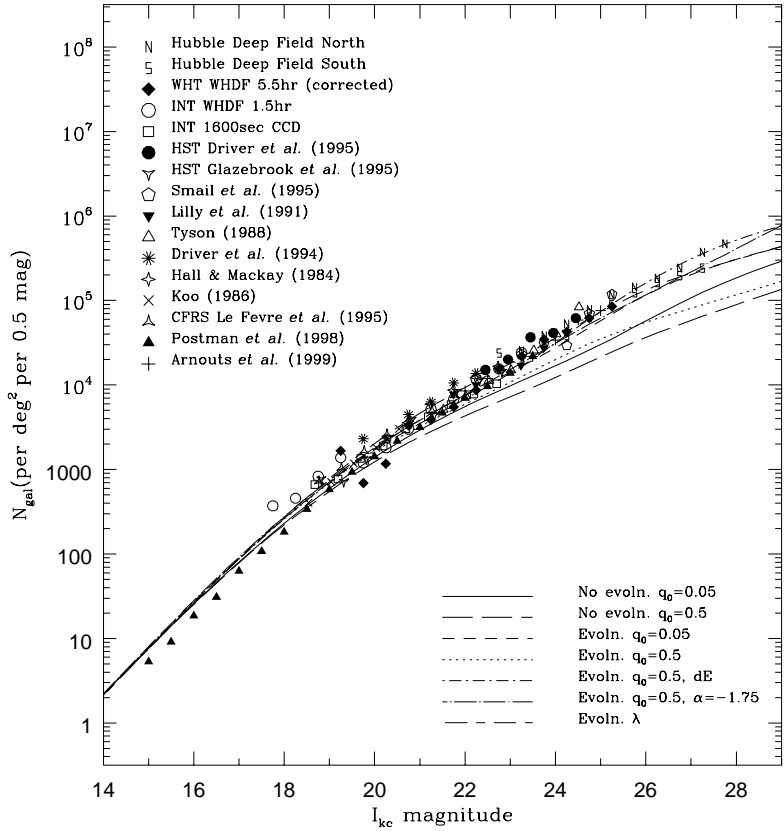}}
\caption{I-band differential count compilation. Again, the models discussed 
in the text are shown, calculated for our ground-based $i_{ccd}$ band.}
\label{fig:i counts}
\end{center}
\end{figure*}

The counts are plotted in Fig. \ref{fig:i counts}. Note that as well as our
HDF counts and the WHT $I$-counts, we also show in this diagram for the first
time two sets of ground-based $I$-band CCD counts taken at prime focus of the
Isaac Newton telescope. The first set, to $I\sim23$, is based on 10 of
the fields in Paper II (total area $6.3\times10^{-3}$ sq. deg.). These data
consisted of $4\times400$s exposures on each field taken with an RCA CCD. The
data reduction and analysis of these frames were identical to those for the $B$
and $R$ frames in Paper II. The second set are taken off a 1.5hr total exposure
with a Tek $1024\times1024$ CCD (total area $2.4\times10^{-2}$ sq. deg.)
centred on the WHDF (see also section 3.2). These reach
$I\sim23.75$. The data reduction and analysis are similar to that
described here for the WHT data. Once again, all magnitudes have been adjusted 
to be on the standard ground-based system.

The ground-based counts have a slope of
$d\/log(N)/dm\sim0.33$ for $21\la I\la 25$, which appears to flatten slightly
to $d\/log(N)/dm\sim0.27$ for the fainter HDF data.

\begin{table}
\caption{$i$-band differential galaxy counts from
the WHT data. Corrections are based on artificial stars.}
\halign to\hsize{%
\hfil\rm#\hfil&\hskip 10pt\hfil\rm#&\hskip 10pt\rm\hfil#&
\hfil\rm#\hfil\cr
Magnitude&\multispan2\hskip 10pt\hss Raw N$_{gal}\hss$&Corrected
N$_{gal}\hss$\cr
($i_{ccd}$)&(per frame)&($\deg^{-2}$)\hss&($\deg^{-2}$)\hss\cr
\noalign{\vskip10pt}
20.0-20.5&17\hskip 10pt&1175\hskip 10pt&1175\cr
20.5-21.0&48\hskip 10pt&3320\hskip 10pt&3320\cr
21.0-21.5&56\hskip 10pt&3870\hskip 10pt&3870\cr
21.5-22.0&79\hskip 10pt&5460\hskip 10pt&5460\cr
22.0-22.5&125\hskip 10pt&8640\hskip 10pt&8640\cr
22.5-23.0&205\hskip 10pt&14170\hskip 10pt&15500\cr
23.0-23.5&283\hskip 10pt&19560\hskip 10pt&21900\cr
23.5-24.0&413\hskip 10pt&28550\hskip 10pt&34700\cr
24.0-25.5&490\hskip 10pt&33870\hskip 10pt&42700\cr
24.5-25.0&623\hskip 10pt&43070\hskip 10pt&61700\cr
25.0-25.5&585\hskip 10pt&40440\hskip 10pt&85100\cr
}
\label{tab:i counts}
\end{table}

\begin{table}
\caption{Differential galaxy counts from the HDF-N and HDF-S $F814W$ 
fields.}
\halign to\hsize{%
\hfil\rm#\hfil&\hskip10pt\hfil\rm#\hfil&\hskip10pt\hfil\rm#\hfil&
\hskip10pt\hfil\rm#\hfil&\hskip10pt\hfil\rm#\hfil&\hskip10pt\hfil\rm#\hfil\cr

Magnitude&\multispan2 \hskip15pt\hfil N$_{gal}\hfil$&
Magnitude&\multispan2 \hskip10pt\hfil N$_{gal}\hfil$\cr

($F814_{vega}$)&(total)&($\deg^{-2}$)&
($F814_{vega}$)&(total)&($\deg^{-2}$)\cr
HDF-S&&&HDF-N\cr
22.5-23.0&31&23900 & 22.5-23.0&22&18800\cr
23.0-23.5&27&20800 & 23.0-23.5&30&25600\cr
23.5-24.0&39&30100 & 23.5-24.0&45&38500\cr
24.0-24.5&50&38600 & 24.0-24.5&62&53000\cr
24.5-25.0&83&64000 & 24.5-25.0&92&78600\cr
25.0-25.5&117&90200 & 25.0-25.5&136&116200\cr
25.5-26.0&161&124100 & 25.5-26.0&169&144400\cr
26.0-26.5&212&163500 & 26.0-26.5&214&182900\cr
26.5-27.0&255&196600 & 26.5-27.0&282&241000\cr
27.0-27.5&315&242900 & 27.0-27.5&437&373400\cr
27.5-28.0&335&258300 & 27.5-28.0&547&467400\cr
28.0-28.5&194&149600 & 28.0-28.5&573&489600\cr
}
\label{tab:ihst counts}
\end{table}

\subsection{Colours}

As noted in sections 3.3 and 4.1 we use fixed aperture colours. However,
the difference in this aperture between the WHT and HDF data  ($1.5''$ radius 
as opposed to $0.35''$) certainly means that we are measuring colours 
(on average) over a different metric aperture from the two data-sets. 
This should be borne in mind when interpreting the data.

\begin{figure}
\begin{center}
\centerline{\epsfxsize = 3.5in
\epsfbox{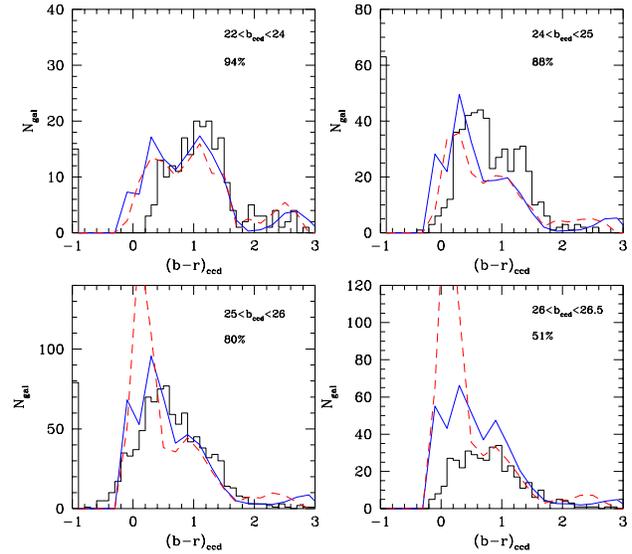}}
\caption{$b_{ccd}$-limited $(b-r)_{ccd}$ colour-magnitude histograms for the 
WHDF data, split into four magnitude bins. The percentages indicate the
colour completeness for that magnitude range.  The predictions of the 
$q_0=0.05$ (solid line) and $q_0=0.5$ dwarf (dashed line) models are also 
shown.}
\label{fig:b-r wcmhist}
\end{center}
\end{figure}

\begin{figure}
\begin{center}
\centerline{\epsfxsize = 3.5in
\epsfbox{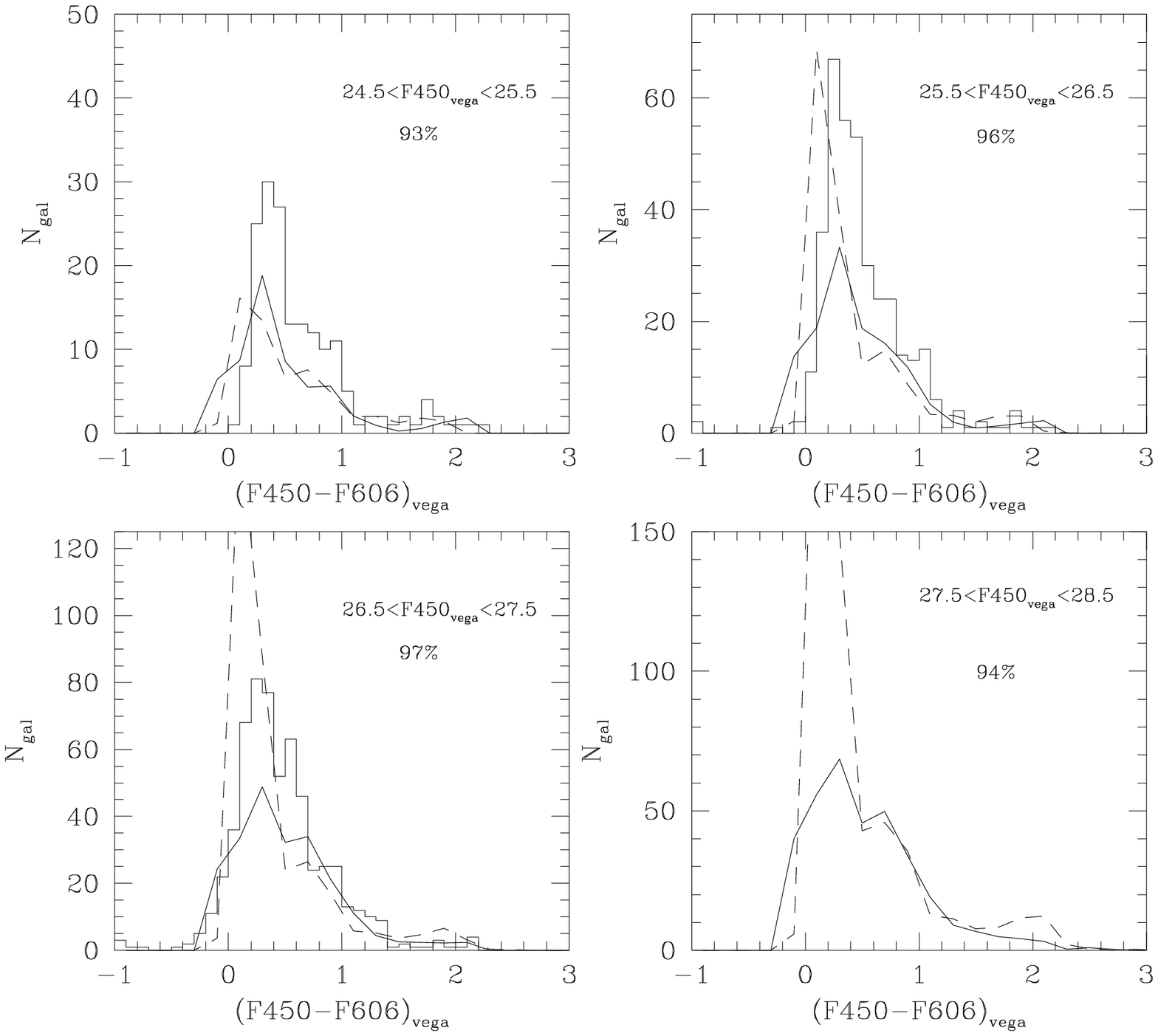}}
\caption{As Figure \ref{fig:b-r wcmhist} but now for the HDF-N and HDF-S 
$F450_{vega}$-limited $(F450-F606)_{vega}$ data.}
\label{fig:b-r hcmhist}
\end{center}
\end{figure}

Fig \ref{fig:b-r wcmhist} displays the $b_{ccd}$ versus $(b-r)_{ccd}$
colour magnitude histograms for our whole $7'\times7'$ field, split into
four magnitude ranges. Fig \ref{fig:b-r hcmhist} shows the equivalent
$F450_{vega}$-limited $(F450-F606)_{vega}$ data for the HDF-N and HDF-S. The 
percentage colour completeness is indicated on each of the histograms. 
This is particularly severe in the faintest ground-based bin due to the
comparatively bright limit of the $r_{ccd}$ data. Also shown are the
predicted histograms for two of our evolutionary models (these are 
discussed in sections 6.3 and 6.4).
Figs
\ref{fig:u-b wcmhist} and \ref{fig:u-b hcmhist} show similar plots, but now 
for $b_{ccd}$-limited $(u-b)_{ccd}$ and $F450_{vega}$-limited 
$(F300_F450)_{vega}$ colours.
Figs \ref{fig:r-i wcmhist} and \ref{fig:r-i hcmhist} show the plots for
$(r-i)_{ccd}$ and $(F606-F814)_{vega}$ colours.

The colour-colour diagrams for the WHDF, $(r-i)_{ccd}:(b-r)_{ccd}$ and 
$(u-b)_{ccd}:(b-r)_{ccd}$, are shown in Fig. \ref{fig:colour tracks}(a),(b). 
Fig. \ref{fig:cc} shows the equivalent plots for the spaced-based
data, split into bright and faint magnitude ranges. We discuss these in more 
detail in the next section. Note, however,
that due to the different K-corrections between the HST and WHT passbands,
especially in the ultra-violet, where the filters differ by $\sim600$\AA~, the
tracks of galaxies in these plots are not expected to be the same.

\section{Galaxy Evolution Models}
\subsection{Overview}

In our previous work (Shanks 1990, Paper II, Paper III) we have noted 
that if the $B$-band models are normalised at $B\sim18$ rather than $B\sim15$
then non-evolving models give a reasonable representation of the $B$ counts
and redshift distributions in the range $18\la B\la 22.5$. Some support for this
high normalisation comes from work on HST galaxy counts subdivided by
morphology, where non-evolving models with high normalisation give an
excellent fit to both spiral and early-type counts with $17\la I\la 22$ 
(Glazebrook
et al. 1995, Driver et al. 1995). In Paper IV we have also shown that these
same non-evolving models fit the $K$-band counts as faint as $K\sim23$.
The $B$ counts now have the work of Bertin \& Dennefeld (1997) added at
the bright end, comprising photographic measurements over 145 deg$^2$.
Again these show a steeper slope than the models at $B<17$. 
The same authors' R counts show
a similar effect at $R<16$. In the $I$ band the recent counts of Postman
et al. (1998) also show a steep slope at $I<17$. Although these counts
cover only 16 deg$^2$, they are made using a CCD. Finally, the 2MASS
counts (Cutri \& Skrutskie 1998) which cover 158 deg$^2$ to $H=15$ and $K=14$
are also found to be steeper than the no evolution models 
(see McCracken et. al. in preparation). 
This effect therefore seems to be present 
over a wide range
of passbands. If number count steepness were caused by evolutionary
changes in star formation rate (SFR) at low redshifts, $z\la0.1$, then it 
would be surprising if the near-IR counts were
affected as much as the $B$ counts. We therefore continue to believe that
the most likely interpretation is that this steepness at bright magnitudes
is caused by large scale inhomogeneities in the galaxy distribution on
$\sim150$h$^{-1}$Mpc scales (Shanks et al. 1990, Paper III). Further evidence
for this has come from preliminary luminosity function data in the 2dF
Galaxy Redshift Survey which shows a LF with the local M$^*$ and slope
$\alpha$ (Efstathiou et al. 1988, Loveday et al. 1992, Ratcliffe et al.
1998), but a $\phi^*$ which is 50\% higher than the local value and in
good agreement with the value that is used here. We therefore believe that
these developments justify our normalising  the count models at $B\sim18$,
corresponding to an overall value of $\phi^*\sim2.4\times10^{-3}$. The
exact normalisation ($\phi^*$) we use is given in Table \ref{tab:model lf}
for the evolutionary models. In the case of no evolution we increase these
$\phi^*$'s by 10\% to give the same normalisation at $B=18$ as in the 
evolutionary models.

Metcalfe et al. (1996)  presented two evolutionary models with this
high normalisation which gave reasonable fits to the counts at all the
wavelengths for which they had data, and to the HDF colours. These were a
simple, $q_0=0.05$ luminosity evolution model with exponentially decaying
star formation rates, and a $q_0=0.5$ model, similar but for the addition of
a population of dwarf (hereafter `dE') galaxies with a constant 
star-formation rate until $z\sim1$ and zero thereafter.

As in Metcalfe et al. (1996), for simplicity we only consider two
basic forms of evolution; a $\tau$=2.5Gyr exponentially decaying star formation
model for E/S0 and Sab galaxies, and a $\tau$=9 Gyr model for Sbc/Scd/Sdm, both
from Bruzual \& Charlot (1993). This assumes that the evolution of the
different morphological types is simply governed by their SFR history. For
q$_0$=0.05 and q$_0$=0.5, the present day galaxy ages are assumed to be 16Gyr
and 12.7Gyr, implying  formation redshifts of $z_f=6.3$ and $z_f=9.9$. We also
consider a model with a cosmological constant, zero spatial curvature and
$\Omega_0$=0.3, with a present day age of 18 Gyr and a formation redshift of
$z_f=7.9$. Details  of the present day luminosity functions and colours used in
our models are given in Table \ref{tab:model lf}.  All models take account of
Lyman-$\alpha$ forest/break absorption as described by Madau (1995).
Similar models have been used by Pozzetti et al. (1998).

One feature of these pure luminosity evolution (PLE) models is that they
predict that $z>1$ galaxies should be detectable at reasonably bright
magnitudes and this prediction has been broadly confirmed by $B\sim24$
redshift surveys (Cowie et al. 1995, 1996) undertaken on the Keck 10-m
telescope. However, to provide a detailed fit to the $B$-band $n(z)$
distribution it is necessary to moderate the number of high $z$ spiral
galaxies predicted by the $\tau$=9 Gyr model by including the effect
of internal dust extinction, with $A_{\lambda}\propto 1/\lambda$
(Metcalfe et al, 1996, cf.Campos \& Shanks 1997 and refs. therein). 
We have, at least initially, assumed that the extinction law and the
star-formation history/metallicity evolution can be treated independently.
Edmunds \& Phillipps (1997) and Edmunds \& Eales (1998) have investigated
the effects of dropping this assumption and indeed in Section 7 of this
paper we discuss possible evidence for evolution of the extinction law
with redshift.
A similar problem exists in the $K$-band $n(z)$ with the $\tau$=2.5Gyr
model for early-type galaxies. Here, essentially passive evolution
overpredicts the high $z$ tail at $K\sim19$. 
We have overcome this by
adopting a dwarf dominated IMF, with a slope $x=3$, for early-type
galaxies (see Paper IV). In the optical bands this produces results
very similar to the more widely used Salpeter or Scalo IMF's.  
There is some evidence that passively evolving models overpredict
the HST morphological counts of early-type galaxies (see e.g. 
Driver et. al. 1998 and references therein), but our model does
a reasonable job of matching these counts, except 
at the faintest magnitudes, where morphological identification becomes 
difficult.
One problem for the x=3 model is that it predicts an
M/L$_B\sim$120M/L$_\odot$ for early-type galaxies as opposed to
M/L$_B$=20 for x=1.35 and compared to the observed value of
M/L$_B$=10hM/L$_\odot$. This disagreement in the x=3 case is caused by
the excess of low-mass stars. We have therefore also considered a model
still with $\tau$=2.5Gyr but with an x=3 IMF which cuts off below 
0.5M$_{\odot}$. The isochrone synthesis code for this case was specially
supplied by G. Bruzual. This model then predicts M/L$_B$=5M/L$_\odot$,
in much better agreement with observation. The effect on the galaxy
colour predictions is reasonably small; for example, at 16Gyr the
predicted rest $(V-K)$ colour is 3.27 for the x=3+0.5 M$_\odot$ cutoff case
compared to $(V-K)=3.67$ in the x=3 uncut case and to $(V-K)=3.25$ in the
Salpeter case.  Values observed for giant ellipticals are in the range
3.1-3.5. Vazdekis et al. (1997) have further shown that dwarf dominated
IMF's with a low mass-cut, solar metallicities and a 16Gyr age do
produce consistent population synthesis model fits to nearby
early-type galaxy spectra. Another problem for a dwarf-dominated IMF
is that it contains too few giant stars to generate the observed
metallicity of present-day early-type galaxies. However, this is also
a problem for a Salpeter IMF, and a short initial burst with a much
flatter IMF slope usually has to be invoked to solve this
problem. (e.g Vazdekis et al. 1997). Indeed, Vazdekis et al. find that
their chemical evolution, `closed-box' model with a flat IMF during an
initial short burst and then a steeper IMF(x=2-3) plus low mass
cut-off, produces an excellent fit to the spectra, and hence the
observed metallicities, of local giant ellipticals.

\begin{table*}
\begin{minipage}{140mm}
\caption{Details of the luminosity evolution models
($H_0=50 kms^{-1}Mpc^{-1}$). The dE type is only employed in the $q_0=0.5$
case. $\phi^*$s are given for our evolving models; 10\% higher $\phi^*$s are
used for the no-evolution models.}
\halign{\rm#\hfil&\hskip 10pt\hfil\rm#\hfil&\hskip 10pt\rm\hfil#\hfil&
\hskip 10pt\rm\hfil#\hfil&\hskip 10pt\rm\hfil#\hfil&
\hskip 10pt\rm\hfil#\hfil&\hskip 10pt\rm\hfil#\hfil\cr

&E/S0&Sab&Sbc&Scd&Sdm&dE\cr
\cr
$\phi^*$(Mpc$^{-3}$)&$9.27\times10^{-4}$&$4.63\times10^{-4}$&$6.2\times
10^{-4}$&$2.73\times10^{-4}$&$1.36\times10^{-4}$&$1.9\times10^{-2}$\cr
$\alpha$&$-0.7$&$-0.7$&$-1.1$&$-1.5$&$-1.5$&$-1.2$\cr
$M^*_{b_{ccd}}$&$-20.85$&$-20.88$&$-21.22$&$-21.37$&$-21.38$&$-15.85$\cr
$M^*_{F450_{vega}}$&$-21.00$&$-21.00$&$-21.32$&$-21.44$&$-21.45$&$-16.0$\cr
$(u-b)_{ccd\ z=0}$&$0.52$&$0.08$&$-0.10$&$-0.18$&$-0.16$&$0.52$\cr
$(b-r)_{ccd\ z=0}$&$1.59$&$1.38$&$1.16$&$0.84$&$0.75$&$1.59$\cr
$(r-i)_{ccd\ z=0}$&$0.70$&$0.70$&$0.59$&$0.53$&$0.47$&$0.70$\cr
$(F300-F450)_{vega\ z=0}$&$1.44$&$0.44$&$0.00$&$-0.24$&$-0.20$&$1.44$\cr
$(F450-F606)_{vega\ z=0}$&$1.16$&$1.01$&$0.85$&$0.62$&$0.54$&$1.16$\cr
$(F606-F814)_{vega\ z=0}$&$1.21$&$1.21$&$1.02$&$0.92$&$0.81$&$1.21$\cr
$A{_B}(z=0)$&$0.00$&$0.00$&$0.30$&$0.30$&$0.30$&$0.00$\cr
}
\label{tab:model lf}
\end{minipage}
\end{table*}

\subsection{Galaxy Count Models}

Galaxy counts generally constrain particular {\it combinations} of
evolutionary models and cosmological parameters. 
Figs \ref{fig:u counts} - \ref{fig:i counts} show how our various models fit to
the counts in the four bands. The HDF-S STIS count models (see 
Shanks et al. in prep.)  are 
shown inset into Fig. \ref{fig:r counts}. As well as the two evolving models 
described above, we also show their non-evolving counterparts, with the 
exception of the $\Lambda$ model which is very similar to the q$_0$=0.05 
no-evolution model. Also shown is the  $q_0=0.5$ evolutionary model 
{\it without} the extra dE component and a new q$_0$=0.5 evolutionary model 
which has a steeper luminosity function slope for Scd/Sdm galaxies 
(see below). As can be seen, even at the depths reached by STIS, our 
preferred evolutionary models fit the counts reasonably well over a wide 
range of wavelengths. This agreement continues into the infra-red (Paper IV).

Note that both the evolution and no-evolution  models in the range 
18$\la U\la $24 have different slopes when expressed in the ground-based $U$
and HST $F300W$ passbands, with the slope in the $F300W$ band being steeper.
This is due to the HST band peaking at 3000\AA~ and the ground-based band 
peaking at 3800\AA~.

\begin{figure}
\begin{center}
\centerline{\epsfxsize = 3.5in
\epsfbox{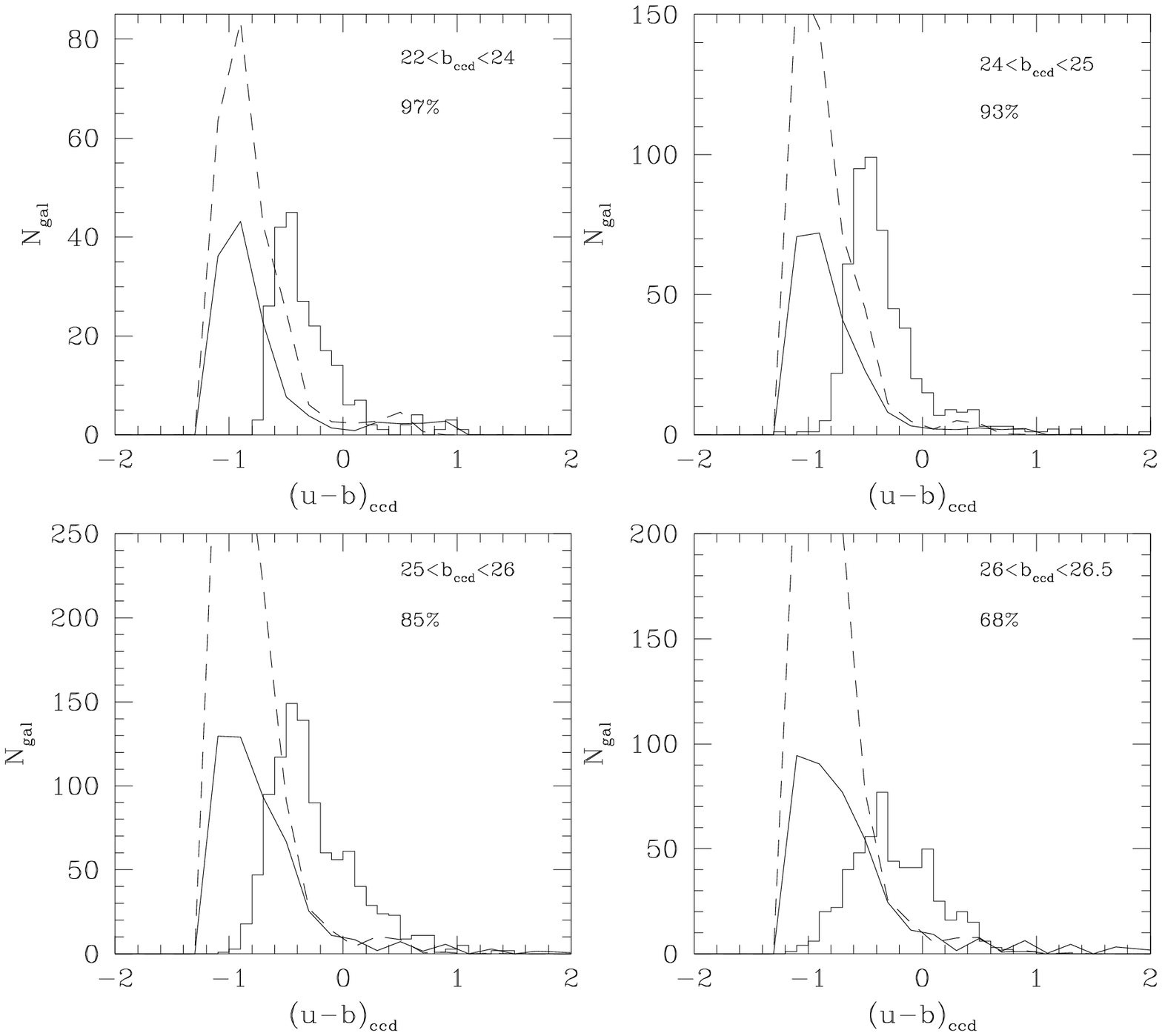}}
\caption{$b_{ccd}$-limited $(u-b)_{ccd}$ colour-magnitude histograms for the
WHDF data. The percentages indicate the colour completeness for 
each magnitude range. The predictions of the $q_0=0.05$ (solid line)
and $q_0=0.5$ dwarf (dashed line) models are also shown.}
\label{fig:u-b wcmhist}
\end{center}
\end{figure}

\begin{figure}
\begin{center}
\centerline{\epsfxsize = 3.5in
\epsfbox{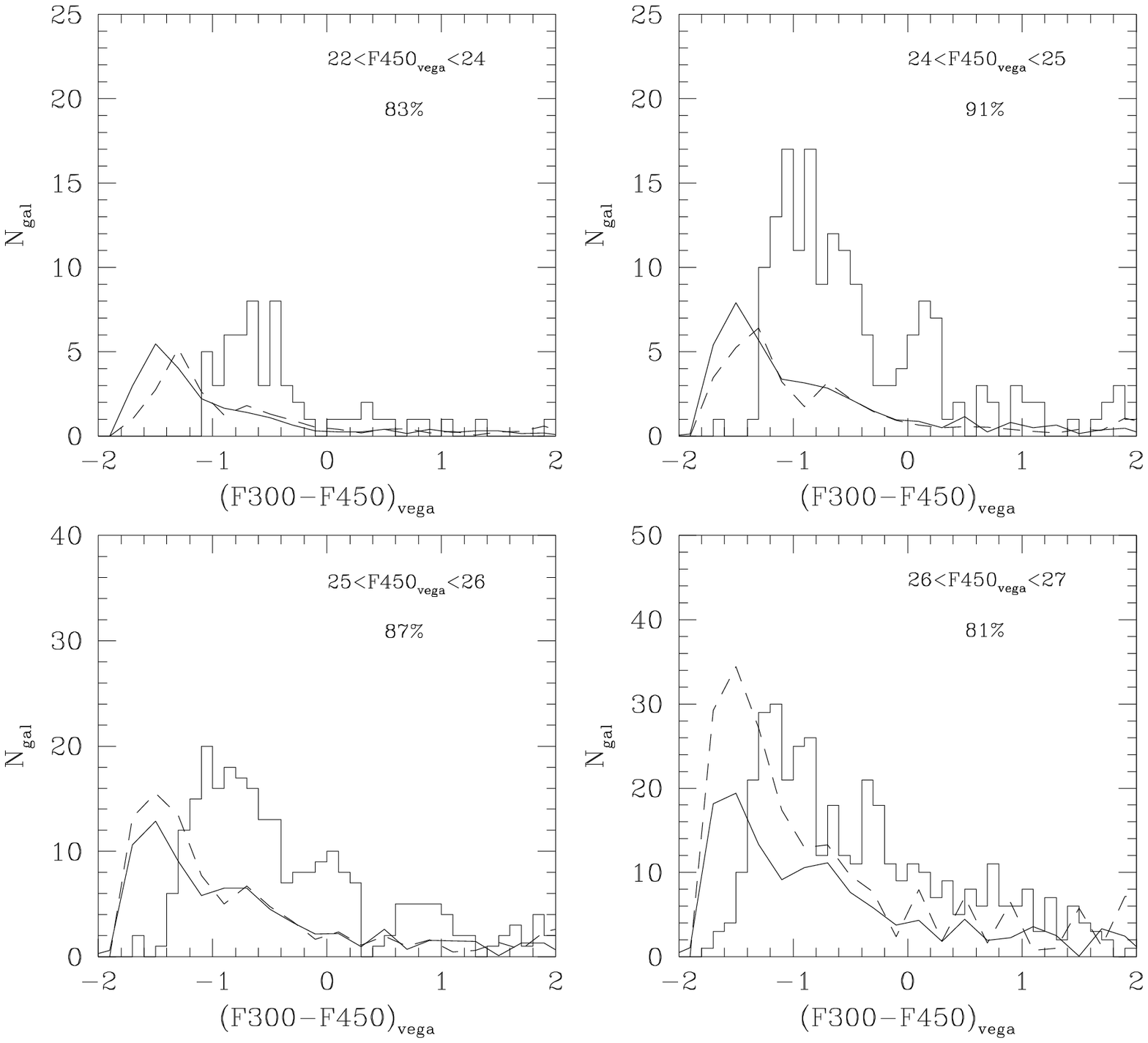}}
\caption{As Figure \ref{fig:u-b wcmhist} but now for the HDF-N and HDF-S 
$F450_{vega}$-limited $(F300-F450)_{vega}$ data.}
\label{fig:u-b hcmhist}
\end{center}
\end{figure}

\begin{figure}
\begin{center}
\centerline{\epsfxsize = 3.5in
\epsfbox{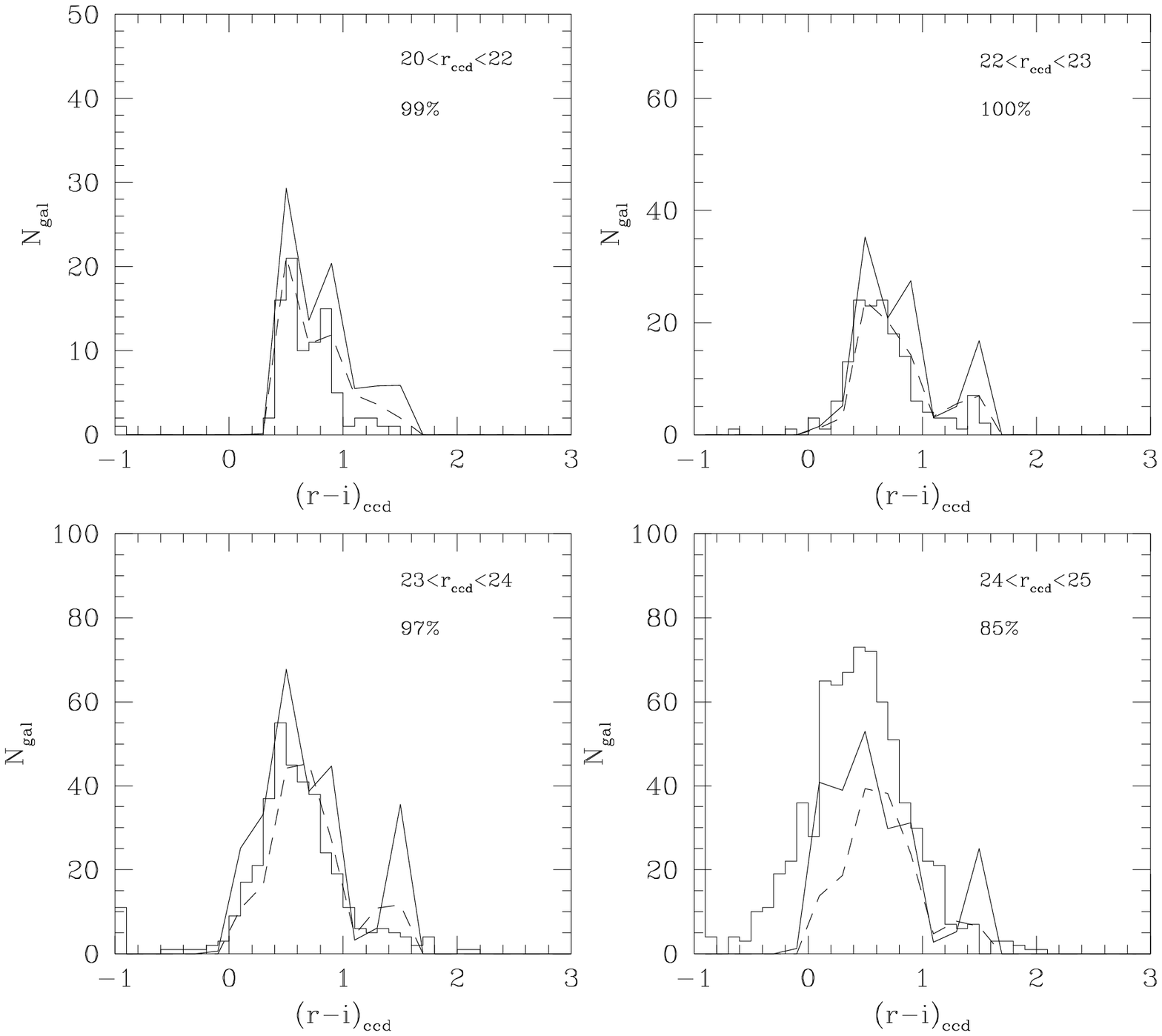}}
\caption{$r_{ccd}$-limited $(r-i)_{ccd}$ colour-magnitude histograms for the 
WHDF data. The percentages indicate the colour completeness 
for each magnitude range. The predictions of the $q_0=0.05$ (solid line) 
and $q_0=0.5$ dwarf (dashed line) models are also shown.}
\label{fig:r-i wcmhist}
\end{center}
\end{figure}

As the counts move bluewards from $I$ to $B$, the pattern is that the
evolutionary excess of galaxies sets in at increasingly lower space
densities of galaxies, indicating the increased sensitivity of the
bluer bands to the increased SFR. However, at $U$ the evolutionary
excess appears smaller than at $B$; it also sets in no earlier in
terms of galaxy number density. The reason is that the $U$ band is
unique in that it lies shortward of the 4000\AA~ break even at $z=0$.
Thus the K-corrections for late types are less than in $B$ by about a
magnitude at $z>0.5$ and indeed are negative for $z>1$. This means
that the no-evolution $U$ models are intrinsically steeper than in $B$
at $U\sim24$ and lie closer to the $U$ data, which in fact has a similar
slope to the $B$ data. Thus less $U$ evolution might, in principle, be
thought to be needed. The Bruzual \& Charlot models, however, predict
a similar amount of evolution in $U$ as in $B$, about 1 magnitude at $z>1$.
This leads to the $U$ evolution models being slightly higher than the
data at $U\sim24$ whereas the $B$ evolution models are slightly lower
than the data. This results in a problem for the models in fitting the
detailed $(U-B)$ colours at $B\ga24$, as we discuss further below with
regard to the $n(u-b)_{ccd}$ histograms in Fig. \ref{fig:u-b wcmhist} and the
$(u-b)_{ccd}:(b-r)_{ccd}$ distribution in Fig. \ref{fig:colour tracks}(b).

\begin{figure}
\begin{center}
\centerline{\epsfxsize = 3.5in
\epsfbox{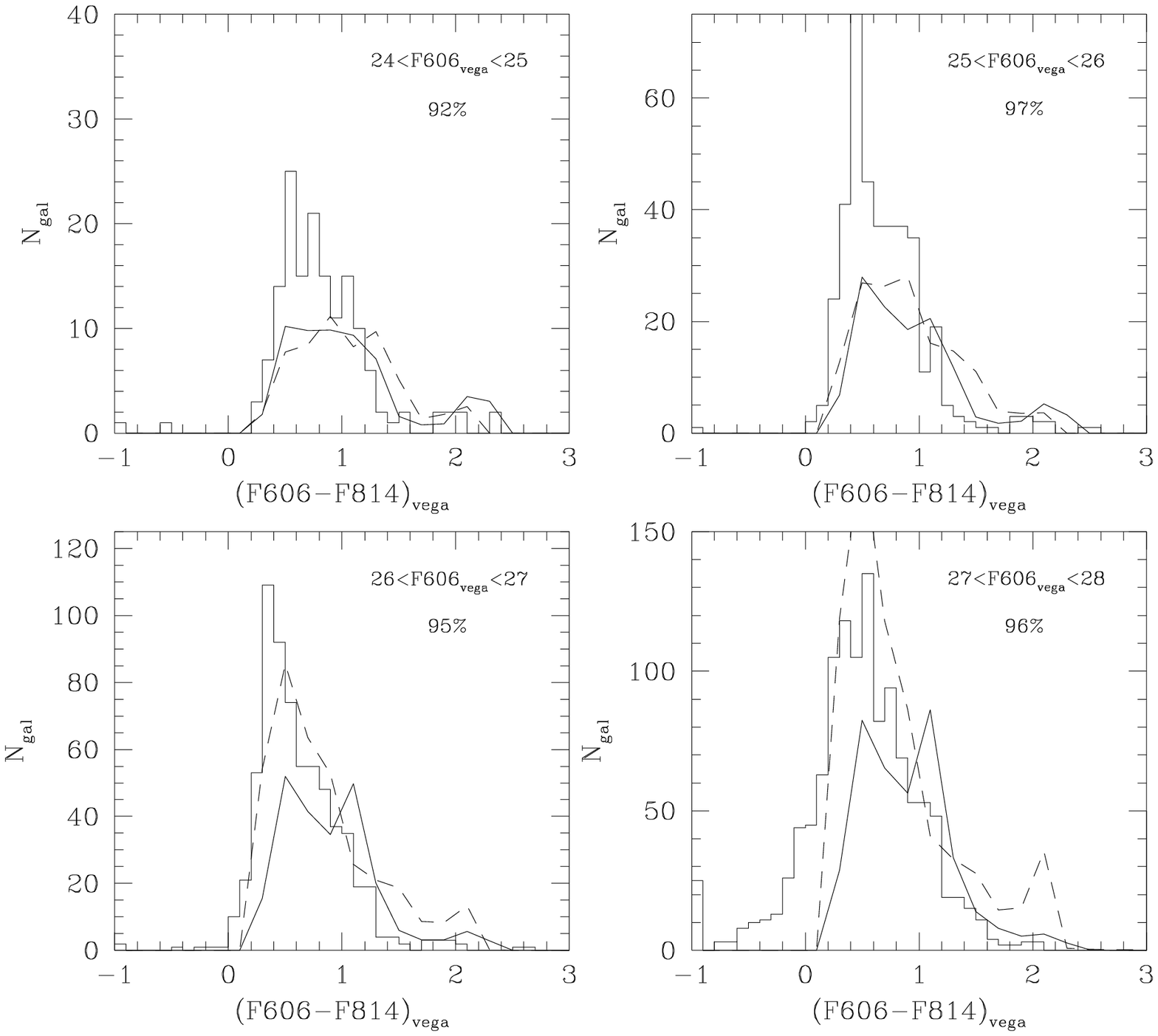}}
\caption{As Figure \ref{fig:r-i wcmhist} but now for the HDF-N and HDF-S 
$F606_{vega}$-limited $(F606-F814)_{vega}$ data.}
\label{fig:r-i hcmhist}
\end{center}
\end{figure}

The q$_0$=0.5 evolutionary model (without the added dE component) fits the 
counts in the range
$18\la B\la 25$ whereas the q$_0$=0.05 and the similar $k=0$,
$\Omega_\Lambda$=0.7, $\Omega_0$=0.3 models extend the fit to
$18\la B\la 27$. It is well known that the 4$\times$ smaller volume
available to $z=4$ in the q$_0$=0.5 case makes it impossible to fit
the faintest optical data with a PLE model (Yoshii \& Takahara 1988,
Guiderdoni \& Rocca Volmerange 1990, Koo et al. 1993). Metcalfe et al.
(1996) therefore invoked the extra dE population at high redshift to
improve the fit of this model. The parameters for this population are
also given in Table \ref{tab:model lf}. Essentially, 
these galaxies have a constant
star-formation rate between $z_f=9.9$ and $z=1$ when the SFR cuts
off. This causes the galaxies to dim dramatically in brightness, till they
end up at the present day with an  M$^*$ some 5 magnitudes fainter 
than the normal M*($z=0$) and with the
colour of an E/SO. This is in the spirit of a
disappearing dwarf model (e.g. Babul \& Rees 1992, Babul \& Ferguson 1996, see
also Phillipps \& Driver 1995 for a discussion)
built in the Bruzual
\& Charlot `single-burst' framework. Clearly it is not unique and if
q$_0$=0.5 then other models such as dynamical merging may also produce
the increased numbers of galaxies required at faint magnitudes.

A further q$_0$=0.5 model that we consider explicitly here for the
first time is a model where we use a steeper faint end slope for the
Scd and Sdm luminosity functions, $\alpha=-1.75$ rather than
$\alpha=-1.5$. Although this slope is steeper than seen in local
luminosity functions, if the local under-density is real then it could
be that the local luminosity function is deficient in faint
galaxies due to biasing. Then even cluster-free luminosity function
estimates may give misleading results, assuming as they do that the
luminosity function form is independent of galaxy
environment. Alternatively, the late type galaxy luminosity function
may be assumed to evolve to steepen from $\alpha=-1.5$ to
$\alpha=-1.75$ at $z>1$. Possible evidence for such steepening at
lower redshift has come from Lilly et al. (1995). Such a model gives a
good fit to the optical $UBRI$ counts in Figs. \ref{fig:u counts} -
\ref{fig:i counts}. We also find that it is in reasonable agreement
with the galaxy $n(z)$ data at $B<22.5$ and $22.5<B<24$. McCracken et.
al. (in prep.) find that the $H$ counts are only slightly over-predicted
by such a model at $H=28$. As they point out, the flat slope of the
$H$ counts seems to disallow any such steepening of the luminosity
function in the q$_0$=0.05 cosmologies.

\subsection {Galaxy Colour Distributions}

The matching of both the q$_0$=0.05 and  the  q$_0$=0.5 (dE and steep LF)
evolutionary models to the count data over a wide wavelength range does not
imply that the colour distributions predicted by the models will agree with the
data, as the colours are more sensitive to model details. In Figs. \ref{fig:u-b
wcmhist} - \ref{fig:r-i hcmhist} we show in the form of histograms the 
comparison between our models and
the data in both the HDF and WHDF. Generally we regard the agreement in
all bands as good, at least in terms of where the broad body of the galaxy
colours lie. The exception is in the UV, where both the HDF 
$(F300-F450)_{vega}$ and the WHDF $(u-b)_{ccd}$ colour data peak some 0.5 
magnitudes redder than do the models. Recall that
the spiral types in our models are already assumed to have dust absorption,
A$_B=0.3$ mag at $z=0$. If we added more dust in the context of these models then
we would have difficulty in matching the extended $n(z)$ in the Keck
redshift surveys at $B<24$ (Metcalfe et al. 1996). We shall also argue later
that simply invoking dust extinction laws with a 2200\AA~ dust feature as 
seen in the Galaxy is not supported by the $(u-b)_{ccd}:(b-r)_{ccd}$ 
colour-colour data. The only other parameter which can potentially affect 
the UV colour of a star-forming galaxy 
is the IMF. We shall look further at what changes might improve the fit 
of the $(u-b)_{ccd}$ colours in Sect 6.4.

Metcalfe et al. (1996), Madau et al. (1996) and Shanks et al. (1998) all 
assumed on the basis of the Bruzual \& Charlot models that in the HDF 
$(F300-F450)_{vega}>0$
corresponded to $z\ga2$ galaxies. However, the above problem, together with the
sensitivity of $U-B$ colours to both dust and Lyman absorption encourages us to
use the extensive spectroscopic data in the HDF to check  the reliability of
our models. We shall see in Fig. \ref{fig:ccreds}(b) that 
$(F300-F450)_{vega}$=0 clearly
discriminates spectroscopic galaxy redshifts, above and below $z=2$. Fig.
\ref{fig:uvdrop} also shows that the HDF UV dropout galaxies accurately fall in
their predicted place for $z\ga2$ galaxies in the 
$(F450-F606)_{vega}:(F606-F814)_{vega}$
diagrams, which strongly suggests that $z\ga2$ galaxies can be discriminated from
$z<2$ galaxies  on the basis of their ground-based $B-R:R-I$ colours (see also
fig. 4a of Metcalfe et al. 1996). Thus it seems that, whatever the model
uncertainties at $z=1$, by $z=2$ the effects of Lyman absorption dominate,
leaving the model $(F300-F450)_{vega}$ colours roughly correct at $z=2$. 
Metcalfe et al.
then went on to determine the fraction of HDF UV dropouts at $27<B<28$ and
found $\sim$47$\pm$7\%, a fraction which compares well with that predicted by
the models. The results of calculating the $z\ga2$ UV luminosity density with
these criteria are discussed in Section 6.5 below (and by Shanks et al. 1998).

In the WHDF a simple $(u-b)_{ccd}$ cut is not adequate to discriminate
high $z$ galaxies, partly due to the difference in passbands, but mainly 
due to the brighter magnitude limit and hence lower average redshift of 
the sample. As a result it is necessary to apply a simultaneous cut 
in $(b-r)_{ccd}$, in order to eliminate low $z$ early type galaxies 
(especially for $r$-limited samples). In order to compare with previous
work in the literature we take as a starting point  UV dropout criteria 
which match those of 
Steidel et al. (1999) as closely as possible; these transform to a cut with 
approximately $(u-b)_{ccd}>$1 (and $-0.1<(b-r)_{ccd}<1.6$) in our magnitude 
system, and a limit of $r_{ccd}<$25.25.
We also assign $(u-b)_{ccd}$ colours (which are really blue limits) to those 
galaxies which were not detected
in $u$ by assuming their $u_{ccd}$ magnitude to lie 
at the $3\sigma$ limit ($u_{ccd}=26.8$). Given the relative 
depths of our $u$-band data and Steidel et al.'s, this is roughly 
equivalent to the $1\sigma$ magnitude they adopted for their 
non-detections. Many of these galaxies then fail the selection criteria due to
being too blue in $(u-b)_{ccd}$. 
The main purpose of this procedure is to remove faint, 
low-$z$ early-type galaxies which might otherwise be mistaken for high-$z$ 
dropouts. However it should be noted that it also loses potential 
high-$z$ candidates in the process. We shall see how some of these might be
recovered in section 6.4.

With these criteria (and an eyeball check to exclude photometry problems such
as mergers) we detect 32 UV dropouts in 49 arcmin$^{2}$, 
i.e. 0.65$\pm$0.12 arcmin$^{-2}$, to
$r_{ccd}<25.25$ which is comparable to the 1.18$\pm$0.04 arcmin$^{-2}$ sky 
density for $z\sim3$ galaxies found by Steidel et al. (1999) at 
$R_{AB}<25.5$ (or $r_{ccd}<$25.25) and the upper limit of
3.3$\pm$0.3 arcmin$^{-2}$ for UV dropouts set by Guhathakurta et al. (1990) 
at a similar $R$-band limit.  Applying  the same blending 
corrections assumed by Steidel et al. (1999) for the brighter galaxies 
would increase these three observed sky densities by a factor of 1.3. 

\subsection{WHDF Colour-Colour Distributions}

\begin{figure*}
\begin{center}
\centerline{\epsfxsize = 7in
\epsfbox{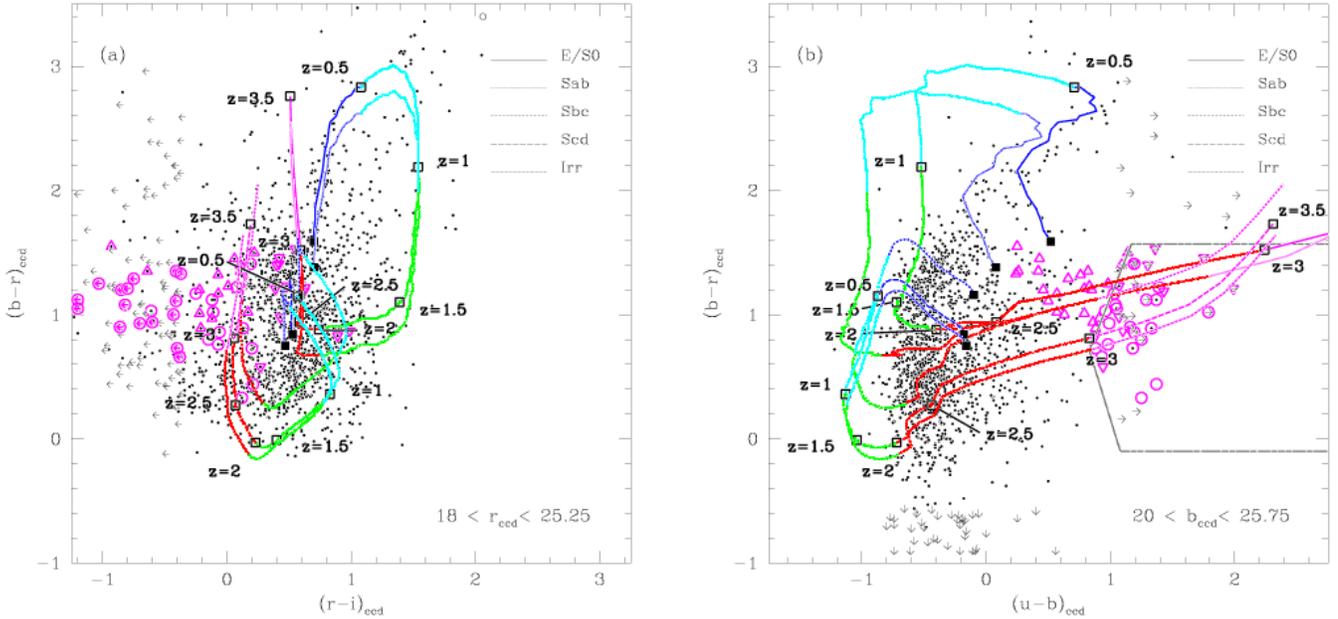}}
\caption{(a) $b_{ccd}$-limited $(r-i)_{ccd}:(b-r)_{ccd}$ and (b)
$(u-b)_{ccd}:(b-r)_{ccd}$ colour-colour
diagrams for the WHDF data, showing the evolutionary tracks predicted
from our Bruzual \& Charlot models for the five galaxy types for $0<z<3.5$.
Boxes indicate 0.5$z$ intervals. The tracks are colour coded according
to redshift; $0<z<0.5$, blue;
$0.5<z<1$, cyan; $1<z<2$, green; $2<z<3$, red; $z>3$, magenta.
Arrows show limits to colours for galaxies
not detected in $i$, $r$ or $u$, assuming the missing object has a magnitude
at the $3\sigma$ limit for each band. The enclosed, dotted area on the
$(u-b)_{ccd}:(b-r)_{ccd}$ plot shows the $u$-dropout selection of
Steidel et al. (1999) transformed into our passbands. The locations of our
initial sample of 32 high-$z$, UV-dropout galaxies based on the criteria
of Steidel et al. (1999) are shown on both plots as circles and inverted
triangles, the latter symbol indicating candidates which pass
the Steidel et al. criteria but are too red in $(r-i)_{ccd}$. The extra
candidates which increase our final sample to 43 dropouts are shown as upright
triangles (note that all the candidates are
selected from an $r$-limited sample and so may not have corresponding data
points or arrows on these plots).}
\label{fig:colour tracks}
\end{center}
\end{figure*}

We first discuss the WHDF $(r-i)_{ccd}:(b-r)_{ccd}$ colour-colour diagram.
This is shown in Fig. \ref{fig:colour tracks}(a). 
The data shows two unmistakable
features. The first is the plume of red galaxies which moves redwards from
$(b-r)_{ccd}\approx1.6$ to $(b-r)_{ccd}\approx3$ at  $(r-i)_{ccd}\approx0.7$. 
The second is the `hook' feature which 
lies below this red plume and contains the majority of the galaxies.
Interestingly, the models seem to show the same generic features, mostly due to
the redshifting of the H/K break through the passbands. The red plume is
identified with early type galaxies, with their decreasing UV spectrum shortward
of the Ca II H/K break  moving through the $b$ band and making the 
$(b-r)_{ccd}$ colour
redder until $z\sim0.5$. At $0.5\la z\la0.9$ the H/K feature moves through the $r$
band making the $(b-r)_{ccd}$ colour more blue and the $(r-i)_{ccd}$ colour
more red. This
shift of the $(b-r)_{ccd}$ colour to the blue is helped by the 
$\tau$=2.5 Gyr evolution
whose effect on the UV spectrum is increasingly felt in the $b$ band at
$z\ga0.5$. The hook feature is identified with spiral galaxies, 
with again the H/K
break moving through the $b$ band until $z=0.25$ when the increasing spiral UV
spectrum  is redshifted into the $b$ band. Then, helped by the H/K break moving
through the $r$ band at $z\ga0.5$, $(b-r)_{ccd}$ continues to become 
bluer until the
Lyman $\alpha$ forest enters the $b$ band at $z\sim2.3$. The H/K break entering
the $r$ band at $z\sim0.5$ causes $(r-i)_{ccd}$ to become redder before 
the same feature entering the $i$ band at $z>0.8$ causes $(r-i)_{ccd}$ to 
become bluer again.

\begin{figure*}
\begin{center}
\centerline{\epsfxsize = 7in
\epsfbox{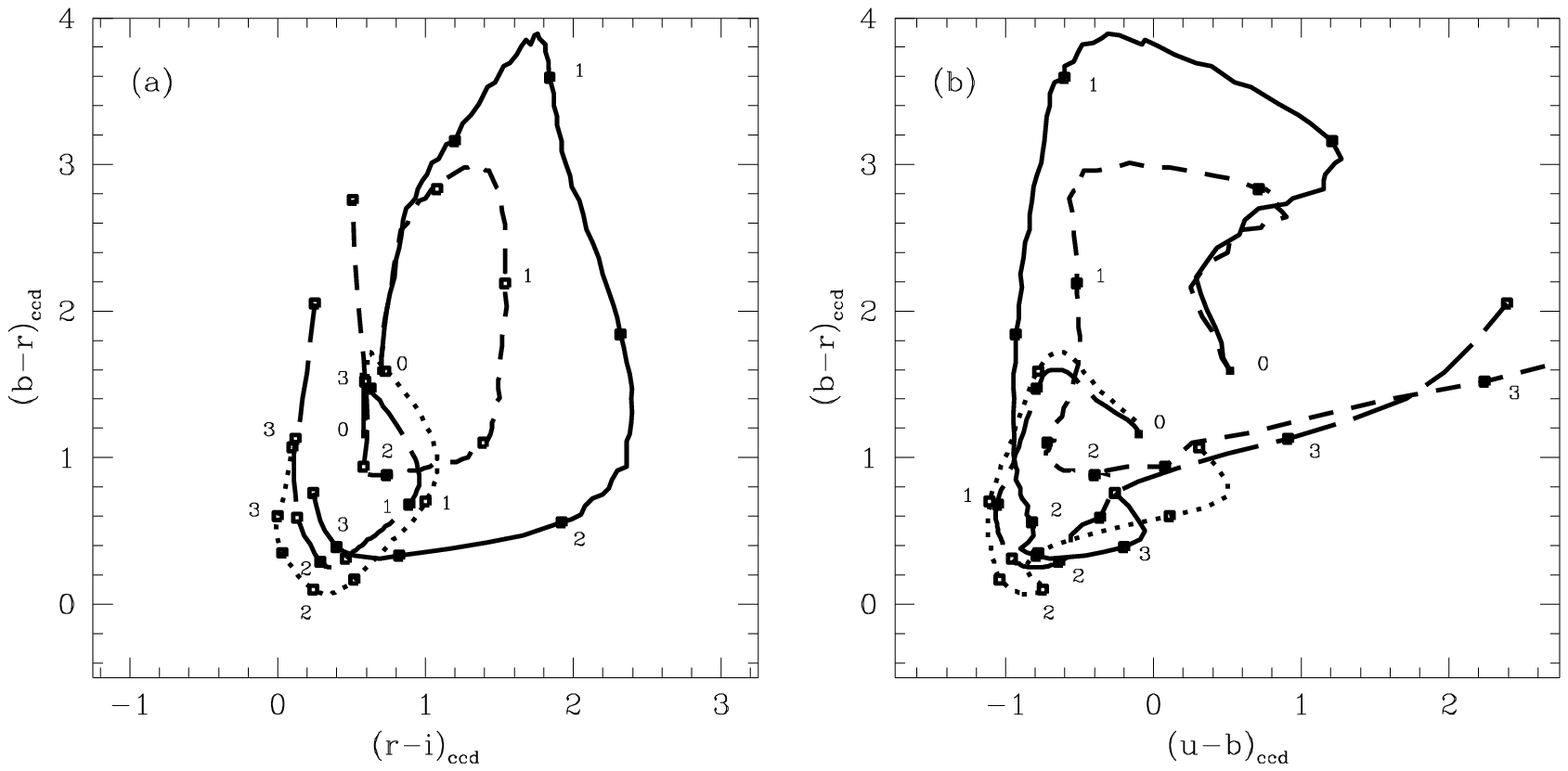}}
\caption{The K-corrections compared to the evolutionary tracks plotted in both
$(r-i)_{ccd}:(b-r)_{ccd}$ and $(u-b)_{ccd}:(b-r)_{ccd}$ planes.
Solid lines - E/S0 K-correction;
short dash - E/S0 $q_0=0.05$ evolution; dots - Spiral K-correction; long
dash -  Spiral $q_0=0.05$ evolution. Boxes indicate $0.5z$ steps for
$0<z<3.5$. Small numerals indicate redshift.}
\label{fig:k-corr}
\end{center}
\end{figure*}

The fact that the models show these features suggest that they are broadly
correct.  Indeed the evolutionary models show increased agreement with the
data than simply using the K-corrections, especially in the case of the
early-types beyond $z\sim0.8$. This is shown in Fig. \ref{fig:k-corr} where we
compare the colour tracks from our evolutionary models to the K-corrections
calculated from our Bruzual models. In the case of the spirals, the K-corrections
provide a better first-order description of the colour-colour tracks than
for the early-types but evolution at low redshift and dust reddening at high
redshift also help; evolution makes the spirals slightly bluer than the 
K-corrections in $(b-r)_{ccd}$ and $(r-i)_{ccd}$ at $z\la1$ but then 
the K-corrections and
evolutionary model colours become very similar in the $1\la z\la3$ range. The
effect of spiral dust actually makes the evolutionary model redder in 
$(b-r)_{ccd}$ at
$z\ga1.5$ and here it is the effect of evolution and dust combined which
improves the fit of the Sbc model to the data.

However, in detail there remain differences between the models and the
$(r-i)_{ccd}:(b-r)_{ccd}$ data. The most serious problem in 
Fig. \ref{fig:colour tracks}(a) is that the  early-type galaxies 
appear to be too blue in $(r-i)_{ccd}$ at $z\ga0.5$, indicating more flux
shortwards of the H/K break than predicted in the models. This problem is a
feature of all models tested including those with a Salpeter IMF. Increasing
$\tau$  helps by extending the period of active star-formation to lower
redshifts but then the $(b-r)_{ccd}$ colours tend to turn to the blue 
at $z=0.5$ too
quickly in comparison with the data. It is possible that models with a less
sharp H/K break at high redshift may help because of decreased metallicity
but so far we have not tested this hypothesis further.

\begin{figure*}
\begin{center}
\centerline{\epsfxsize = 7in
\epsfbox{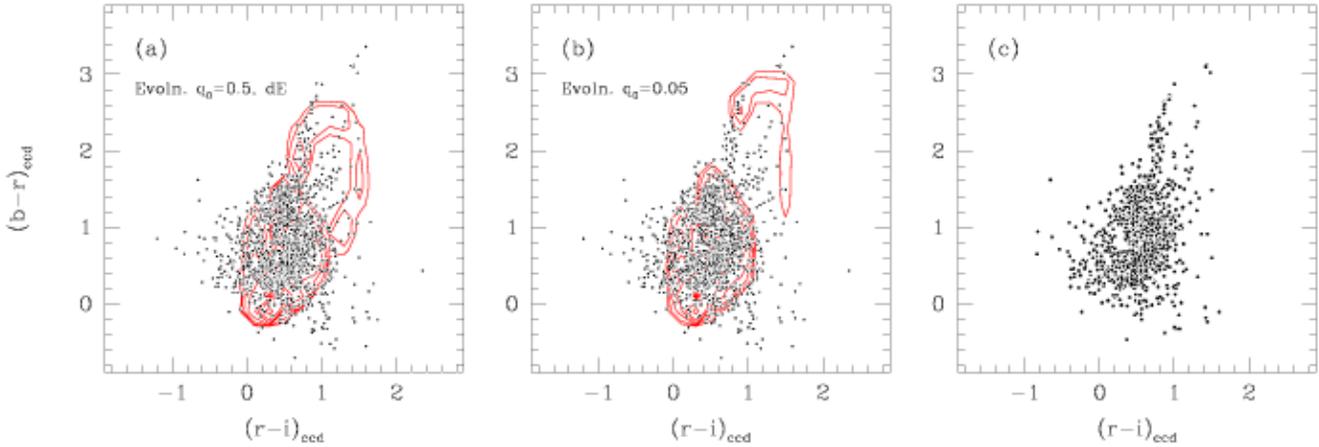}}
\caption{(a) the $(r-i)_{ccd}:(b-r)_{ccd}$ colour-colour diagram for 
the WHDF data limited at $b_{ccd}<25.75$ compared with the predicted 
contours for the $q_0=0.5$ dE  model. The contours are equally spaced in 
log surface density.(b) as (a) but for the $q_0=0.05$ model. (c) the data
points shown without the contours for clarity - note the prominent `hook' of 
spirals and the `finger' of early-types.
}
\label{fig:cc-contour}
\end{center}
\end{figure*}

Fig. \ref{fig:cc-contour} compares the  predicted number density of galaxies at
each point in the WHDF $(r-i)_{ccd}:(b-r)_{ccd}$ colour-colour diagram with 
the data. It 
can be seen that the q$_0$=0.05 and q$_0$=0.5 dE 
evolutionary models both generally give a good fit. In both cases the 
spirals are
well fitted. As in Fig. \ref{fig:colour tracks}(a), the evolutionary models 
predict
early-type colours which are too red in $(r-i)_{ccd}$ for $z>0.5$. It should 
be noted that in the case of the early-types the 
q$_0$=0.5+dE model  appears to give more low redshift early types and thus 
better agreement with the data than the low q$_0$ model. These galaxies are 
estimated to lie at $z\sim0.15$ from their  colours and thus will have 
M$_B\sim-16$. These low luminosity galaxies are therefore excellent candidates 
for  a disappearing dwarf' population and spectroscopy is required to 
confirm their redshifts and investigate the nature of these  galaxies.

The 32 UV dropout, $r_{ccd}<25.25$, candidate $2.5\la z\la3.5$ galaxies 
based on the Steidel et al. (1999) selection criteria
discussed in the previous section have been circled in the 
$(r-i)_{ccd}:(b-r)_{ccd}$ plane in  Fig. \ref{fig:colour
tracks}. However, it is clear that we can use $(r-i)_{ccd}$ to 
refine further our candidate selection, as some of the dropout galaxies have 
$(r-i)_{ccd}$ colours that are too red for them to be at high redshift. At 
first glance this would seem to reduce our numbers even further. However, if
we re-examine those galaxies with no measurable $u$ magnitudes which were 
assigned $u_{ccd}$ magnitudes at the $3\sigma$ limit and
subsequently failed Steidel et al.'s selection criteria (see section 6.3), we
find that many of these have $(r-i)_{ccd}$ colours which indicate they are
$2.5\la z\la3.5$ galaxies ($(r-i)_{ccd}<0.2$). The net result is that our 
number of dropout candidates increases from 32 to 43, or 
$0.88\pm0.13$ armcmin$^{-2}$. It is this sample that
we take forward to our discussions in Sections 6.7 and 6.8.
The q$_0$=0.05 and 0.5 evolutionary models predict a sky density of 2.6-2.9
arcmin$^{-2}$ for $2.5\la z\la3.5$ galaxies at $r_{ccd}<$25.25.
Thus the sky densities predicted by both models
are in reasonable agreement with the sky density of $z\sim3$ galaxies found
in the WHDF and by Steidel et al. (see section 6.3), especially if we apply 
Steidel et al.'s bright-end 30\% blending correction (which should be relevant 
to both datasets). This suggests that Lyman break galaxies
may be interpreted as the high luminosity tail of normal spiral galaxies
as their SFR increases exponentially back towards
their formation epoch. This conclusion applies both in the q$_0$=0.05 case
where 99\% of $z\ga3$ galaxies are Sbc/Scd/Sdm and even in the q$_0$=0.5
`disappearing dwarf' dE case where 86\% of
the $z\ga3$ galaxies are Sbc/Scd/Sdm and the remainder mostly E/SO/Sab with
less than a 1\% contribution from dE.

It can also be seen from Fig. \ref{fig:colour tracks}(a) that many other 
galaxies appear in the same $(r-i)_{ccd}:(b-r)_{ccd}$ location as our
UV-dropout candidates but with no UV dropout. 
Clearly there is the possibility that these are also $z\sim3$ candidates. 
Madau (1995) has used the incidence of  Lyman limit systems in QSO spectra 
to argue that almost all ($\sim$ 80\%) $z\ga3$ galaxies should show UV 
dropout. This argument works well at
$z\ga3.5$, because there is a relatively large redshift range between 
$3\la z\la3.5$ for a Lyman limit system to enter the $U$ band. However, the 
empirical mean redshift of the UV dropout galaxies of Steidel et al. (1999) 
lies at $z=3.04\pm0.24$ and here
the above argument reverses, because there is then only a very limited redshift
range in which a Lyman limit system can intervene in the $U$ band. 
Thus at $z\sim3$ there is the possibility that 80\% of galaxies do {\it not} 
show UV dropout. Taking all the galaxies with similar 
$(r-i)_{ccd}:(b-r)_{ccd}$ colours  to the UV dropout
galaxies i.e. those with $-0.1<(b-r)_{ccd}<1.6$ and $(r-i)_{ccd}<0.2$, 
we find $\sim220$ new candidate $z\approx3$ galaxies. Spectroscopy of these 
candidates are required
to determine how many of them are actually at high redshift.

The WHDF $(u-b)_{ccd}:(b-r)_{ccd}$ diagram in Fig. \ref{fig:colour tracks}(b) 
shows the same two populations as in $(r-i)_{ccd}:(b-r)_{ccd}$ although 
slightly less distinctively. The plume of early types lies at 
$(b-r)_{ccd}>1.6$ with the spirals now occupying a broad strip below them. 
Again the E/SO model seems to fit the E/SO
plume to $z\sim0.5$ with the H/K break exiting the $b$ band at $z\sim0.25$ 
and then the evolved spectrum shortwards of 2500\AA~ entering the $u$ band 
at $z\ga0.5$ which takes the $(u-b)_{ccd}$ colour blueward until $z\sim0.8$, 
when, with both $u$ and $b$ on the steep but constant slope of the UV `upturn' 
spectrum, a constant $(u-b)_{ccd}=-0.4$ colour is maintained until the 
spectrum reddens again when Lyman
$\alpha$ absorption enters the $u$ passband at $z\ga1.5$. The strong bluewards 
shift
in $(b-r)_{ccd}$ between $0.7\la z\la1.5$ is caused both by the H/K break being
redshifted through $r$ and the UV `upturn' spectrum at $\lambda<2500$\AA~ being
redshifted through $b$.
The spiral $(u-b)_{ccd}:(b-r)_{ccd}$ tracks again first move bluer in
$(u-b)_{ccd}$ and redder in $(b-r)_{ccd}$ due to H/K moving through the 
$b$ band at $0<z\la0.3$. Then $(b-r)_{ccd}$ becomes bluer at $z\ga0.5$ when the 
decreasing spectrum shortwards of the H/K break is redshifted through $r$ 
until the UV `upturn' reaches $r$ at $z=1.5$. Both $u$ and $b$ lie on the 
steeply increasing UV `upturn' spectrum in the range $0.3\la z\la1.5$ which 
maintains the $(u-b)_{ccd}$
roughly constant at $(u-b)_{ccd}\sim-0.8$ before the spectrum reddens 
as dust and then Lyman $\alpha$ increasingly affects the $u$ band at 
$z\ga1.5$.

Also shown on Fig. \ref{fig:colour tracks}(b) (enclosed by the dotted lines) 
is the the area of the 
$(u-b)_{ccd}:(b-r)_{ccd}$ colour-colour plane corresponding to the UV-dropout
selection criteria of Steidel et al. (1999), as discussed in Section 6.3.
The locations of our final sample of 43 $u$-band dropouts is indicated, with
distinct symbols indicating the original 32 satisfying the criteria of
Steidel et al., and those lost and gained from invoking the $(r-i)_{ccd}$
selection. Note that, as the dropouts are chosen from an $r$-limited 
sample, not all have corresponding data points on this $b$-limited plot. It can
be seen that there is good agreement with the location of $z\approx3$ 
galaxies implied by our models, for both early-types and spirals.

As shown by Fig. \ref{fig:k-corr}, in $(u-b)_{ccd}:(b-r)_{ccd}$ the effects of
evolution are again bigger for the early-type model than the late-type model. 
At $0.5\la z\la1.8$ the $(b-r)_{ccd}$ colours for early-type evolution model 
become bluer than the K-corrections  and at $0.5\la z\la0.9$ the $(u-b)_{ccd}$ 
colours for this model also become bluer than the K-correction. 
However, for $z>0.9$ in $(u-b)_{ccd}$ and $z>1.8$ in $(b-r)_{ccd}$ the 
colours for the evolution model are
{\it redder} than the K-correction and this trend is confirmed when intervening
Lyman $\alpha$ absorption enters both $u$ and $b$. For the spirals the
K-corrections seem to give a similar result everywhere to the evolutionary 
model, evolution making the colours slightly bluer than the K-correction 
at low redshift and dust making the evolutionary model slightly redder at 
high redshift. We note
that the K-corrections of Pence (1976) are similar to the Bruzual \& Charlot
K-corrections for $z\la0.5$ and, if anything, slightly bluer in $(u-b)_{ccd}$ at
$z\ga0.5$.

Thus in the redshift range $0.3\la z\la1$ the predicted $(u-b)_{ccd}$ colours 
for the spiral model are much bluer than is observed; the spiral model reaches 
$(u-b)_{ccd} =-1.0$ whereas the data only reach $(u-b)_{ccd}=-0.5$. The
discussion of the $u_{ccd}$ and $b_{ccd}$ photometry zeropoints in 
section 3.2 indicates that errors in the colour zero-point of $0.1$ mag or 
larger are unlikely. In any case we shall see below that the 
same effect is also seen in the independent HDF data. This discrepancy is 
surprising since the UV spectra of  star-forming galaxies are only governed 
by two parameters, dust absorption and, more weakly, the IMF. We postpone to 
Section 7  further  discussion of possible explanations of  
this result.

\subsection{HDF Colour-Colour Distributions}

\begin{figure*}
\begin{center}
\centerline{\epsfxsize = 6in
\epsfbox{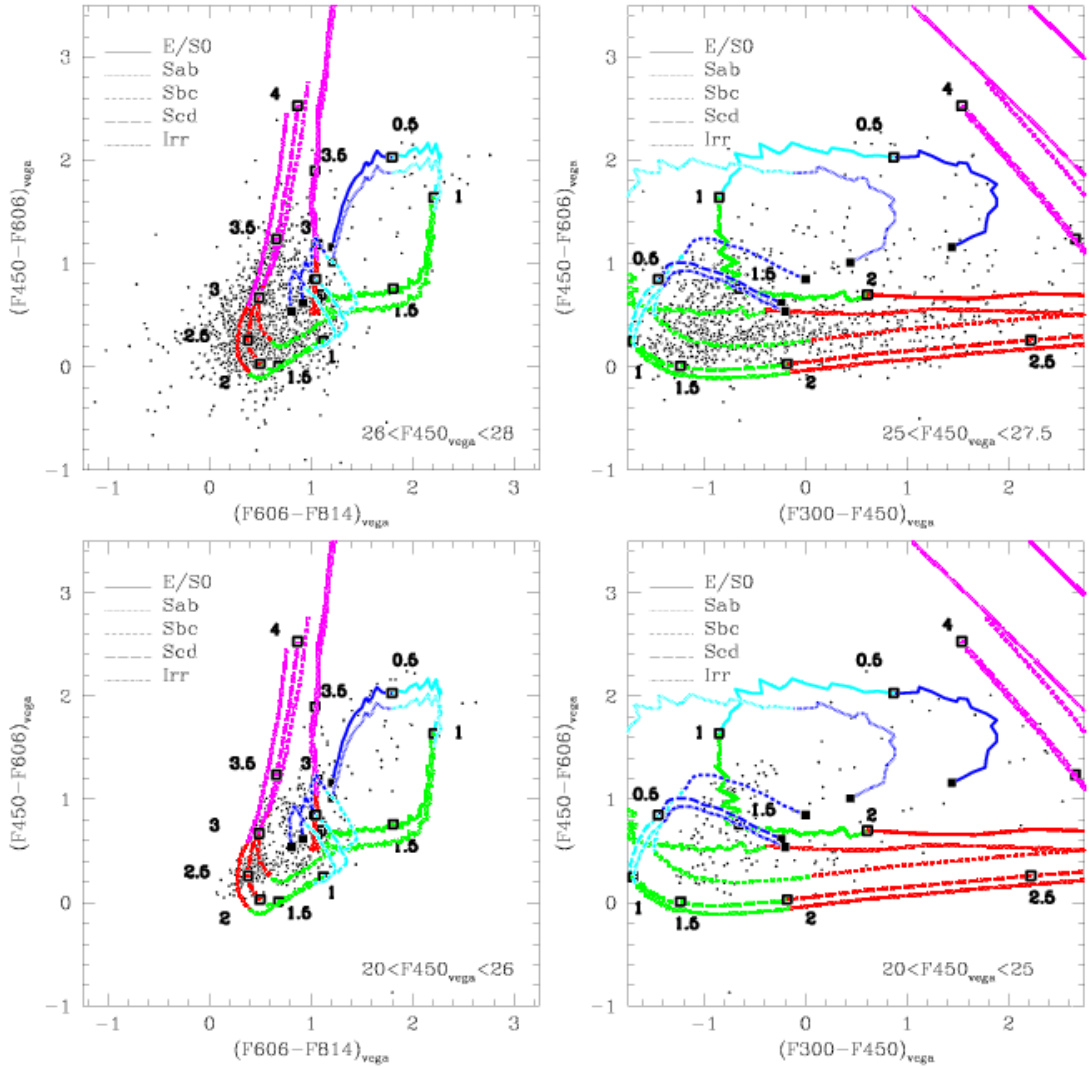}}
\caption{$b$-limited $(F606-F814)_{vega}:(F450-F606)_{vega}$ and
$(F300-F450)_{vega}:(F450-F606)_{vega}$ colour-colour diagrams for the
HDF-N and HDF-S data,
split into bright and faint magnitude ranges. The evolutionary tracks predicted
from our Bruzual \& Charlot models are shown out to $z=4$ for the
five galaxy types. Filled squares show $z=0$, open squares with numerals
indicate $0.5z$ steps.}
\label{fig:cc}
\end{center}
\end{figure*}

\begin{figure*}
\begin{center}
\centerline{\epsfxsize = 7in
\epsfbox{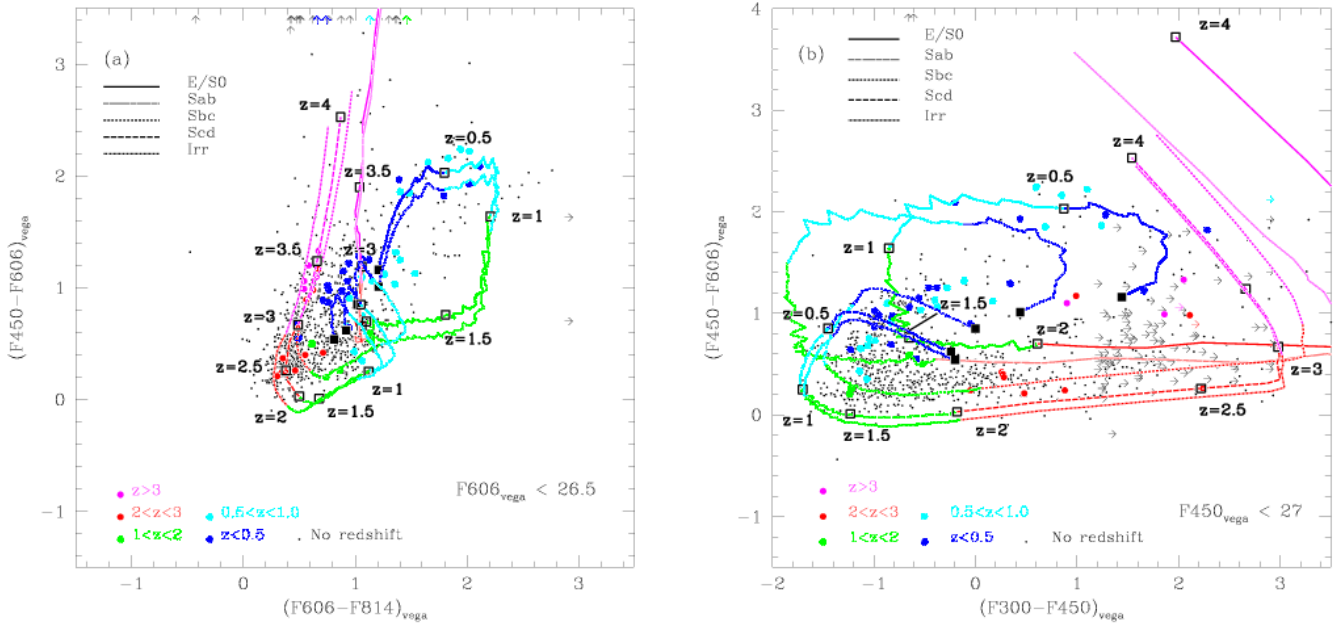}}
\caption{(a) $(F606-F814)_{vega}:(F450-F606)_{vega}$ and 
(b) $(F300-F450)_{vega}:(F450-F606)_{vega}$colour-colour diagrams
for the HDF-N and HDF-S
data showing the location of galaxies with known redshifts (coloured dots,
with redshifts as indicated in the legend) compared with the model tracks out
to $z=4$. The
tracks are colour-coded according to redshift range in the same fashion as for
the galaxies. Arrows indicate limits to colours for galaxies not detected in
one band, assuming a magnitude at the $3\sigma$ limit for the missing objects.}
\label{fig:ccreds}
\end{center}
\end{figure*}

\begin{figure*}
\begin{center}
\centerline{\epsfxsize = 4in
\epsfbox{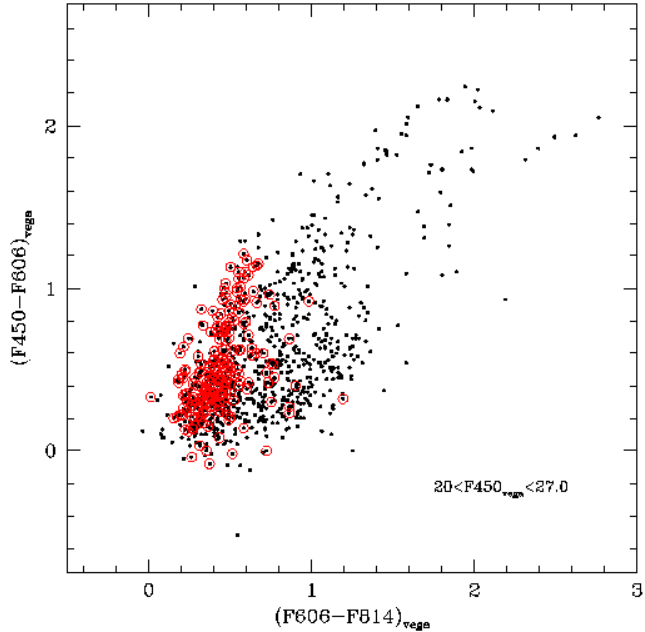}}
\caption{The location of $(F300-F450)_{vega}$ drop-out galaxies in the HDF
$(F606-F814)_{vega}:(F450-F606)_{vega}$
plane. Circled dots indicate galaxies with no measured $(F300-F450)_{vega}$. By
comparing with Fig. \ref{fig:ccreds}(a) it can be seen that these drop-out
galaxies have colours very close to those predicted by the models for
galaxies with $z>2$ in this colour plane.}
\label{fig:uvdrop}
\end{center}
\end{figure*}

\begin{figure*}
\begin{center}
\centerline{\epsfxsize = 7in
\epsfbox{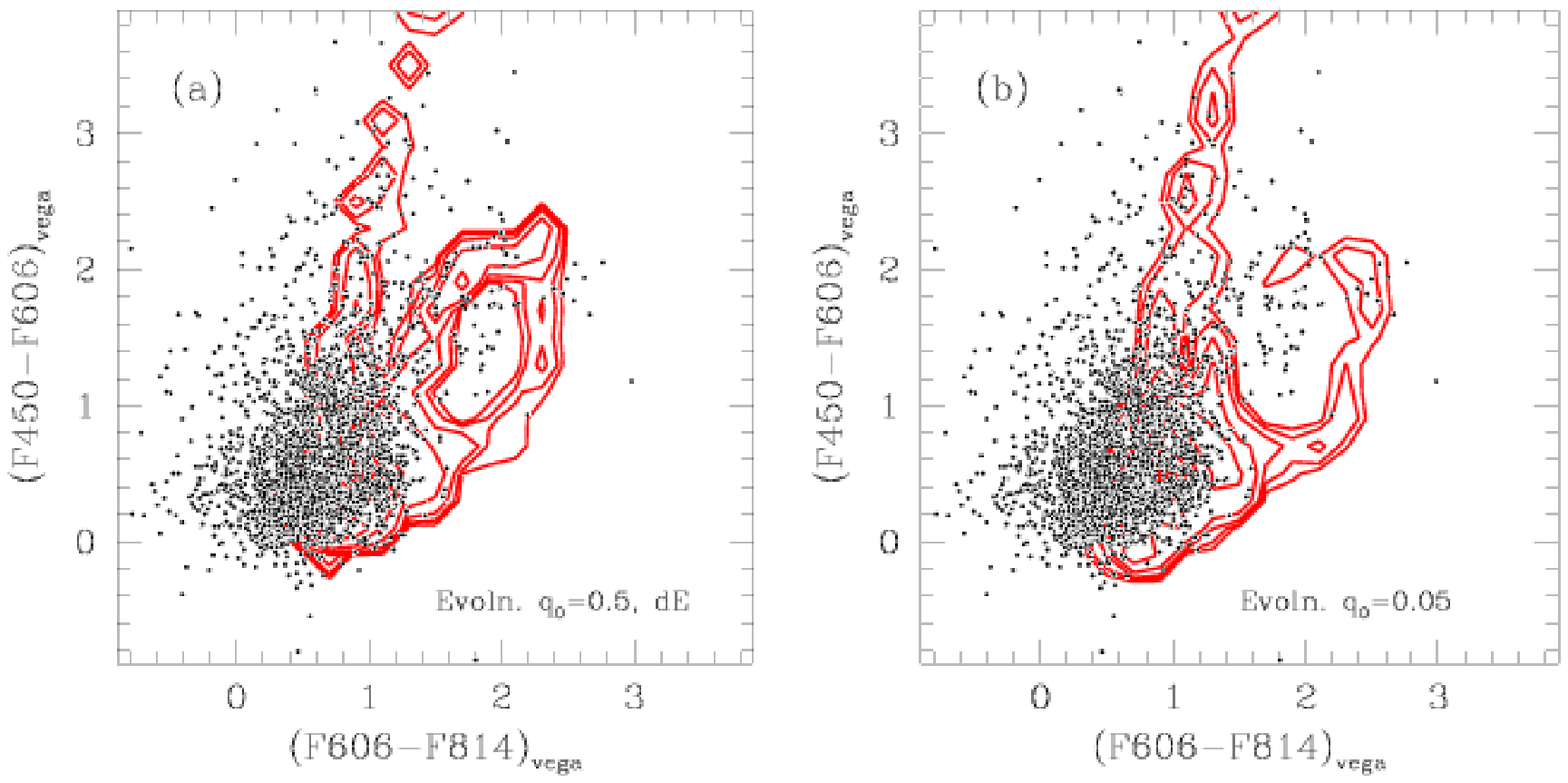}}
\caption{(a) the $(F606-F814)_{vega}:(F450-F606)_{vega}$ colour-colour
diagram for the HDF-N and HDF-S data, limited at $F606_{vega}<28$
compared with the predicted density contours of the $q_0=0.5$ dE models. The
contours are equally spaced in log surface density. (b) as (a) but for the 
$q_0=0.05$ model.}
\label{fig:hdfcontour}
\end{center}
\end{figure*}

An HDF-N colour-colour comparison with similar  models has previously been
briefly discussed by Metcalfe et al. (1996). We now add in the HDF-S data and
look at the combined $(F606-F814)_{vega}:(F450-F606)_{vega}$ and 
$(F300-F450)_{vega}:(F450-F606)_{vega}$ diagrams at both a bright and a
faint limit in Figs. \ref{fig:cc}. The most prominent feature remains the line
of galaxies moving redward in $(F300-F450)_{vega}$ while remaining blue 
in $(F450-F606)_{vega}$. These are
the UV dropouts galaxies expected to be at $z\ga2$ as discussed above.  In
general, the tracks appear broadly consistent with the models and as such
provide strong support for the approach used here. The 
$(F606-F814)_{vega}:(F450-F606)_{vega}$ tracks again
seem good fits; the Sbc track in particular gives an excellent fit to many
of the bluer galaxies, although when shifted to the rest colours of Scd and
Sdm's gives slightly too blue a colour in $(F450-F606)_{vega}$. 
In the bright data the plume
of early-type galaxies is not seen as clearly as in the WHDF in either
$(F300-F450)_{vega}:(F450-F606)_{vega}$ or 
$(F606-F814)_{vega}:(F450-F606)_{vega}$ but the models again give a reasonable representation of
the redder galaxies. Again, as for the WHDF $(u-b)_{ccd}$, the Sbc model 
predicts too blue
$(F300-F450)_{vega}$ colours for spirals at $z=1$ and the arbitrary shifting 
of the Sbc curve
to the rest colours of Scd and Sdm galaxies leaves their tracks not fitting
well in this redshift range either. At fainter limits, it can be seen 
that galaxies predicted to lie at $z\sim2$ dominate the data.

Next, in Fig. \ref{fig:ccreds}(a),(b) 
we plot the HDF-N and HDF-S galaxies with 
spectroscopic redshifts on the $(F300-F450)_{vega}:(F450-F606)_{vega}$ and
$(F606-F814)_{vega}:(F450-F606)_{vega}$ diagrams.
The points are colour-coded
according to the galaxy redshift range and the model tracks are similarly
colour-coded. It can be seen there is good general agreement between the
predicted and observed redshift ranges, again firmly supporting the idea that
these simple models do an excellent job of describing the colours of faint
galaxies.

Fig. \ref{fig:uvdrop} shows how the HDF UV dropouts populate the correct part
of the HDF $(F606-F814)_{vega}:(F450-F606)_{vega}$ diagram for 
$z\ga 2$ galaxies. Again this indicates that
ground-based $(r-i):(b-r)$ colour-colour combinations may be effectively 
used to discriminate $z\ga 2$ galaxies without the need of a UV band.

Fig. \ref{fig:hdfcontour} compares the  predicted number density of galaxies
at each point in the $(F606-F814)_{vega}:(F450-F606)_{vega}$ colour-colour 
diagrams with the data at $F606<28$.
It can be seen that the q$_0$=0.05 and q$_0$=0.5 dE
evolutionary models both give a good fit with the density peaking in both the
models and the data at the 
colours of $z\sim2$ spirals. The fact that the spiral 
models fit so well throughout the range $0<z\la3$ is again a strong argument
that the evolutionary history of the SFR is well described by the $\tau$=9Gyr
exponentially increasing SFR used here. The early-type galaxies are not so
clearly seen in this HDF data compared to the equivalent WHDF data shown in
Fig. \ref{fig:cc-contour}.

\subsection{Sizes}

\begin{figure}
\begin{center}
\centerline{\epsfxsize = 3.5in
\epsfbox{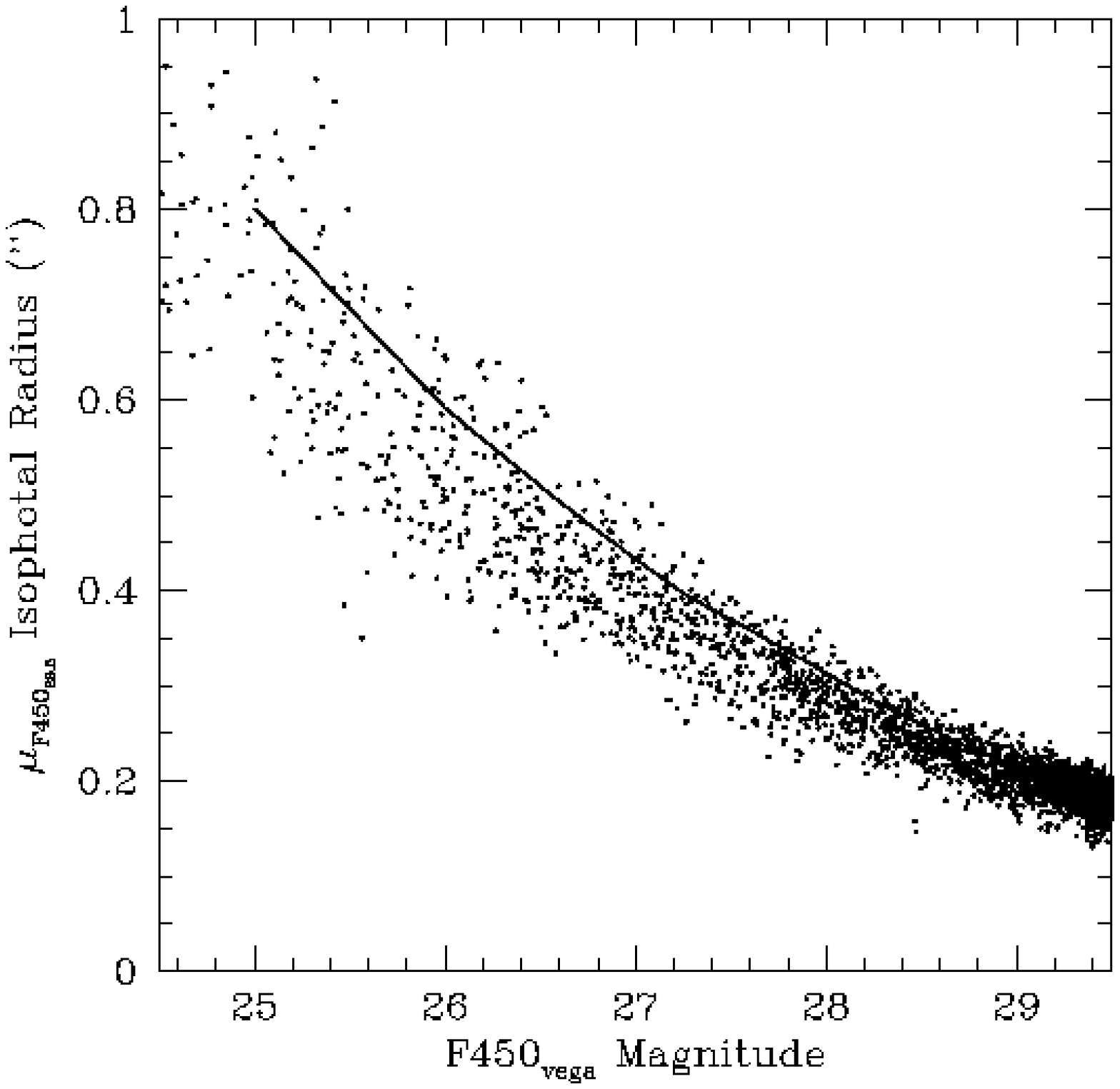}}
\caption{
The isophotal radii versus $F450_{vega}$ magnitude relation for the 
HDF-N data (dots) compared with the prediction for the mean radii from 
a simulation of an HDF frame (solid line) based on our q$_0=0.05$ 
evolutionary model.
}
\label{fig:sizes}
\end{center}
\end{figure}

It has been claimed that the angular sizes of faint HDF galaxies are small
compared to local galaxies (Roche et al. 1998 and references therein),
suggesting there has been size as well as luminosity evolution. We have
examined this question in the light of our simple evolutionary models by
adapting the simulation procedures used to replicate the WHT data in section
3.6 to produce mock HDF frames based on our models. We are then able to apply
our data-analysis procedures to these frames and compare the angular sizes
measured for the artificial galaxies with the real HDF data. As before (see
Paper III), we assume the Freeman (1970) law to relate exponential disk size to
absolute magnitude and the diameter-magnitude relations of Sandage \&
Perelemuter (1990) for $r^{1/4}$ bulges. Fig. \ref{fig:sizes} shows the results
for our $q_0=0.05$ model. The solid line represents the mean isophotal radius
(at $\mu_{F450}=28.5$) of our artificial galaxies, 
whilst the dots show the same
quantity for the $F450$-band HDF-N data. Although small differences exist,
the agreement between model and data is remarkably good. It should be noted
that the apparent isophotal angular radii for the faintest galaxies
($\sim0.2''$) correspond to metric radii of only $\sim2$ kpc ($H_0=50$
kms$^{-1}$Mpc$^{-1}$). This is much smaller than local $L^*$ galaxies, but
demonstrates both the effect of isophotal dimming and the fact that, at
$F450_{vega}\sim29$, in the models we are looking down the luminosity 
function to galaxies with low intrinsic luminosities and small sizes.

We find similar results for the $q_0=0.5$ dE model, if we assume the 
``disappearing dwarf'' population follows the same Sandage \& Perelemuter law
as local bulge dominated galaxies.

\subsection{Star Formation History}

Our evolutionary models imply that the SFR density rises to beyond
$z=2$, according to the $\tau$=9Gyr Bruzual \& Charlot model. This is
at odds with the conclusions of Madau et al. (1996) who used UV fluxes
as an estimate of the SFR and, on the basis of the same HDF-N data,
concluded that the SFR peaks at $z\sim1$ and decreases rapidly
thereafter. Following Shanks et al. (1998) we first discuss the reasons
for the differences seen between our SFR history and the original SFR
estimates of Madau et al. (1996).

\begin{table*}
\begin{minipage}{140mm}
\caption{UV luminosity  and SFR densities measured in the HDF and WHDF
($H_0=50 kms^{-1}Mpc^{-1}$). LF indicates luminosity function incompleteness corrected. A$_B=0.3$ indicates dust corrected. V$_{eff\
}$ indicates that the
effective volumes of Steidel et al. (1999) have been assumed, rather than the
boxcar volume, V. In the WHDF case, factors are V/V$_{eff}$ for 
q$_0$=0.5,0.05 respectively. }
\halign{\rm#\hskip10pt\hfil&\hfil\rm#\hskip5pt\hfil&
\rm\hfil#\hskip5pt\hfil&\rm\hfil#\hskip5pt\hfil&
\rm\hfil#\hskip5pt\hfil&\rm\hfil#\hskip5pt\hfil&
\rm\hfil#\hskip5pt\hfil&\rm\hfil#\hskip5pt\hfil\cr

&$\rho_{1500} $&log $\rho^* $ &log $\rho^* $&log $\rho^* $&log $\rho^* $&log $\rho^* $&log $\rho^* $\cr
&$(q_0=0.5)$&$(q_0=0.5)$ &$(q_0=0.05)$&$(q_0=0.5,$&$(q_0=0.05,$&$(q_0=0.5, LF,$&$(q_0=0.05, LF,$\cr
&&&&$LF)$&$LF)$&$A_B=0.3)$&$A_B=0.3)$\cr
\cr
WHDF B$<26$ $(1\la z\la2)$                     &$8.9\times10^{25}$ &$-2.09$&$-2.29$&-1.80&$-2.13$&$-1.48$&$-1.80$\cr
WHDF R$<25.5$ $(1\la z\la2)$                   &$1.0\times10^{26}\dagger$&$-2.02$&$-2.22$&-1.74&$-2.06$&$-1.55$&$-1.87$\cr
WHDF  U dropouts$(2.5\la z\la3.5)$             &$3.0\times10^{25}$ &$-2.55$&$-2.80$&-1.94&$-2.18$&$-1.61$&$-1.85$\cr
WHDF  $''$  (V$_{eff}$, $\times$2.38,3.15) &$7.1\times10^{25}$ &$-2.17$&$-2.30$&-1.56&$-1.64$&$-1.23$&$-1.31$\cr
HDF-N U dropouts$(2.0\la z\la3.5)$   &$2.1\times10^{26}$ &$-1.70$&$-1.95$&-1.40&$-1.73$&$-1.07$&$-1.40$\cr
HDF-N $''$  (V$_{eff}$, $\times$1.13)&$2.4\times10^{26}$ &$-1.65$&$-1.90$&-1.35&$-1.68$&$-1.02$&$-1.35$\cr
HDF-S U dropouts$(2.0\la z\la3.5)$   &$2.5\times10^{26}$ &$-1.62$&$-1.87$&-1.32&$-1.65$&$-0.99$&$-1.32$\cr
HDF-S $''$  (V$_{eff}$, $\times$1.13)&$2.8\times10^{26}$ &$-1.57$&$-1.82$&-1.27&$-1.60$&$-0.94$&$-1.27$\cr
WHDF  B dropout$(3.5\la z\la4.5)$              &$2.2\times10^{25}$ &$-2.68$&$-2.94$&-1.98&$-1.92$&$-1.65$&$-1.59$\cr
WHDF $''$   (V$_{eff}$, $\times$1.94,2.68) &$4.2\times10^{25}$ &$-2.39$&$-2.51$&-1.69&$-1.49$&$-1.36$&$-1.16$\cr
HDF-N B dropout$(3.5\la z\la4.5)$    &$8.1\times10^{25}$ &$-2.11$&$-2.36$&-1.89&$-2.17$&$-1.56$&$-1.84$\cr
HDF-N $''$  (V$_{eff}$, $\times$1.19)&$9.6\times10^{25}$ &$-2.03$&$-2.28$&-1.81&$-2.09$&$-1.48$&$-1.76$\cr
HDF-S B dropout$(3.5\la z\la4.5)$    &$7.6\times10^{25}$ &$-2.14$&$-2.33$&-1.92&$-2.14$&$-1.59$&$-1.81$\cr
HDF-S $''$  (V$_{eff}$, $\times$1.19)&$9.0\times10^{25}$ &$-2.06$&$-2.25$&-1.84&$-2.06$&$-1.51$&$-1.73$\cr
$\dagger\/\rho_{2800} $&&&&&&\cr
}
\label{tab:sfrz}
\end{minipage}
\end{table*}

The approach of Madau et al. (1996), following Cowie et al. (1988) and
Lilly et al.(1995,1996), uses the fact that the galaxy UV flux density
is proportional to the SFR density to obtain estimates of the SFR
history of the Universe. They use the CFRS spiral evolving LF of Lilly
et al. (1995) to determine the 2800\AA~ flux density in the range
$0<z\la1$ and the HDF $F300W$ and $F450W$ dropouts to obtain the 1620\AA~
flux density in the ranges $2\la z\la3.5$ and $3.5\la z\la4.5$. 
Their conclusion
of a peak in the SFR at $z\approx1$ is in contradiction
with our $\tau$=9~Gyr exponentially increasing SFR for spirals, which
we have shown fits a variety of faint galaxy count, colour and
redshift data over a wide range of passbands. The SFR evolution rate
of Madau et al. increased markedly faster than our exponential in the
$0<z\la1$ range and then quickly decreased below our rate at $z\ga1$ (see
fig. 4 of Shanks et al. 1998 and also Fig. \ref{fig:sfr} below).

At $z=0$, Shanks et al. suggested that the reason our SFR density lies
above that of Gallego et al. (1995) may be related to the bright $B$-band
count normalisation issue, discussed above. Supported by the results
of Glazebrook et al. (1995) and Driver et al. (1995), we explicitly
ignore the low galaxy count at $B\la17$ mag as being possibly contaminated
by the effects of local large-scale structure. Further supporting
evidence comes from the $B<19.5$ mag 2dF galaxy redshift survey results
of e.g. Colless (1999) where the luminosity function at low $z$
appears to move in the density rather than the luminosity direction, as
would be expected on the basis of the large-scale structure hypothesis.

In considering possible reasons first for the differences in the
$0.2\la z\la1$ range, fig. 5 of Shanks et al. showed that our exponential
models reasonably fit the rest $B$-band luminosity functions determined from
the CFRS data. However, the worst fit of our evolving luminosity
function to the CFRS data was in the $0.2<z<0.5$ spiral bin and this
probably accounts for most of the discrepancy between our model SFR
density and the CFRS data in this range. Further uncertainty is
introduced because the 2800\AA~ flux density in this bin has to be
extrapolated from the observed $B$-band data (Lilly et al. 1996). Other
measures of the SFR from the OII measurements of Tresse \& Maddox (1998)
and Treyer et al. (1998) at z$\sim$0.2 agree much better with our
exponential model at low redshift.

In the $2\la z\la3.5$ range, the reasons that the SFR density estimate of Madau et
al. lies a factor of $\approx$3 below our model in fig. 4 of Shanks et. al.
(1998) are also clear. First, if we repeat the Madau et al. measurement in the
HDF-N, then we obtain a factor of 1.56$\times$
bigger luminosity density at 1620\AA~ for $z\ga2$ UV drop-out galaxies than
these authors, mostly due to the fact that we measure brighter magnitudes for
individual galaxies (see Fig. \ref{fig:stsci comparison}). Second, if Madau et
al. were to include our $A_B($z=0$)=0.3$ mag dust absorption for spirals, then
this would effectively increase their 1620\AA~ luminosity density by another
factor of 2.12. Combined, these two effects give a factor of $3.3\times$
overall which explains the above discrepancy of approximately a factor of 3.
In the $3.5\la z\la4.5$ bin the same two effects apply. However, the discrepancy
here with our model is so large that we shall suggest below that 
they may in fact be in genuine disagreement.

In Fig. \ref{fig:sfr}(a,b) we illustrate some of these points by
comparing various observational estimates of the SFR density with our
evolutionary models for both the q$_0$=0.05 ($\tau$=9Gyr,
t$_{galaxy}$=16Gyr, $h_{50}=1$) and the q$_0$=0.5+dE ($\tau$=9Gyr,
t$_{galaxy}$=12.7Gyr, $h_{50}=1$) models. In (a) the discontinuity arises
from the contribution of the `disappearing dwarfs with their constant
SFR at z$>$1. The $z=0$ normalisation used in both cases was obtained by
first estimating the local luminosity density in the $F300W$ passband
from the Sbc/Scd/Sdm Schechter parameters in Table \ref{tab:model lf}. This
gave $\rho_{3000A}=9.8\times10^{25} h_{50}^{-3}$ erg s$^{-1}$
Hz$^{-1}$ Mpc$^{-3}$. We then converted to SFR density using the factors
given by Madau et al. (1996) (corrected downwards by a factor of 1.25
according to the extrapolation to 3000\AA~ from 1500-2800\AA~ and then
downwards again by the factor 1.6 for the reasons suggested by
Connolly et al. 1997) to obtain $\rho_*=0.00735 M_{\odot} h_{50}^{-3}$
yr$^{-1}$ Mpc$^{-3}$. Finally we corrected this SFR(z) normalisation
for our inclusion of A$_B=0.3$ mag dust absorption. Assuming our
1/$\lambda$ dust model this amounts to an upwards correction by a
factor of 1.54.

\begin{figure}
\begin{center}
\centerline{\epsfxsize = 3in
\epsfbox{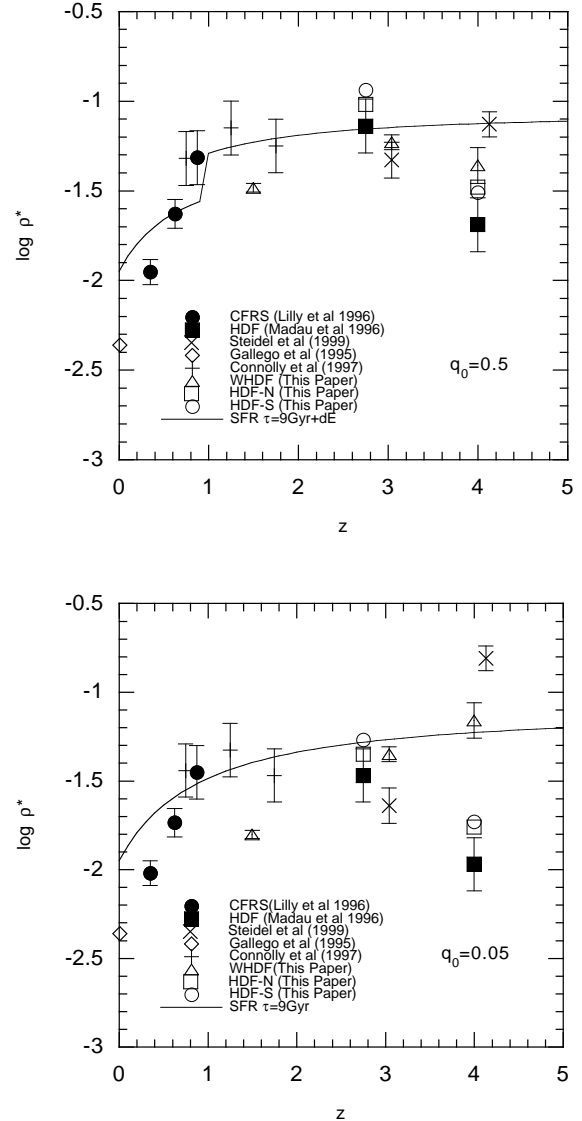}}
\caption{(a) Star formation rate versus redshift in the case
q$_0$=0.5. The solid line represents our $\tau=9$Gyr spiral
evolutionary model, normalised to include our dust absorption
correction, with the contribution to the SFR density from the dE
population added in at z$>1$. This model is compared to data points
from Lilly et al. (1996), Madau et al. (1996). Steidel et al. (1999),
Gallego et al. (1995), Connolly et al. (1997) and from our estimates
from the WHDF and the HDF. All data are corrected for LF
incompleteness and A$_B=0.3$ mag dust absorption. Where appropriate the SFR
densities use the effective volumes of Steidel et al. (1999) (see Table \ref{tab:sfrz}).  
(b) As for (a) for the case q$_0$=0.05. This time the solid line simply
represents the $\tau=9$Gyr spiral evolutionary model.
}
\label{fig:sfr}
\end{center}
\end{figure}

All observational points based on $\rho_{2800A}$ are first corrected
downwards by the factor of 1.6 recommended by Connolly et al
(1997). They are then corrected using the A$_B=0.3$ mag dust absorption
as used in our model. Further, we have LF corrected each point either
by taking the published correction or, where these are absent, as in
the case of the HDF-N data of Madau et al. (1996), by running the
q$_0$=0.05/ q$_0$=0.5+dE evolutionary models in each redshift range
and integrating the galaxy luminosity density to convergence. We also
apply these LF corrections to our own data. Thus if our model and the
data are fully consistent, then they should agree in this Figure. As
well as the points of Gallego et al. and the CFRS points of Lilly et al.
(1996) (both dust/LF corrected), we have also plotted the dust/LF
corrected HDF-N points of Connolly et al. (1997) in the $1\la z\la2$ range.

Also shown in Fig. \ref{fig:sfr} are our new WHDF estimates
of the SFR density at $1\la z\la2$. These were obtained by integrating the
luminosity density of galaxies with $(b-r)_{ccd}<0.6$ and $(r-i)_{ccd}>0.4$ 
which are
approximately where the $1\la z\la2$ galaxies lie in Fig. \ref{fig:colour
tracks}. The 343 galaxies with $b_{ccd}<26$ and the 267 galaxies with
$r_{ccd}<25.5$ give the values for the UV luminosity density and SFR
densities shown in Table \ref{tab:sfrz}; these are somewhat lower than
given by Connolly et al. particularly in the q$_0$=0.05 case. However,
the colour cuts we have used are relatively crude and this may be responsible
for any difference with the HDF results and our own data.

In the redshift range $2.5\la z\la3.5$, Fig. \ref{fig:sfr} shows the HDF-N UV 
dropout data of Madau et al. (1996), corrected for our A$_B=0.3$ mag dust 
absorption and our LF incompleteness corrections. Also shown is the Keck UV 
dropout galaxy data of Steidel et al. (1999), with our dust (A$_B=0.3$ mag 
with $1/\lambda$ giving A$_{1500}=0.9$ mag) and LF incompleteness corrections.  
With their own dust ($E(B-V)_{Calzetti}$=0.15 giving A$_{1500}=1.7$ mag) and 
LF incompleteness corrections, Steidel et al. (1999) find their $z=3$ SFR 
is a factor of 3.4 higher than shown in Fig. \ref{fig:sfr} (a).  
Fig. \ref{fig:sfr} also gives new estimates of the SFR density at 
$2.5\la z\la3.5$ from WHDF $u$-band dropouts to $r_{ccd}<$25.25
together with the estimates from the data of Steidel et al. (1999).  
Here we have taken the 43 UV dropout candidates we identified in Section 6.4
to give the value of $\rho_{1500}$ shown 
in Table \ref{tab:sfrz}. Again using the conversion factors stated in the 
paper of Madau et al. this gives the SFR densities shown in 
Table \ref{tab:sfrz}. The dust/LF corrected estimate is
plotted in Fig. \ref{fig:sfr}. We have corrected our WHDF data upwards by 
a factor of 2.38 in the case of q$_0=0.5$, 3.15 in the case of q$_0=0.05$, 
to use the effective volumes (including their colour 
incompleteness/deblending correction) of Steidel et al. (1999).  The observed WHDF 
SFR density estimate is close to the observed value from 
Steidel et al. (1999), when our corrections are
consistently used.  We similarly have plotted our estimates of the SFR at 
$2\la z\la3.5$ from the HDF-N and HDF-S UV dropouts using the same criteria as 
used by Madau et al. (1996).  All HDF SFR's at $z=2.75$ have been corrected 
upwards by a factor of 1.19 to account for colour incompleteness/blending as 
estimated by Steidel et al (1999). 
The HDF points agree well amongst themselves and with the model and are close
to the WHDF and Steidel et al. points which have bigger LF incompleteness
corrections.

Further, we have plotted estimates of SFR density at $3.5\la z\la4.5$
using $F450W$ dropout galaxies from the HDF-N analysis by Madau et al. (1996),
from the HDF-N and HDF-S analyses by ourselves using identical selection criteria, 
and from the WHDF, using the criteria of Steidel et al. (1999), transformed
into our magnitude systems. These are all corrected using our dust and LF
incompleteness corrections (see Table \ref{tab:sfrz}). The WHDF data
is also corrected upwards by the factor 1.94 in the case of q$_0=0.5$, 2.68
in the case of q$_0=0.05$, to use the effective 
volume of Steidel et al. (1999).  All HDF data has also been
corrected by a factor of 1.19 to use the effective volume recommended by
Steidel et al. (1999).  Also plotted are the SFR densities from Keck $B$
dropout galaxies from Steidel et al. (1999); these are based on 49
$3.5\la z\la4.5$ galaxies with spectroscopic confirmation. Again the
points shown assume our dust and LF corrections. Using their own LF
and dust absorptions would result in a factor of 1.7 increase in the SFR at
$z=4$ over the value shown in Fig. \ref{fig:sfr} (a).  Although Steidel et al.
include $1.7$ magnitudes of dust absorption at 1500\AA~ as opposed to our 
0.9 magnitudes, their use of
the effectively gray extinction law of Calzetti (1997) means that they
include $1.2$ magnitudes of dust at 2800\AA~ compared to our $0.5$ magnitudes,
so the effect
of their higher dust absorption is mainly on the normalisation of the
SFR:z plot rather than its form (compare their fig. 9 to Fig. 
\ref{fig:sfr} (a)). In Fig. \ref{fig:sfr}(a),(b) the WHDF points 
are close to both models at $z=4$ but the Steidel et al. point is higher 
than the
q$_0$=0.05 model at $z=4$ due to a large LF correction. However, in
all cases the HDF points at $z=4$ are lower, even though we have applied
both the dust and LF incompleteness corrections consistently. We shall
return to discuss this discrepancy between the HST and ground-based
SFR estimates at $z=4$ in Section 6.8.

We conclude that, within the various errors and correction factors,
there is reasonable self-consistency with the fully corrected SFR
density data and our model assuming an exponentially increasing SFR
with look-back time, at least to $z\sim3$. At $3.5\la z\la4.5$, the
ground-based data of Steidel et al. and ourselves is generally in much
better agreement with our exponential model than the HDF data.
Steidel et al. (1999) have suggested that this may be because the HDF-N
is too small to be representative. However, we note that the $B$ dropout
data from the HDF-S show good agreement with the HDF-N
SFR density. Therefore the picture at high redshift based on the
integrated measure of UV luminosity and SFR density remains
unclear. We note that because of the large size of the LF
incompleteness and dust corrections, even where our model and the SFR
density data agree, this only really checks the internal consistency
of our model and the SFR data. In Sect. 6.8 we shall make a more
direct test of our models than afforded by Fig. \ref{fig:sfr}.

Probes of the SFR at high $z$ using SCUBA sub-millimetre observations
(Smail et al. 1997, Hughes et al. 1998, Barger et al. 1998) have
also suggested that dust may re-radiate absorbed UV luminosity and thus 
provide an explanation for the sub-mm counts. 
Busswell \& Shanks (2000) have shown
that dust re-radiated UV flux in our evolving spirals, even though the
$B$-band absorption is as low as 0.3 magnitudes, can also make a significant
contribution to the steep sub-mm counts. Thus our spiral evolution
models which include dust may be enough in themselves to explain the
sub-mm counts, at least at  faint, $\sim 1$mJy fluxes.

\subsection{Counts of UV and $B$ dropout galaxies and the epoch of galaxy formation}

\begin{table}
\caption{WHDF Galaxy counts for UV dropouts $2.5<z<3.5$ 
(from section 6.4). 
These are the raw data and have to be multiplied by a factor of 1.3 to 
account for colour incompleteness-blending after Steidel et al. (1999).}
\halign to\hsize{%
\hfil\rm#\hfil&\hskip 10pt\hfil\rm#&\hskip 10pt\rm\hfil#\hss\cr
Magnitude&\multispan2\hskip 10pt\hss N$_{gal}$\hss\cr
($r_{ccd}$)&(per frame)&(arcmin$^{-2}$)\hss\cr
\noalign{\vskip10pt}
23.0-24.0&1\hskip 15pt &$0.01\pm0.01$\cr
24.0-24.5&8\hskip 15pt&$0.16\pm0.06$\cr
24.5-25.0&22\hskip 15pt&$0.45\pm0.10$\cr
25.0-25.25&12\hskip 15pt&$0.49\pm0.14$\cr
}
\label{tab:whdfuvdrop counts}
\end{table}

\begin{table}
\caption{HDF-N \& HDF-S $F606_{vega}$ galaxy counts for UV dropouts 
$2<z<3.5$ 
(from Metcalfe et al. 1996). These are the raw data and have to be multiplied
by a factor of 1.19 to account for colour incompleteness/blending after 
Steidel et al. (1999).}
\halign to\hsize{
\hfil\rm#\hfil&\hskip5pt\hfil\rm#\hfil&\hskip5pt\hfil\rm#\hfil&
\hskip15pt\hfil\rm#\hfil&\hskip5pt\hfil\rm#\hfil&\hskip5pt\hfil\rm#\hfil\cr

Magnitude&\multispan2\hfil N$_{gal}\hfil$&
Magnitude&\multispan2\hfil N$_{gal}\hfil$\cr

($F606_{vega}$)&(N)&($arcmin^{-2}$)&
($F606_{vega}$)&(N)&($arcmin^{-2}$)\cr

HDF-S&&&HDF-N\cr
23.0-23.5& 3&0.56$\pm$0.19&23.0-23.5& 1&0.19$\pm$0.19\cr
23.5-24.0& 0&0.00 &23.5-24.0& 0&0.00\cr
24.0-24.5& 4&0.75$\pm$0.38&24.0-24.5& 4&0.75$\pm$0.38\cr
24.5-25.0&14&2.63$\pm$0.70&24.5-25.0& 8&1.50$\pm$0.53\cr
25.0-25.5&19&3.56$\pm$0.82&25.0-25.5&24&4.50$\pm$0.92\cr
25.5-26.0&39&7.31$\pm$1.17&25.5-26.0&35&6.56$\pm$1.11\cr
26.0-26.5&19&3.56$\pm$0.82&26.0-26.5&28&5.25$\pm$0.99\cr
26.5-27.0& 4&0.75$\pm$0.38&26.5-27.0& 7&1.31$\pm$0.50\cr
}
\label{tab:hdfuvdrop counts}
\end{table}

\begin{figure}
\begin{center}
\centerline{\epsfxsize = 3in
\epsfbox{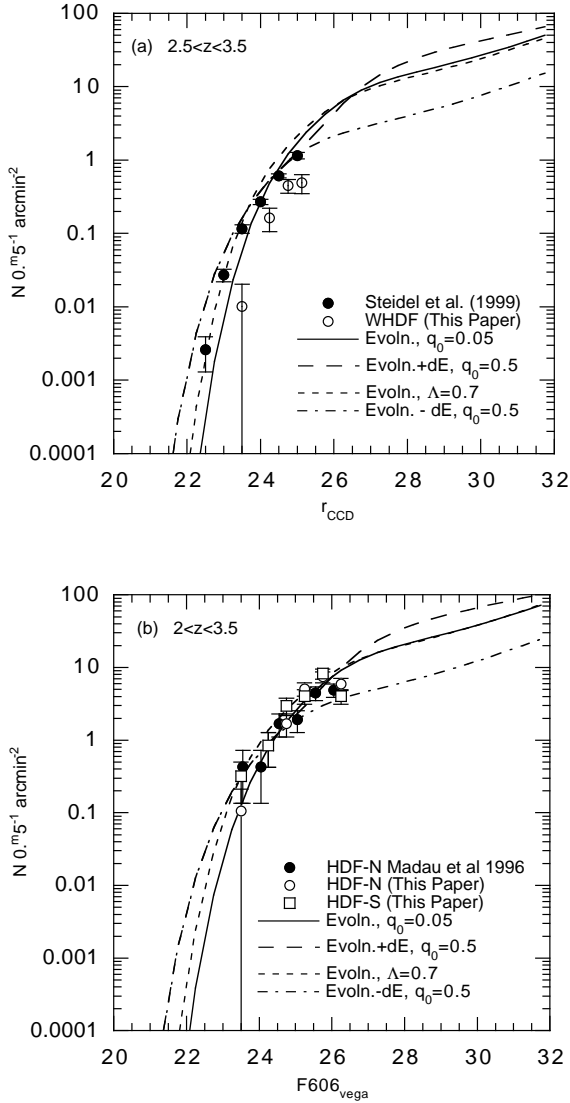}}
\caption{(a) The $r_{ccd}$ number count of $2.5<z<3.5$ ground-based UV dropout
galaxies, from Steidel et al. (1999) and from the WHDF, compared to our 
evolutionary model predictions with q$_0$=0.05, 0.5 and $\Omega_\Lambda$=0.7.
In the q$_0$=0.5 case the contribution of the dE population can be seen at 
$r_{ccd}>25$.  A colour incompleteness/blending factor of 1.3 has been 
applied to 
the WHDF data, the same factor used to correct the data of 
Steidel et al. (1999).  (b) The $F606_{vega}$ number count of
$2<z<3.5$ HDF UV dropout galaxies compared to these same models. A colour
incompleteness/blending 
factor of 1.19 has been applied to  all the HDF data as 
suggested by Steidel et al. (1999). In all cases the agreement with  
the models is good.  }
\label{fig:nmuvdrop}
\end{center}
\end{figure}

\begin{figure}
\begin{center}
\centerline{\epsfxsize = 3in
\epsfbox{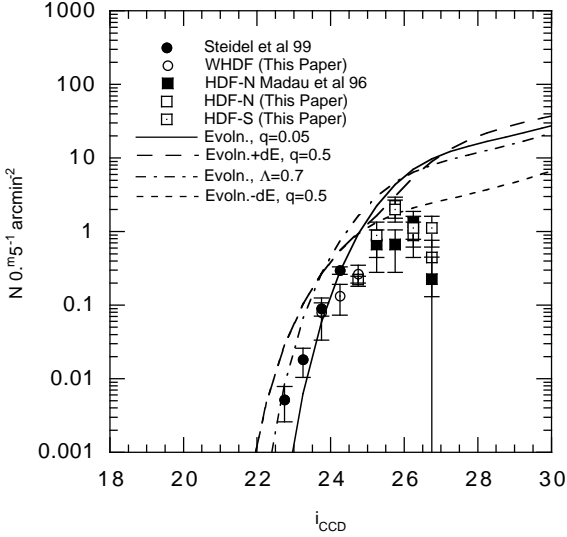}}
\caption{The $i_{ccd}$ number count of $3.5<z<4.5$ $B$-band dropout galaxies 
from the ground-based data of Steidel et al. (1999) and from the WHDF and 
also from the HDF-N data of Madau et al. (1996) and the HDF-N and HDF-S 
data from 
this paper.  A colour incompleteness/blending factor of 1.3 has been
applied to the ground-based data and a factor of 1.19 to the HDF data. 
These data are compared to our evolutionary model predictions 
with q$_0$=0.05, 0.5 and $\Omega_\Lambda$=0.7. 
In the q$_0$=0.5 case the contribution of the dE population can be seen 
at $i_{ccd}>25$ and the prediction for the q$_0$=0.5 model without the 
dE population is also shown.  The observed
faint galaxy LF appears to fall below our model predictions by a factor 
of $\approx$5, indicating that while bright L$^*$ galaxies may have been 
in place at $z\approx$4, the faint galaxy population may have formed 
between $z\approx3$ and $z\approx4$. } 
\label{fig:nmbdrop}
\end{center}
\end{figure}

In order to clarify the status of our model predictions at $2.5\la z\la4.5$ we now
consider Figures \ref{fig:nmuvdrop} and \ref{fig:nmbdrop} where we have 
plotted the number-magnitude counts of UV dropout (as defined in section 6.4) 
and B dropout (see section 6.7) galaxies 
and compared them to our model predictions. The count predictions  for 
the q$_0$=0.05, q$_0$=0.5(+dE) and $\Lambda$ evolutionary models are shown 
and their shapes are similar to the galaxy luminosity functions at high 
redshift. The upturn at M$^*$ in the q$_0$=0.5 case is
due to the extra contribution from the disappearing dwarf population at high
redshift. Taking Fig. \ref{fig:nmuvdrop}(a) first, which applies to the 
counts of ground-based UV dropouts in the redshift range $2.5\la z\la3.5$, plotted 
as a function of $r_{ccd}$ magnitude, we see that, although the new WHDF 
count (Table \ref{tab:whdfuvdrop counts}) is slightly lower than the Keck 
data from Steidel et al. (1999), overall the data agree well with the model 
predictions at bright magnitudes. Similarly, in Fig. \ref{fig:nmuvdrop}(b), 
the HDF-N UV dropout data from Madau et
al. (1996) and our own HDF-N and HDF-S analyses 
in the $2\la z\la3.5$ range agree well 
with the models at fainter $F606_{vega}$ magnitudes. In 
Fig. \ref{fig:nmuvdrop} we have
adopted the average colour incompleteness/blending factors suggested by 
Steidel et al. (1999) which  means that both the Steidel et al. and the 
WHDF number counts are increased by a factor of 1.3. We conclude that the 
luminosity function of galaxies in the
redshift range $2\la z\la3.5$ appears reasonably consistent with the local galaxy
luminosity function of Sbc/Scd/Sdm galaxies (which dominate at
high redshifts), as modified by the effects of $\tau$=9Gyr evolution plus dust.
In the particular case of the q$_0$=0.5 model,
the data seem more consistent with a model which 
includes the dE population rather than the same model without the dE 
population (also shown in Fig. \ref{fig:nmuvdrop}). Thus
there is strong evidence that the space density of galaxies at low redshift and
$z\sim3$ may be very similar, whatever the underlying model, and this may be a
stronger conclusion than any drawn from Fig. \ref{fig:sfr} above which 
frequently have to rely on large LF incompleteness corrections at high 
redshift to compare SFR densities with low redshift measurements. 
Fig. \ref{fig:nmuvdrop} suggests that not
only may the galaxy luminosity density be constant in the range $1\la z\la3$ but the
space density and luminosity function of galaxies may also be reasonably 
unchanging in this redshift range. It also suggests that the Lyman-break 
galaxies detected by Steidel et al. may be interpreted as the evolved bright 
tail of the local Sbc/Scd/Sdm
galaxy luminosity function. These conclusions would not be altered by 
assuming the Calzetti (1997) dust law with $E(B-V)=0.15$ since the 
differential 
extinction between 1500\AA~ at $z=3$ and the $B$ band at $z=0$ is similar to 
that  found in our $1/\lambda$ dust model with $A_B=0.3$ mag.

We now turn to Fig. \ref{fig:nmbdrop}  where the counts of $B$-band dropout
galaxies from Steidel et al. (1999), Madau et al. (1996) in the HDF-N and 
ourselves in HDF-N and HDF-S and in the WHDF in the range $3.5\la z\la4.5$ are plotted 
as a function of $i_{ccd}$ magnitude. Again, the data is compared to the
same four evolutionary models as in Fig. \ref{fig:nmuvdrop}. Although the three
basic models still fit the brighter data of Steidel et al. (1999), they  
overpredict the fainter WHDF and HDF data by about a factor of 5.  At 
$z\sim4$, the data seem more similar to the fourth model shown, which is the 
q$_0$=0.5 model without the extra dE
population.  Thus the overall conclusion seems to be that at $z\approx4$  the
observed galaxy luminosity function appears to drop significantly below our 
basic models at absolute magnitudes fainter than M$^*$.  

One interpretation would be that the assumption of a constant space 
density of galaxies in our models had failed by $z\sim4$ and that we were 
possibly detecting the epoch of galaxy formation. It might also be suggested 
in the context of the q$_0$=0.5 model that the results might imply that the 
bright galaxies were in place even as early as $z\approx4$ and the dE 
population appeared between $3\la z\la4$ and then
dimmed to become the dwarf galaxy population possibly  detected at $b<25$ 
in the WHDF $(r-i)_{ccd}:(b-r)_{ccd}$ colour-colour diagram. 
This may be similar to 
the `downsizing' scenario claimed by other observers at  lower redshifts 
(e.g. Cowie et al. 1996).

\begin{table}
\caption{$i$-band differential galaxy counts for $b$-band dropouts on
the WHDF. These are the raw data and have to be multiplied
by a factor of 1.3 to account for colour incompleteness/blending
 after Steidel et al. (1999).}
\halign to\hsize{%
\hfil\rm#\hfil&\hskip 10pt\hfil\rm#&\hskip 10pt\rm\hfil#\cr
Magnitude&\multispan2\hskip 10pt\hss N$_{gal}$\hss\cr
($i_{ccd}$)&(per frame)&($arcmin^{-2}$)\hss\cr
\noalign{\vskip10pt}
23.0-23.5& 0\hskip 10pt&$0.00$\cr
23.5-24.0& 3\hskip 10pt&$0.061\pm0.04$\cr
24.0-25.5& 5\hskip 10pt&$0.10\pm0.05$\cr
24.5-25.0&10\hskip 10pt&$0.20\pm0.06$\cr
}
\label{tab:whdfbdrop counts}
\end{table}

\begin{table}
\caption{Differential galaxy counts for $F450W-band$ dropouts from the 
HDF-N and HDF-S fields.These are the raw data and have to be multiplied
by a factor of 1.19 to account for colour incompleteness/blending
after Steidel et al. (1999).}
\halign to\hsize{%
\hfil\rm#\hfil&\hskip5pt\hfil\rm#\hfil&\hskip5pt\hfil\rm#\hfil&
\hskip15pt\hfil\rm#\hfil&\hskip5pt\hfil\rm#\hfil&\hskip5pt\hfil\rm#\hfil\cr

Magnitude&\multispan2\hfil N$_{gal}$\hfil&
Magnitude&\multispan2\hfil N$_{gal}$\hfil\cr

($F814_{vega}$)&(N)&($arcmin^{-2}$)&
($F814_{vega}$)&(N)&($arcmin^{-2}$)\cr
HDF-S&&&HDF-N\cr
24.0-24.5& 0&0.00   & 24.0-24.5& 3&0.56$\pm$0.32\cr
24.5-25.0& 1&0.19$\pm$0.19 & 24.5-25.0& 0&0.00\cr
25.0-25.5& 4&0.75$\pm$0.38 & 25.0-25.5& 0&0.00\cr
25.5-26.0& 9&1.69$\pm$0.56 & 25.5-26.0&10&1.88$\pm$0.59\cr
26.0-26.5& 5&0.94$\pm$0.42 & 26.0-26.5& 4&0.75$\pm$0.38\cr
26.5-27.0& 5&0.94$\pm$0.42 & 26.5-27.0& 2&0.38$\pm$0.27\cr
}
\label{tab:HDFbdrop counts}
\end{table}

\section{Discussion of Model Successes and Failures.}

Our basic conclusion is that the galaxy counts in the optical bands can be
well fit by simple Bruzual \& Charlot models with an exponentially
increasing rate of star-formation to $z\sim3$. In the low q$_0$ case, this type
of model can fit the data to $B\sim27$  whereas in the 
high q$_0$ case the model
only fits to $B\sim25$, just beyond the current limit for galaxy redshift
surveys. In this case a model with either a disappearing dwarf dE population or
an increased luminosity function slope at high redshifts must be invoked to
increase the counts at faint magnitudes. We have also shown that the models
produce a magnitude-size relation, which for either value of q$_0$, is
consistent with the data.

We have also presented new results from  the WHDF and  the HDF-N and HDF-S 
for the SFR
density of galaxies over the redshift range $1\la z\la3.5$ and the results from
this approach are broadly consistent with both previous data  and the above
results if effects such as internal dust absorption in spirals are included
in the models. However, other systematic effects may still dominate this
route to the SFR history. Also, with this approach the poor fit to the faint 
counts of the q$_0$=0.5 model using the local galaxy luminosity function
would be missed.

At high redshifts, the number of galaxies predicted by the simple models 
at $z\approx3$ agrees well with the number of UV dropout galaxies
detected at $R\sim25$ by ourselves and by Steidel et al. (1999).  Thus the
PLE models seem to continue to fit the data as far as
z$\approx$3. However, at $z\approx4$, the PLE models start to
overpredict the faintest data and this may suggest that dwarf  galaxies may 
have
formed at $z\approx4$. We regard the agreement between the spiral
models and the bright, Lyman break galaxy data as another indication 
that the spiral model is broadly correct in the range $1\la z\la3$ and therefore 
that the
star-formation rate follows our exponential model throughout this redshift
range.

We have seen that the agreement between these models and the WHDF and HDF
galaxy colours is particularly good in $(R-I):(B-R)$, where  two faint galaxy
populations are clearly seen corresponding to early- and late-type galaxies.
The features seen in the data match those predicted by the models so well that
this forms a strong argument that these models must be a reasonably accurate 
description of the evolving spectra of faint galaxies. Even in $(U-B):(B-R)$ 
they at least provide a good qualitative description of the data.

The outstanding issue for the early-type galaxy models is the need to use a
dwarf-dominated IMF (with a 0.5M$_\odot$ cut-off) for early-types in order
to reduce the number of $z>1$ galaxies predicted in $K<19$ redshift surveys.
The effect of using this IMF on the predicted optical colours is small. A
slightly more extended initial period of star-formation is required with our
dwarf dominated IMF($\tau$=2.5Gyr) than with Salpeter ($\tau$=1Gyr) to match
the move towards bluer $(B-R)$ colours at $z\sim0.5$ seen in the $(R-I):(B-R)$
diagram. However, we note that the Salpeter/$\tau$=1Gyr model would then
also overpredict the numbers of $z>1$ galaxies detected in optical ($B<24$)
redshift surveys too. Thus, over a wide range of wavelengths, the case 
against the Salpeter plus $\tau$=1Gyr model for early-types from the 
redshift surveys is strong. The alternative solution to the dwarf-dominated IMF
is a merging model as considered by other authors 
(e.g. Kauffmann \& Charlot 1998).
It remains to be seen whether the merging model will produce as good a fit
to the $(R-I):(B-R)$ data as the simple PLE model discussed here.

For the spiral model, the outstanding issue is that in the redshift range 
$0.5\la z\la1$
the predicted $(u-b)_{ccd}$ colours are up to 0.5 magnitudes too blue 
compared to the data. Since the $u$ and $b$ bands peak close together 
at 3800\AA~ and 4400\AA~ this implies quite a sizeable discrepancy in the 
redshifted spectral slope between 1500-2500\AA~, where the Bruzual and Charlot 
model predicts approximately $f_\lambda\propto\lambda^{-2.5}$ and 
the data would prefer $f_\lambda\propto\lambda^{0.5}$. $(u-b)_{ccd}$ 
is more affected than $(b-r)_{ccd}$
because the small width of the $u$ band means there is less smoothing 
and because the wide wavelength separation of $b$ and $r$ means that 
they simultaneously lie in
the range 1500-2500\AA~ for a shorter redshift span (at $z\sim1.7$). The
$(u-b)_{ccd}$ discrepancy is puzzling because the UV colours of star-forming
galaxies are governed by only two parameters, dust and the IMF.  We already have
A$_B=0.3$ mag dust in the model and because the reddening vector runs parallel to the
$(u-b)_{ccd}:(b-r)_{ccd}$ track at $z=1$, it would take a significant increase in
the amount of spiral dust to improve the fit and this would spoil the $n(b)$ and
$n(z)$ fits even at $b_{ccd}\sim25$.

We have experimented with including the Galactic 2200\AA~ dust absorption
feature but although this tends to help while it moves through $u$, it
then moves through $b$ making $(u-b)_{ccd}$ sharply redder and this tends to
make the agreement with the data even worse at $z\ga0.7$.  We also tried
using an x=3 IMF as for the early-types with $\tau$=9Gyr to redden the
$(u-b)_{ccd}$ colours but although this helped redden $(u-b)_{ccd}$ in 
the $0.5\la z\la1$
range, $(b-r)_{ccd}$ also became too red for any agreement with the data even
at low redshift. We conclude that the original parameters of
A$_B=0.3$ mag, $1/\lambda$, dust absorption and a Salpeter IMF remains
closer to the $(u-b)_{ccd}:(b-r)_{ccd}$ data than these alternatives. 
The fact that
the K-corrections seem to demand a steep UV slope for galaxies locally
might be thought to indicate an explanation for the $(u-b)_{ccd}$ problem
where the IMF might evolve with redshift. However, the UV spectra of
local spirals has much scatter and it has long been known that only
the bluest local galaxies (e.g. NGC4449) give spectra as steep as
predicted by the Bruzual \& Charlot models with an IMF that is a
single power-law.

G. Bruzual has suggested (priv. comm) that an upper mass cut-off of 5-6
M$_\odot$ in the spiral IMF rather than the 125 M$_\odot$ which we
conventionally use will flatten the spiral s.e.d. shortwards of 2000\AA~
and help take the colours closer to the data. He has supplied models which
show this effect on the $(u-b)_{ccd}:(b-r)_{ccd}$ diagram and 
it clearly helps the $(u-b)_{ccd}$
colours at $z=1$, while continuing to produce reasonable colours for a
spiral at $z=0$. A 5-6 M$_\odot$ cut corresponds to an IMF with no stars
earlier than about B2-B3 which might be plausible, particularly if more
massive stars are assumed to be preferentially cocooned in dust.  A
potential  problem with this or any other explanation which effectively
reduces the slope of the UV s.e.d. for spirals is  that the K+evolution
correction  at a given SFR may produce less bright galaxies at high
redshift, causing our evolutionary model to underpredict the $B$ counts, for
example. Increasing the SFR at early times to account for this might not be
possible because the galaxy would evolve to be too red for a spiral at $z=0$.

Flat UV spectra have also  been observed in the spectra of local starburst
galaxies. For example, Gonzalez Delgado et al. (1998) have obtained UV spectra
of four local starbursts with the Hopkins Ultraviolet Telescope and the
Goddard High Resolution Spectrograph on HST. These authors show that the
spectral energy distribution follows $f_\lambda \propto \lambda^{\beta}$
with $\beta=-0.84 \pm0.26$. This is much flatter from the $\beta=-2.6$ to
$-2.2$ they expect on similar grounds to our argument above. They, 
however, believe
that the prediction for $\beta$ is so stable that any deviation can be
immediately interpreted as being caused by dust absorption. They obtain an
average value of A$_B=0.73\pm0.11$ for their four starburst galaxies using
an LMC extinction law. They also  obtain the relation $E(B-V)=0.11\Delta\beta$
where $\Delta\beta$ is the difference between the observed and predicted
slope. In our case $\Delta\beta\sim$3 and so $A_B \sim1.3$. This is so large
that the dust would redden  the  galaxy colour predicted for an Sbc spiral
by 0.6 magnitudes 
and a prime motivation  for adopting the $\tau=9Gyr$ model, that it
gave a blue enough colour for a present day spiral, would then be lost.
Gonzalez Delgado et al. also detect SiIV and CIV absorption lines with
P Cygni profiles which is a signature of O-type and early B-type stars. They
suggest that this argues against a bright IMF cut-off earlier than B2. 

\begin{figure}
\begin{center}
\centerline{\epsfxsize = 3in
\epsfbox{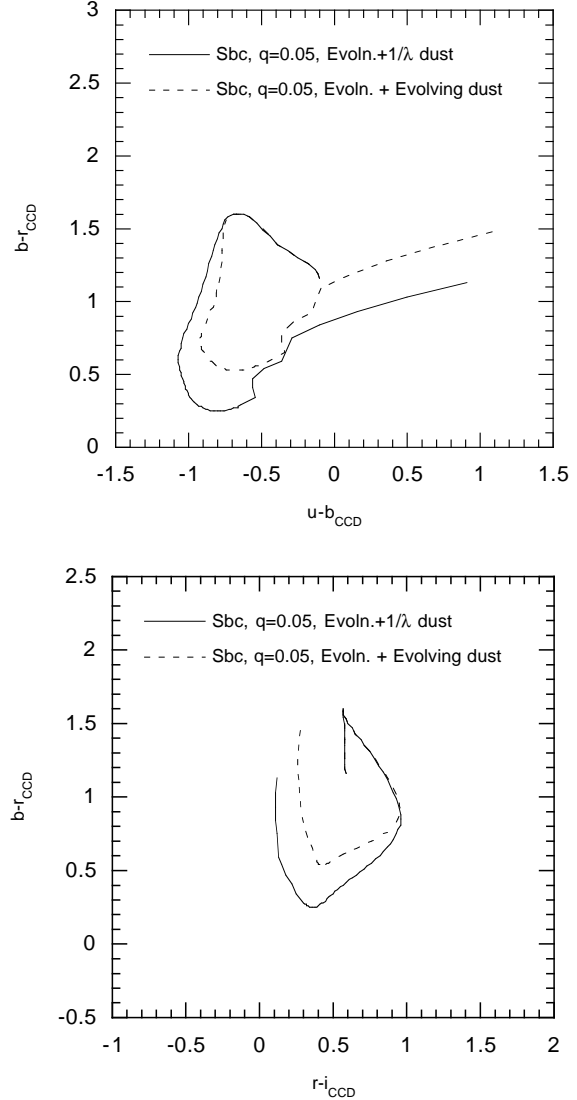}}
\caption{(a) The effect on our Sbc evolutionary model (q$_0$=0.05) for
$(u-b)_{ccd}:(b-r)_{ccd}$ of a dust reddening law 
which evolves from a Galactic law
exhibiting a 2200\AA~ absorption feature at low redshifts ($z<0.6$) to a
$1/\lambda$ extinction law at high redshifts ($z>0.6$). The evolution
helps to redden the model $z=1$ to agree better with the data.  (b) As
for (a) showing the effect of the evolving dust model on the model
predictions for $(r-i)_{ccd}:(b-r)_{ccd}$.  }
\label{fig:dustev}
\end{center}
\end{figure}

Steidel et al. (1999) have fitted reddenings to their Lyman-break
galaxies and find values very similar to those we use here,
finding an average $E(B-V)=0.15$ for an average $A_B\sim0.5$ mag using
the Calzetti (1997) absorption law. However, $E(B-V)=0.15$ would only
produce a spectral slope change of $\Delta\beta=1.2$, and this is less
than half what is needed to redden spiral $(u-b)_{ccd}$ colours to what is
observed in the WHDF. Pettini et al. (2000) have also detected P Cygni
profiles in the spectra of the lensed Lyman-break galaxy MS 1512-cB58
which again may indicate the presence of O-stars and argue against the
existence of an upper mass cut-off. In this galaxy the spectral index
implies $E(B-V)=0.10$ again close to what we have assumed in our spiral
models. However, this dust absorption implies a far higher dust-gas
ratio than seen in comparably low-metallicity galaxies such as the LMC
and SMC.

Our view is that it may be unreasonable to invoke dust absorption as
high as $E(B-V)=0.3$ or A$_B=1.3$ mag to explain the $(u-b)_{ccd}$ colours 
of the $z=1$
spirals in the WHDF. On the other hand it is clear that P-Cygni
profiles of SiIV and CIV are seen in both Lyman-break galaxies and nearby
starbursts. These features are seen in stars as late as B0 and a 5-6
M$_\odot$ cut would correspond to a cut at B2-B3. We are therefore
forced to consider another model where the evolution of the dust
extinction law is now invoked to explain the red $(u-b)_{ccd}$ colours of the
WHDF data. Noting that the 2200\AA~ feature helped redden $(u-b)_{ccd}$ as it
passes through the $u$ band, we suggest that the high metallicity
galaxies found at low redshift may therefore show a 2200\AA~ feature as
seen in the Milky Way. Then, since the predicted sharp feature caused
by the 2200\AA~ feature passing through the $b$ band is not observed, we
hypothesise that at higher redshifts than $z\approx0.6$ the
extinction law evolves to look more like the extinction law seen in
the low metallicity SMC, which is close to our 1/$\lambda$ law. In this
way, the strength of the 2200\AA~ feature produces increased reddening as
the $u$ (and $b$) bands move towards 2200\AA~ than in $(b-r)_{ccd}$ at 
higher redshift
when the 2200\AA~ feature has disappeared.  We have tested this
evolutionary model using the following simple evolutionary prescription to
mimic this effect in the $u$ band.

$A_\lambda=A_B\times4400/\lambda,\ \ \ \ \lambda > 2500\AA$

$A_\lambda=A_B\times(4400/\lambda+(9886/\lambda-3.9544)\times(1-exp(-t_{exp}/t) )),\ \ \ \ \lambda < 2500\AA$

\noindent where $t_{exp}$=8Gyr and $t$ is the look back time. This ensures a
smooth transition from the 2200\AA~ feature dominated law at low $z$ to the
$1/\lambda$ law at higher redshifts. This dust evolution then produces
a much redder $(u-b)_{ccd}$ at $z=1$ while not disturbing unduly the fit in
$(r-i)_{ccd}:(b-r)_{ccd}$ at either this or higher 
redshift (see Fig. \ref{fig:dustev}).

\section{Conclusions}

We have produced $u_{ccd},b_{ccd},r_{ccd}$ and $i_{ccd}$ galaxy number 
counts and $(u-b)_{ccd}$, $(b-r)_{ccd}$ 
and $(r-i)_{ccd}$ colours from one $7'\times7'$ field, on which we have the 
deepest ground-based $B$-band image yet taken. We also have deep $H$ and 
$K$-band photometry on this field (Paper IV, McCracken et al. in prep).

We have compared these data with our analysis of the HDF-N and HDF-S, produced
using the same software. For the UV and blue-bands, although the HDF's are
deeper for unresolved objects, in terms of surface brightness
signal-to-noise our WHT image is $\sim0.4$ mag deeper, and also covers 5
times the area. The counts from the datasets agree well, after the
ground-based data has been corrected for confusion. All are still rising at
the faintest magnitudes, with a slope $d\/log(N)/dm\sim0.25$. The HDF $F606W$
counts are over 2 magnitudes deeper than the previous equivalent 
ground-based limits.
They also show a still-rising count at the faint end with a slope
$d\/log(N)/dm\sim0.25$. The ground-based $U$ counts show an almost constant
slope of $d\/log(N)/dm\sim0.4$ to $U\sim25.5$. However the HDF $F300W$ counts
flatten to a slope $d\/log(N)/dm\sim0.15$.

The deepest counts are the HDF-S STIS data, which reach to $STIS_{AB}\sim30$, 
which is roughly equivalent to $R\sim30$. Our evolutionary models 
continue to fit to the limit of these deepest optical number count data.

In terms of the comparison with evolutionary models, we summarise our
conclusions  as follows:

1. We have fitted simple exponential star-formation history models to
our data, and found that low $q_0$ cosmologies give a good fit from
$U$ to $K$ over a range of $\sim10$ magnitudes. To enable high $q_0$
models to fit, an extra component must be added to the galaxy
population at high ($z\ga1$) redshift. One way of doing this is by
adding a `disappearing dwarf' dE population which is bright and blue at
high redshift and faint and red at low redshift. 

2. At very bright magnitudes ($B\la 17$) our count models fail and overpredict
the counts. Since the steep count at $B\la 17$ seems to be reproduced at the
same level at $R$, $I$, $H$ and $K$, which is unlikely for an evolutionary
explanation, this  represents strong evidence that the high normalisation we
have adopted for the counts is substantially correct.

3. The models provide a good fit to the distributions of colours in 
$(b-r)_{ccd}$ and $(F450-F606)_{vega}$ and in $(r-i)_{ccd}$ and 
$(F606-F814)_{vega}$, although in both $(u-b)_{ccd}$ and 
$(F300-F450)_{vega}$ the spiral models are too blue by $0.5$ 
magnitudes at $z=1$.

4. The same models produce elliptical and spiral tracks in the colour-colour
planes with shapes very similar to the loci of the data. The fact that 
the fit of our models is so good, particularly in $(r-i)_{ccd}:(b-r)_{ccd}$ 
for the ground-based data and $(F606-F814)_{vega}:(F450-F606)_{vega}$ for 
the HDF,
data lends strong support to the underlying assumption that the star-formation
rate increases exponentially to beyond $z=2$ (Metcalfe et al. 1996).
It is clearly possible to use $(R-I):(B-R)$ to probe the star-formation rate
directly over a wide redshift range $0.5\la z\la3$, emphasising the power of
this colour-colour diagram as a diagnostic tool for galaxy evolution.

5. In the WHDF $(r-i)_{ccd}:(b-r)_{ccd}$ plane, we detect an excess of low
redshift($z\approx 0.1$), low luminosity early-type galaxies which may
have parameters consistent with the `disappearing dwarfs' invoked in
our q$_0$=0.5 evolutionary model. 
 
6. We have shown how the discrepancy with the $z=1$ SFR peak of Madau
et al. (1996) comes from the different treatment of dust and photometry
in the HDF. When we repeat the UV dropout SFR density measurement made
in the $2\la z\la3.5$ range by Madau et al. we find good
self-consistency with the prediction of our exponentially increasing
SFR ($\tau$=9Gyr) spiral model.

7. We have made new measurements of the SFR density in the WHDF and in 
the HDF-N and HDF-S. In the redshift range $1\la z\la2$ the WHDF SFR density lies 
lower than the measurement of Connolly et al. (1997) in the HDF. These 
measurements bracket the prediction of our $\tau$=9Gyr spiral model. In 
the $2\la z\la3.5$ range our observed SFR density based on
UV dropouts, corrected for dust and LF incompleteness, and the ground-based
estimates of Steidel et al. (1999) lie close to the HDF estimates and our 
model. At $3.5\la z\la4.5$ there is a wider spread, with the brighter ground-based 
SFR estimates from the WHDF and Steidel et al. generally being higher and  
closer to the model than the HDF SFR estimates which are significantly 
lower than the model.

8. We have detected a population of galaxies that occupy the same part 
of the $(r-i)_{ccd}:(b-r)_{ccd}$ plane as $z\sim3$ UV dropout galaxies, but
which do have measurable $u$-band fluxes. We have suggested that these may 
be candidates for high redshift galaxies that do not show UV dropout 
and that statistics
of QSO absorption lines may not rule out such a population at $z\approx3$.
The increase in SFR density at $z\approx3$ will depend on the fraction 
which prove to be at high redshift.

9. We have created simulated CCD frames to check the apparent sizes
of galaxies predicted by our evolutionary models and found they are
quite consistent with the sizes of galaxies measured on the HDF-N.

10. For the early-type models, the main remaining problem is that
although they produce reasonable fits to the data at optical
wavelengths, they predict too many high redshift ($z\ga1$) galaxies at
$K\sim19$ unless they have a dwarf-dominated stellar IMF. This reduces
the evolutionary brightening at infra-red wavelengths whilst still
allowing sufficient optical evolution.  Dynamical merging may
constitute an alternative explanation and this possibility has been
explored by other authors (eg Kauffmann \& Charlot 1998).

11. The main outstanding question for the spiral models is why do they predict
$(u-b)_{ccd}$ and $(F300-F450)_{vega}$ colours which are too blue in the range
$0.5\la z\la1$? We have suggested that this problem is unlikely to be solved by the
addition of more dust to the models because the amount of dust needed is so 
great (A$_B\sim1.3$ mag) that it would lead to the model predicting $(B-R)$ 
colours that are $\sim$0.5 magnitudes too red for a $z=0$ spiral. 
One possibility is that the spiral
IMF is truncated above 5-6 M\sun. It is the subject of continuing 
research to see if such a model can produce enough evolution to continue 
to fit the steep
$B$-band number-count slope. Another possible explanation, suggested by the
dust laws in nearby galaxies with different metallicities, is that the dust
extinction law evolves, with the 2200\AA~ dust absorption feature 
appearing in high
metallicity galaxies  at $z\la0.5$ and disappearing in lower metallicity 
galaxies at higher redshift.

12. The evolutionary q$_0$=0.05 and q$_0$=0.5 models correctly predict 
the sky density of Lyman break $z\approx3$ galaxies detected by 
Steidel et al. (1999) from
Keck spectroscopy. The models suggest that these galaxies may be spiral 
galaxies
which have undergone an exponentially increasing SFR with look-back time. At
$z\approx4$ these models generally over-predict the space density of sub-L$^*$
galaxies at faint limits in the HDF-N and HDF-S and WHDF by a factor of 5, 
suggesting that
we may have reached an epoch of significant galaxy formation. The $z\approx4$
luminosity function matches best our q$_0$=0.5 model with the dE population 
removed,
suggesting that one interpretation might be that  bright L$^*$ galaxies 
were already in
place at $z=4$ but many fainter galaxies formed at $3\la z\la4$. 
These then quickly dimmed to become the red, dwarf population possibly 
detected in the WHDF $(r-i)_{ccd}:(b-r)_{ccd}$ data at M$_B=-15$ at 
$z\approx0.1$.

\section*{Acknowledgments}

HJMCC and NM acknowledge financial support from PPARC. The INT and WHT
are operated on the island of La Palma by the Isaac Newton Group at
the Spanish Observatorio del Roque de los Muchachos of the Instituto
de Astrof\'\i sica de Canarias. Data reduction facilities were
provided by the UK STARLINK project.

\label{lastpage}

\end{document}